\documentclass[12pt]{iopart}
\usepackage{amsfonts}
\usepackage{amssymb}

\newcommand{\no}{\nonumber\\}
\newcommand{\be}{\begin{equation}}
\newcommand{\ee}{\end{equation}}
\newcommand{\ba}{\begin{eqnarray}}
\newcommand{\ea}{\end{eqnarray}}
\newcommand{\ci}[1]{\cite{#1}}
\newcommand{\la}[1]{\label{#1}}

\def\gl#1{(\ref{#1})}

\usepackage{fancyhdr}

\begin{document}
\title[Nonlinear Supersymmetric]{Nonlinear Supersymmetric Quantum Mechanics: concepts and realizations }
\author{A A Andrianov\dag\S \ and M V Ioffe\dag}
\address{\dag\ V A Fock Department of Theoretical Physics, Saint-Petersburg
State University,
198504 Sankt-Petersburg, Russia}
\address{\S\ Departament d'Estructura i Constituents
de la Mat\`eria and Institut de Ci\`encies del Cosmos (ICCUB) ,
Universitat de Barcelona,  08028 Barcelona, Spain}

\ead{andrianov@icc.ub.edu;\quad m.ioffe@pobox.spbu.ru}

\begin{abstract}
Nonlinear  SUSY approach to spectral problems in Quantum Mechanics is reviewed. Its building from the chains (ladders) of linear SUSY systems is outlined and different one-dimensional and two-dimensional realizations are described.
For various two-dimensional models it is elaborated how Nonlinear SUSY provides two new methods of supersymmetric separation of variables.
In the framework of these methods partial and/or complete solution of some two-dimensional models becomes possible. The full
classification of ladder-reducible and irreducible chains of SUSY
algebras in one-dimensional QM is given. Emergence of hidden symmetries and spectrum generating algebras is elucidated in the context of Nonlinear SUSY in one-dimensional stationary and non-stationary as well as in two-dimensional QM.
\end{abstract}

\submitto{Journal of Physics A: Mathematical and Theoretical}

\maketitle


\newpage
\tableofcontents
\clearpage

\section{Introduction}
\hspace*{3ex} The concept  of Supersymmetric Quantum Mechanics (SUSY QM)
 embodies an algebraic form of transformations of (complete or partial)  spectral equivalence between different dynamical systems
and in this sense it gives the algebraic tools for spectral design of new quantum systems from a given set with controllable energy spectra. At present, there is a number of reviews
\cite{genkri}--\cite{ghosh} devoted to development and various applications, mostly of the {\it linear} SUSY QM.
The very isospectral transformations realizing SUSY represent the Darboux (Darboux-Moutard) transformations
 \cite{darb}--\cite{leble}, which
 are known in the theory of ordinary differential equations for a long time. According to \cite{berk} these transformations must be referred as Euler-Imshenetsky-Darboux-Moutard ones having their origin in the paper by Leonard Euler \cite{euler}. We thank our referee for drawing our attention to this chain of old papers.

The Supersymmetric Quantum Mechanics in its  form was formulated
for one-dimensional space by E.Witten \cite{witt} about thirty years ago(see also \cite{nico} -- \cite{coop}).
This approach has two predecessors: Factorization Method \cite{schr}
in Quantum Mechanics and Darboux transformation \cite{darb}--\cite{matv} for the Sturm-Liouville equation in Mathematics.

The first one was proposed by E.Schr\"odinger in 1940-ties \cite{schr} (see the excellent detailed review \cite{inf}),
and its basic relation is the representation of Hamiltonian in a factorized form,
\begin{equation}\label{factorization}
  h^{(0)}(x)=-\partial^2+V^{(0)}(x)=q^+q^-,
\end{equation}
where operators $q^{\pm}$ are,
\begin{equation}\label{q}
  q^{\pm}=\mp\partial + (\partial W )(x); \quad q^+=(q^-)^{\dagger }\quad \partial\equiv\frac{d}{dx}.
\end{equation}
The factorization (\ref{factorization}) is in general possible for an arbitrary potential $V(x)$
(see details below). The main idea of Factorization Method is to consider
together with the Hamiltonian
(\ref{factorization}) the partner Hamiltonian, also of Schr\"odinger form,
\begin{equation}\label{partner}
 h^{(1)}(x)=-\partial^2+V^{(1)}(x)=q^-q^+,
\end{equation}
with the same operators $q^{\pm}.$ By construction, these Hamiltonians satisfy the intertwining relations,
\begin{equation}\label{intertw}
  h^{(0)}q^+=q^+h^{(1)};\quad q^-h^{(0)}=h^{(1)} q^-,
\end{equation}
which play the key role in the method under discussion. Just the intertwining relations (\ref{intertw})
allow to obtain the simple relations between wave functions $\Psi^{(0)}_n$ and
$\Psi^{(1)}_n$ of $h^{(0)}$ and $h^{(1)},$ correspondingly. Indeed, according to (\ref{intertw}) an arbitrary
wave function $\Psi^{(0)}_n (x)$ of $h^{(0)}$ gives the wave function $\Psi^{(1)}_n$ of $h^{(1)},$
and vice versa,
\begin{equation}\label{connection}
q^-\Psi^{(0)}_n \propto \Psi^{(1)}_n; \quad q^+\Psi^{(1)}_n \propto \Psi^{(0)}_n .
\end{equation}
The energy eigenvalues for $\Psi^{(0)}_n$ and $\Psi^{(1)}_n$ are the same, $E^{(0)}_n=E^{(1)}_n.$ There are
three options \cite{abi1} for the spectra of $h^{(0)}$ and $h^{(1)}:$

- spectra coincide completely if neither $q^+,$ no $q^-$ have normalizable zero modes;

- spectra almost coincide: the spectrum of $h^{(0)}$ has one bound state less, if the normalizable zero mode of $q^+$ exists;

- spectra almost coincide: the spectrum of $h^{(1)}$ has one bound state less, if the normalizable zero mode
of $q^-$ exists.

This is an essence of the {\it Spectral Design} method based on the super-Hamiltonian $H$ with two non-singular real potentials. The  supersymmetry results in isospectral relations between the super-Hamiltonian components $h^{(0)}$ and $h^{(1)}$ whereas two non-singular real potentials are different.
Each of the operators $q^{\pm}$ may have one zero mode $\exp{(\pm W(x))}$, and their normalizability properties depend on
the behavior of function $W (x)$: the Hamiltonians $h^{(0)}$ and $h^{(1)}$ are \cite{abi1}--\cite{berez} either isospectral or almost isospectral
(up to one bound state).

The connection between a given potential $V^{(0)}(x)$ and a function $W (x)$  provides a basis for the link of  Factorization Method with Darboux transformation \cite{darb} and represents the analytic tool of Spectral Design.
The substitution of (\ref{q}) into (\ref{factorization}), (\ref{partner}) results in:
\ba
  V^{(0)}(x)&=&(\partial W)^2 (x)- (\partial^2 W) (x);\label{riccati1}\\
  V^{(1)}(x)&=&(\partial W)^2 (x)+(\partial^2 W) (x).\label{riccati2}
\ea
In terms of the theory of ordinary second-order differential Sturm-Liouville equations (adapted to the Schr\"odinger form),
\begin{equation}\label{sturm}
  -\Psi^{(0)\prime\prime}(x) + V^{(0)}(x)\Psi^{(0)} (x)=E^{(0)}\Psi^{(0)} (x),
\end{equation}
the Darboux transformation can be formulated as follows. If one knows
its particular solution $\psi^{(0)} (x)\equiv\exp{(-W(x))}$ for the value $E^{(0)}=0,$
the Eq.(\ref{riccati1}) is satisfied, and an arbitrary solution $\Psi^{(0)} (x; E^{(0)})$ of (\ref{sturm})
can be transformed (Darboux transformation) into a solution of the transformed equation
of the same form and with the same value of parameter $E^{(0)}$,
\begin{equation}\label{darboux}
-\Psi^{(1) \prime\prime}(x) + V^{(1)}(x)\Psi^{(1)} (x)=E^{(0)}\Psi^{(1)} (x).
\end{equation}
The initial and transformed solutions are connected,
\ba
\Psi^{(1)} (x; E^{(0)}) &\propto& \biggl(\partial +(\partial W)(x)\biggr)\Psi^{(0)} (x; E^{(0)});\label{transformed}\\
\Psi^{(0)} (x; E^{(0)}) &\propto& \biggl(-\partial +(\partial W) (x)\biggr)\Psi^{(1)} (x; E^{(0)}).\nonumber
\ea
It is clear that one obtains a one-to-one correspondence between solutions of (\ref{sturm})
and (\ref{darboux}), but up to zero modes of operators $(\mp\partial + (\partial W)(x)).$

The nonlinear first-order differential equation (\ref{riccati1}) (Riccati\cite{ricc} equation) can be solved analytically
for a restricted class of potentials $V^{(0)}(x)$ only. The class of analytically solvable Riccati equations corresponds to the so called exactly solvable models in one-dimensional Quantum Mechanics, and latter ones are related to the so called shape-invariant \cite{gend} potentials.

The above construction can be easily presented in a supersymmetric form. Indeed, the Hamiltonians $h^{(0)},\,h^{(1)}$ and intertwining operators $q^{\pm}$ can be considered as elements of matrix $2 \times 2$ operators - super-Hamiltonian and supercharges,
\be
\fl {H} = \left(
                \begin{array}{cc}
                  h^{(0)} & 0 \\
                  0 & h^{(1)} \\
                \end{array}
              \right);\quad
{Q}^+=\left(
            \begin{array}{cc}
              0 & 0 \\
              q^- & 0 \\
            \end{array}
          \right);\quad
{Q}^-=({Q}^+)^{\dagger}=\left(
            \begin{array}{cc}
              0 & q^+ \\
              0 & 0 \\
            \end{array}
          \right), \label{2times2}
\ee
which obey the (anti)commutation relations - the simplest realization of supersymmetric algebra,
\begin{equation}\label{algebra}
[{H},\, {Q}^{\pm}]=0;\quad \{{Q}^+, \,{Q}^-\}={H};\quad ({Q}^{\pm})^2=0.
\end{equation}
The relations (\ref{algebra}) of the superalgebra express in the compact form the factorization (\ref{factorization}), (\ref{partner}) and the intertwining relations (\ref{intertw}), while the nilpotency of supercharges ${Q}^{\pm}$ follows directly from their matrix structure. By analogy with the Supersymmetric Quantum Field Theory, the component $h^{(0)}$ of $H$ is called the bosonic component, and $h^{(1)}$  fermionic one. Correspondingly, their eigenstates form bosonic and fermionic sectors, although this terminology has no physical sense in the present non-relativistic context.
Nevertheless, the nilpotency of $Q^{\pm}$ can be associated with the fermion property - an analog of the Pauli principle.

Thus in the simplest cases (linear SUSY algebras), intertwining of two differential operators (for instance, Hamiltonians of one-dimensional quantum systems)
by means of Darboux operators of first order in derivatives entails their factorization into differential multipliers which are formed by the same Darboux operators.  However, in general, both interrelation between pairs of dynamical operators ("Hamiltonians") with (almost) equivalent spectra and  structure of operators which generate the spectral equivalence are not that simple
\cite{ais} - \cite{tana2011}. In particular, the closing of a Nonlinear SUSY algebra often produces a hidden symmetry generator which helps to partial solvability of spectral problem for Schr\"odinger Hamiltonians.
This interrelation in the one- and two-dimensional QM  is within the scope of the present review. In particular, we survey well motivated solutions to the following problems:
\begin{itemize} \item
what is the correct self-consistent way to generalize the main relations of one-dimensional SUSY Quantum Mechanics, including factorization and intertwining relations, to the case of multidimensional space?
\item
in what cases the higher-order Darboux-Crum transformations can be constructed with the help of a sequence of intertwining transformations of lower order  SUSY which relate a chain of (almost) isospectral intermediate Hamiltonians with real nonsingular
potentials?
\item
what are elementary blocks for a nonsingular factorization of intertwining
operators?
\item
in what way the irreducibility of elementary blocks of isospectral transformations reveals itself in the SUSY algebra and in the structure of kernels of those transformations?
\item are there more complicated SUSY algebras in one dimensional QM which accept hidden symmetry generators ("central charges")?
\item what is the relationship between shape-invariance and spectrum generating algebras for SUSY QM systems?
\item how SUSY for non-stationary Schr\"odinger operators produces hidden symmetries and spectrum generating algebras?
\item is it possible to find pairs of (almost) isospectral {\it scalar} two-dimensional Hamiltonians intertwined by higher order supercharges,
i.e. to avoid the matrix components of super-Hamiltonian by means of using Nonlinear SUSY algebra?
\item
does the Nonlinear two-dimensional SUSY Quantum Mechanics give the constructive way to obtain new exactly solvable and/or quasi-exactly solvable models not amenable to the conventional separation of variables?

\end{itemize}

 The structure of the paper is as follows. After a short reminder of notation and basic definitions of SUSY theory of isospectral transformations given above in the Introduction, the generalizations of linear SUSY QM for arbitrary space dimensions \cite{abi1}, \cite{abei} are considered in Section 2. All of them can be called as direct generalizations since they are based on the supercharge operators of first order in derivatives. In particular, Subsection 2.1 includes a detailed description of two-dimensional generalization \cite{abi1} in the form similar to the Factorization Method of E.Schr\"odinger \cite{schr}, \cite{inf}, and Subsection 2.2 contains two illustrative examples: Pauli equation for fermion motion on the plane with external magnetic field \cite{pauli-eq} -- \cite{pauli-eq-2} and supersymmetric extension of two-dimensional attractive Coulomb potential \cite{abi1}.  In Subsection 2.3, the well developed formalism of Supersymmetric Quantum Field Theory is used to present the direct (first-order) generalization \cite{abei} of Witten's one-dimensional SUSY Quantum Mechanics onto the space with arbitrary number $d$ of dimensions. The main element of construction is the $d -$component real scalar superfield depending both on time and pair of Grassmann variables.
 This multidimensional generalization allows also the alternative interpretation as a model of $d$ interacting particles on a line \cite{ghosh}. In Subsection 2.4 the structure of obtained super-Hamiltonian is analyzed \cite{abei} in the Hilbert space which is a direct sum of subspaces with fixed fermion numbers. In this basis, the super-Hamiltonian has the block-diagonal form with matrix blocks of different matrix dimensionality. The supersymmetric intertwining relations between different matrix components of the super-Hamiltonian are derived providing their isospectrality. Subsection 2.5 gives a brief description of the illustrative example \cite{abinon1}: $d=3$ SUSY Quantum Mechanics with specific Yukawa-like superpotential provides relations between different sectors of nucleon-nucleon and nucleon-antinucleon interactions.

Section 3 is devoted to building of the Polynomial SUSY algebra for $d=1$ with the help of dressing chain or ladder construction \cite{ais}. It is elucidated how they entail the SUSY algebra representation by differential intertwining operators of higher-order in derivatives. For algebraic formulation its key ingredient is given by SUSY Hamiltonian projections on the formal zero-mode space (the kernel) of supercharge operators which are finite-dimensional
constant matrices \cite{ast00,ansok}. However the basic elements of spectral design in Polynomial SUSY  consist not only of intermediate isospectral transformations of first order in derivative (of linear SUSY's) but also of the transformations of second order in derivative \cite{acdi95,samsirr1,bags}. Three irreducible classes of binary SUSY algebras are revealed and it is shown how they serve to complete the tool kit for spectral design. In particular, they may help to add or remove one or two adjacent levels for excited states. Thus the classification  of irreducible (almost) isospectral transformations and related SUSY algebras is outlined. It was described first in  \cite{Andrianov:2004vz}, then systematically in \cite{anso1,sok1}) and the exhaustive fine classification of irreducible operators
 was completed in \cite{sok11}. In connection to this classification the important problem of a possible redundancy in supercharges which can
be eliminated without any changes in the super-Hamiltonian is analyzed in Subsection 3.4.
With the help of a (redundant) multiplication of supercharge by a polynomial of super-Hamiltonian it is elucidated that certain irreducible second-order supercharges can be embedded into a higher-order SUSY which in turn is decomposable in a chain of linear SUSY's.
In Section 4 we formulate basic theorems which give full description  of the structure of  polynomial one-dimensional SUSY algebras and on minimization of this algebra up to its essential part (proofs of these theorems can be found in
\cite{ansok,anso1,sok1,sok11}). Apart from the above mentioned trivial ways to generate several SUSY algebras for the same super-Hamiltonian there exist  possibilities to obtain two non-minimizable SUSY algebras
with supercharges of even and odd order in derivatives \cite{ansok,acin2000}. Their combination forms a non-trivial hidden-symmetry operator which cannot be expressed as a polynomial of the super-Hamiltonian.
Such systems and algebras are described in Section 5. They are governed by (reflectionless) potentials of a special form arising also as soliton solutions of some non-linear equations \cite{matv,leble}. The related SUSY algebra can be qualified as ${\cal N} =2 $ superalgebra with central charge made of hidden symmetry operators.

In Sect.6 the concept of shape invariance is formulated for $d=1$ polynomial SUSY algebras with the help of intertwining with shift \cite{acin2000}. The arising  spectrum generating algebras represent intrinsic features of shape invariance and they can be considered as deformations of hidden symmetries studied in Sect.5 . It is elucidated on and exemplified for the second- and third-order SUSY shape invariance. The relations with deformed Heisenberg algebras \cite{spirid1,plyus11,ferhus} are described as well. In the case of third-order SUSY
its remarkable connection to the Painleve-IV equations \cite{painl1} is demonstrated (Sect. 6.4) and the typical patterns of  energy spectra are listed. The latter connection has been in focus of recent investigations
\cite{iof2004}--\cite{ferncomp}.

Section 7 is devoted to extension of spectral design methods on nonstationary Schr\"odinger equations \cite{tdi}. We follow the scheme of SUSY intertwining relations and show that addition of time dependence leads to certain restrictions on potentials which are assembled in a super-Hamiltonian. In all cases the closure of the SUSY algebra  reveals a hidden symmetry which becomes a typical rather than exceptional feature for non-stationary Hamiltonians. For linear SUSY (Sect. 7.1)this hidden symmetry entails simply full separation of variables. For the second-order SUSY (Sect. 7.2) one discovers the spectrum generating algebra previously found for the third-order shape invariance and this algebra is completely encoded in time-dependent hidden symmetry operators. Thus hidden symmetries for nonstationary Schr\"odinger super-Hamiltonians give the universal framework for both stationary hidden symmetries and for spectrum generating algebras of shape invariant SUSY systems.

In Section 8, the idea of one-dimensional Polynomial SUSY QM is generalized onto two-dimensional case providing isospectrality of {\it scalar} Schr\"odinger Hamiltonians without any intermediate matrix components \cite{d5} -- \cite{d15}. Subsection 8.1 gives the basic formulas \cite{d5} -- \cite{d2} of this approach for arbitrary form of highest (second-order) terms in supercharge. The most promising case of Lorentz (hyperbolic) form of second-order terms is studied in Subsection 8.2 with special attention to reducible (with twist) and irreducible supercharges. The fundamental difference of two-dimensional Polynomial SUSY QM is that both SUSY-partner Hamiltonians are completely integrable: they commute with symmetry operators which are of fourth order in momenta (Subsection 8.3). Other examples of relation between nonlinear supersymmetry and quasi-exact solvability can be found in
\cite{kliplyu00}. Section 9 is devoted to two new methods \cite{ioffe1}, \cite{ioffe2} of solution of two-dimensional Schr\"odinger equations which are not amenable to conventional method of separation of variables. Both new methods are based on the isospectrality of scalar Hamiltonians established in the previous Section and on the property of shape invariance (Section 6). These methods can be considered as supersymmetric separation of variables. The first of them (Subsection 9.1) provides the quasi-exact solvability of several two-dimensional models, i.e. it allows to find analytically a part of spectrum and corresponding wave functions. The second one (Subsection 9.2) gives the exact analytical solution of the models but it works only for particular values of parameters of the models. Both methods can be applied to several potentials not amenable to standard separation of variables, and here they are illustrated by solution of two-dimensional generalization of Morse model. In Section 10, we list some related problems of SUSY QM which are outside the scope of the present review, and we mention some related open problems and perspectives for future study.

\section{SUSY QM in arbitrary space dimension}

\subsection{Two-dimensional SUSY QM with first-order supercharges}

The generalization of one-dimensional SUSY QM onto higher space dimensionality $d \geq 2$
in the Schr\"odinger equation is of great interest both from the conceptual and practical point of view.
The forthright attempt to preserve the one-dimensional formulas given above but with superpotentials depending on
multidimensional coordinate vector $\vec x=(x_1, x_2, ..., x_d)$ leads to potentials which {\it are amenable}
to standard separation of variables (see details in \cite{kuru}). In what follows such generalizations
will not be considered since the corresponding Schr\"odinger equation is reduced to the simpler problems
of lower dimensionalities.

A more advanced generalization of one-dimensional SUSY QM onto multidimensional
case was made in the framework of  Factorization Method in \cite{abi1} (see also \cite{freedman} and \cite{kirch}). The main idea is
in involving vector intertwining operators $q_l(\vec x);\,\, l = 1, 2,..., d,$ which are still
differential operators of first order in derivatives,
\begin{equation}\label{vectorq}
  q_l^{\pm}(\vec x) = \mp\partial_l + (\partial_l W) (\vec x));\quad \vec x=(x_1, x_2, ..., x_d);\quad \partial_l\equiv \frac{\partial}{\partial x_l}.
\end{equation}
In terms of these operators, the initial Hamiltonian $h^{(0)}(\vec x)$ can be quasifactorized,
\begin{equation}\label{quasifactor}
  h^{(0)}(\vec x)=-\triangle^{(2)}+ V^{(0)}(\vec x) = q_l^+q_l^-; \quad \triangle^{(2)}\equiv\partial_l \partial_l,
\end{equation}
where the sum over repeated index $l$ is implied.
In the framework of this generalization, which can be named as "direct generalization",
the Schr\"odinger operator $h^{(0)}$ is related to the chain of $d$ other Hamiltonians of Schr\"odinger form, but $(d-1)$ of them with matrix potentials (see Subsection 2.5 below).

Now we shall describe this construction \cite{abi1} in the simplest - two-dimensional $d=2$ - case, where the chain of Hamiltonians includes three species, the original $h^{(0)}$ and two others, $h^{(1)}_{ik}$
and $h^{(2)},$ both of the Sch\"odinger form as well. The first of them has $2\times 2$
matrix potential $V^{(1)}_{ik}(\vec x),$ while the second one has  a scalar potential $V^{(2)}(\vec x).$
These partner Hamiltonians (superpartners) are expressible in a quasifactorized form as well,
\ba\fl
h^{(1)}_{ik}=q_i^-q_k^+ + p_i^-p_k^+ = -\delta_{ik}
\partial_l^2 + \delta_{ik}\Bigl((\partial_l W)^2(\vec x) - (\partial_l^2W)(\vec x)\Bigr)
+ 2(\partial_i\partial_k W)(\vec x); \label{h1}\\
\fl h^{(2)}=p_l^+p_l^-=-\partial_l^2 +
V^{(2)}({\vec x}) = -\partial_l^2 + (\partial_l W)^2(\vec x) +
(\partial_l^2 W)(\vec x);\label{h2}
\ea
The operators $q_l^{\pm}$ were defined in (\ref{vectorq}), and the new vector operators $p_l^{\pm}$ are
also of first order in derivatives,
\be
p_l^{\pm}\equiv \epsilon_{lk}q_k^{\mp},\label{2p}
\ee
where $\epsilon_{lk}$ is a standard antisymmetric tensor with $\epsilon_{12}=+1.$

The chain of two-dimensional Hamiltonians $h^{(0)}, \,h^{(1)}_{ik},\, h^{(2)}$ is analogous to
the one-dimensional partnership $h^{(0)},\, h^{(1)},$ since its components are also related by the intertwining relations
with operators $q_l^{\pm}, \,p_l^{\pm}$ playing the role of intertwining operators,
\ba
h^{(0)}q_i^+=q_k^+h_{ki}^{(1)};\quad h_{ik}^{(1)}q_k^- =
q_i^-h^{(0)}; \label{intertw1}\\
h^{(1)}_{ik}p_k^-=p_i^-h^{(2)} ;
\quad p_k^+h^{(1)}_{ki} = h^{(2)}p_i^+ .\label{intertw22}
\ea
By definition, the intertwining operators $q_l^{\pm},\,p_l^{\mp},$ and, correspondingly, two terms in the definition (\ref{h1})
of $h^{(1)}_{ik}$ are mutually orthogonal,
$q_l^{\pm}p_l^{\mp}=0.$ These two terms in $h^{(1)}_{ik}$ were necessary to obtain the Hamiltonian of the Schr\"odinger form, and the orthogonality is important to provide the intertwining relations (\ref{intertw1}), (\ref{intertw22}).

The energy spectra of $h^{(0)}$ and $ h^{(2)}$ are, in general, non-overlapping, but the intertwining relations (\ref{intertw1}), (\ref{intertw22}) provide the
equivalence of energy spectra between a pair of two scalar
Hamiltonians $h^{(0)}$, $ h^{(2)}$ and $2 \times2$ matrix
Hamiltonian $h^{(1)}_{ik}.$ The "equivalence" means coincidence of
their spectra up to zero modes of operators $q_l^{\pm}, p_l^{\pm}.$ Thus,
the supersymmetry (supersymmetric transformation) allows to reduce
the solution of spectral problem for the matrix Hamiltonian
$h^{(1)}_{ik}$ to solutions of the couple of scalar spectral
problems $h^{(0)}$, $ h^{(2)}$ realizing the SUSY diagonalization method. Due to the intertwining
relations, the vector wave functions of matrix Hamiltonian
$h_{ik}^{(1)}$ are also connected (up to a normalization factor)
with the scalar wave functions of scalar Hamiltonians $h^{(0)},
h^{(2)}$,
\ba
\Psi^{(1)}_i(\vec x; E) &=& q^-_i\Psi^{(0)}(\vec x;
E);\quad i=1,2;\quad \Psi^{(0)}(\vec x; E) = q^+_i\Psi^{(1)}_i(\vec
x; E)\nonumber\\
\Psi^{(1)}_i(\vec x; E) &=&
p^-_i\Psi^{(2)}(\vec x; E);\quad\Psi^{(2)}(\vec x; E) =
p^+_i\Psi^{(1)}_i(\vec x; E).\label{connect2}
\ea

The two-dimensional generalization (\ref{quasifactor}), (\ref{h1}), (\ref{h2}), (\ref{intertw1}), (\ref{intertw22}) of Factorization Method can be also formulated compactly \cite{abei} in terms of exactly the same superalgebra (\ref{algebra}). Indeed, three Hamiltonians $h^{(0)},\, h^{(1)}_{ik},\,h^{(2)}$ are combined in the super-Hamiltonian $H$, and intertwining operators $q_l^{\pm},\, p_l^{\pm}$ - in the supercharges $Q^{\pm},$ correspondingly,
\begin{equation}\label{3times3}
H=\left(
             \begin{array}{ccc}
               h^{(0)} & 0 & 0 \\
               0 & h^{(1)}_{ik} & 0 \\
               0 & 0 & h^{(2)} \\
             \end{array}
           \right);\quad Q^+=(Q^-)^{\dagger}=\left(
                           \begin{array}{cccc}
                             0 & 0 & 0 & 0 \\
                             q_1^- & 0 & 0 & 0 \\
                             q_2^- & 0 & 0 & 0 \\
                             0 & p_1^+ & p_2^+ & 0 \\
                           \end{array}
                         \right),
\end{equation}
all matrices are $4 \times 4.$  The components $h^{(0)},\, h^{(2)}$ are bosonic, and $h^{(1)}_{ik} -$fermionic component of the super-Hamiltonian.

A simple rearrangement of this representation gives an idea of SUSY diagonalization of matrix Hamiltonians in terms of the conventional SUSY algebra (\ref{algebra}),
\begin{eqnarray}\label{2times2diag}
&&\widetilde H=\left(
             \begin{array}{cc}
               \tilde{h}^{(0)} & 0 \\
               0 & \tilde{h}^{(1)}  \\
             \end{array}
           \right);\quad \tilde{h}^{(0)}=\left(
             \begin{array}{cc}
               h^{(0)} & 0 \\
               0 & h^{(2)}  \\
             \end{array}
           \right);\quad \tilde{h}^{(1)}=\left(h^{(1)}_{ik}
           \right);\nonumber\\&& \widetilde{Q}^+=(\widetilde{Q}^-)^{\dagger} =\left(
            \begin{array}{cc}
              0 & 0 \\
              \tilde{q}^- & 0 \\
            \end{array}
          \right);\quad \tilde{q}^-=(\tilde{q}^+)^{\dagger}=\left(
                           \begin{array}{cc}
                             q_1^+  & q_2^+\\
                             p_1^+  & p_2^+\\
                           \end{array}
                         \right) .
\end{eqnarray}

\subsection{Examples}

The Schr\"odinger operators with matrix potentials are not something very exotic in Quantum Mechanics. In particular, the described two-dimensional generalization of SUSY QM was successfully used \cite{pauli-eq} -- \cite{pauli-eq-2} to investigate the spectra of stationary Pauli operator for fermion in external nonhomogeneous electromagnetic fields.

1. The Pauli Hamiltonian for nonrelativistic spin $1/2$ particle in three-dimensional space is written as follows,
\begin{equation}\label{pauli}
H_P=(i\partial_a+eA_a)^2-\mu\sigma_aB_a+U,
\end{equation}
where $a=1,2,3,$ $e$ and $\mu$ are the electric charge and magnetic momentum of the particle, $\{A_a(\vec x)\}$ is the vector part of electromagnetic potential, $\vec B(\vec x)=rot \vec A(\vec x)$ is the corresponding magnetic field (in general, nonhomogeneous), $U(\vec x)$ is the scalar potential (which may include the electrostatic term $eA_0(\vec x)$), and $\sigma_a$ are the standard Pauli matrices. Restricting the problem to the case of external potentials which {\it do not} depend on {\it one} of coordinates (on $x_2$ below), the Pauli Hamiltonian (\ref{pauli}), being $2 \times 2$ matrix operator, can be identified with the matrix component $h^{(1)}_{ik};\, (i,k=1,2)$ of the super-Hamiltonian in the previous Subsection. Indeed, the fermionic (two-component) wave function has to be factorized into the trivial plane wave along $x_2$ with some momentum $k$ and nontrivial {\it two-dimensional} wave function,
\begin{equation}\label{wavepauli}
\Psi_P(\vec x)=\exp{(ikx_2)}\Psi(x_1, x_3).
\end{equation}
In the subspace of such functions, Pauli operator can be rewritten as,
\begin{equation}\label{pauli2}\fl
H_P(x_1, x_3)=-(\partial_l+ieA_l)^2+U+(k+eA_2)^2-\mu\sigma_aB_a;\quad l=1,3;\quad A_l=A_l(x_1, x_3).
\end{equation}
Comparing (\ref{pauli2}) and (\ref{h1}), one concludes that their identification (up to replacing $(x_1,x_3)\leftrightarrow (x_1,x_2)$) is possible if external fields in (\ref{pauli2}) are expressed in terms of the only function $W(x_1,x_3),$
\begin{equation}\label{fields}
-\mu \vec B = \Big(2(\partial_1\partial_3W)(x_1,x_3), 0, ((\partial_1^2-\partial_3^2)W)(x_1,x_3)\Big).
\end{equation}
Let for simplicity the external sources of magnetic field are absent,
\begin{equation}\label{sources}
rot \vec B=div \vec B=0
\end{equation}
Then, the function $W$ is a polynomial of fourth order, and magnetic field is of the form,
\begin{equation}\label{B}
-\mu \vec B = ((2ax_1x_3+cx_1+bx_3+g),\, 0,\, (ax_1^2-ax_3^2+bx_1-cx_3+d)),
\end{equation}
with arbitrary constant coefficients $a, b,... .$ The scalar potential $U(x_1, x_3)$ is a polynomial of sixth order in coordinates (see details in \cite{pauli-eq}) -- \cite{pauli-eq-2}). Due to results of the previous Subsection, the spectral problem for {\it matrix} $2 \times 2$ Pauli Hamiltonian is reduced to a couple of spectral problems with {\it scalar} Hamiltonians $h^{(0)},\, h^{(2)},$ which is much simpler. This example demonstrates the opportunity of dynamical supersymmetric diagonalization of a class of matrix problems by means of two-dimensional SUSY Quantum Mechanical approach.
The identification of Pauli operator $H_P$ with matrix component of super-Hamiltonian is possible for much wider class of external fields, if some unitary rotation is preliminary done (more details in \cite{pauli-eq} -- \cite{pauli-eq-2}).

It is instructive to notice that this method works for a wide class of non-homogeneous magnetic fields which can be directed at any angle with respect to plane, including the field parallel to the plane. The magnetic dipole moment $\mu$ of the particle is also arbitrary. It is necessary to mention the different supersymmetric approach to solution of Pauli equation for the fermion in the plane \cite{crom}, \cite{khare-pauli}, \cite{cooper-pauli}, \cite{gend-pauli},
which works only for gyromagnetic ratio equal $2$ and only for (non-homogeneous) magnetic fields directed orthogonally to the plane. The latter condition in itself provides the diagonalizability of the problem, but SUSY allows to study the symmetry properties and the spectrum of Pauli operator. The method of \cite{crom}, \cite{khare-pauli}, \cite{cooper-pauli} basically has the form of one-dimensional SUSY Quantum Mechanics but with matrix $2 \times 2$ supercharge operators $Q^{\pm}$ (see also \cite{IKNN}). The initially diagonalizable Pauli equation for fermions with even integer giromagnetic ratios in some specific magnetic fields orthogonal to the plane was considered also in \cite{0105135} in the framework of nonlinear SUSY.

2. The second illustrative two-dimensional model \cite{abi1} corresponds to the initial potential of Coulomb form,
\begin{equation}\label{coulomb}
V^{(0)}(\vec x)=-\frac{\alpha}{r};\quad r^2=x_1^2+x_2^2;\quad \alpha > 0.
\end{equation}
In this case, the superpotential can be taken as $W = 2\alpha r,$ and therefore, the component $V^{(2)}$ has no bound states at all,
\begin{equation}\label{coulomb2}
V^{(2)}(\vec x)=+\frac{\alpha}{r}.
\end{equation}
These specific circumstances lead to equivalence of the spectra of matrix Hamiltonian operator and scalar component of super-Hamiltonian $h^{(0)}(\vec x).$
Since the discrete spectrum of the latter is well known, the spectrum of the matrix Hamiltonian with potential,
\begin{equation}\label{matrixcoulomb}
V^{(1)}_{ik}(\vec x) = -\frac{\alpha}{r}(2\frac{x_ix_k}{r^2}-\delta_{ik}).
\end{equation}
is known as well. Its wave functions can be obtained from the wave functions of scalar Coulomb problem $V^{(0)}$ by means of intertwining relations (\ref{intertw1}) with operators $q_l^-$.The potential (\ref{matrixcoulomb}) can be physically interpreted in curvilinear (polar) coordinates \cite{aiz}. It corresponds to attractive Coulomb potential for 1/2-spin particle with positive spin projection $s_z=1/2$ and to  repulsive Coulomb - for particle with negative $s_z=-1/2.$

\subsection{SUSY QM in arbitrary dimensionality of space as $(0+1)-$dimensional Quantum Field Theory}

Both one-dimensional Factorization Method and its two-dimensional generalization were shown above to realize the simplest superalgebra (\ref{algebra}) of Supersymmetric Quantum Mechanics. This observation gives the idea how we could generalize this approach further for arbitrary dimensionality $d$ of space: it is necessary to build {\it new realizations} of the same SUSY algebra (\ref{algebra}). For this purpose it is helpful \cite {abei} to exploit some methods and experience of well-developed Supersymmetric Quantum Field Theory \cite{weinberg-3}.  Indeed, new such realizations of the SUSY algebra
was obtained \cite{abei} by study of some simple models of SUSY Quantum Field Theory. The simplest QFT, which can be interpreted as nonrelativistic Quantum Mechanics, is the theory in $(0+1)-$ dimensional space, i.e. fields depend on one variable - the time $t.$ Keeping in mind, that supersymmetry must be a property of the model, we consider the {\it superfield}, which depends both on one bosonic coordinate (time $t)$ and a pair of mutually conjugate Grassman variables $\theta, \bar\theta.$ The dimensionality $d$ of SUSY Quantum Mechanics on target is provided by choosing the fields real, but with $d-$components.

Thus, the main ingredient of the model is \cite{abei} the $d-$dimensional real superfield,
\ba
& &\varphi_l(t; \theta, \bar\theta);\quad l=1,2,...,d;\quad \varphi_l=\bar\varphi_l; \label{phi}\\
& &\{\theta,\,\theta\}=\{\bar\theta,\,\bar\theta\}=\{\theta,\,\bar\theta\}=[t,\,\theta]=[t,\,\bar\theta]=0.\label{theta}
\ea
As it is typical for the genuine superfield theory, the infinitesimal super-transformation in the space $(t; \theta, \bar\theta)$ is defined as,
\be
t \rightarrow t^{\prime}=t-i(\bar\theta\epsilon-\bar\epsilon\theta);\quad \theta\rightarrow\theta^{\prime}=\theta + \epsilon;\quad
\bar\theta\rightarrow\bar\theta^{\prime}=\bar\theta + \bar\epsilon ,
\label{epsilon}
\ee
with a pair of Grassman mutually conjugate parameters of transformation $\epsilon,\, \bar\epsilon .$
The "infinitesimal" (linear in $\epsilon,\, \bar\epsilon $) transformations (\ref{epsilon}) can be generated by differential operators (super-generators) in the superspace $(t; \theta, \bar\theta),$
\be
Q=i\frac{\partial}{\partial \theta}-\bar\theta\frac{\partial}{\partial t};\quad
\bar Q=-i\frac{\partial}{\partial \bar\theta}+\theta\frac{\partial}{\partial t},
\label{barQ}
\ee
which produce the finite transformations by,
\be
U=\exp{[i(\bar\epsilon\bar Q+Q\epsilon)]}.
\label{finite}
\ee
Operators $Q,\,\bar Q$ satisfy the anticommutation relations,
\be
\{Q, Q\}=\{\bar Q, \bar Q\}=0;\quad \{Q, \bar Q\}=i\frac{\partial}{\partial t}.
\label{QQ}
\ee

Due to nilpotency of $\theta,\,\bar\theta ,$ the bosonic superfield can be expanded as,
\be
\varphi_l(t; \theta, \bar\theta)=x_l(t)+i\theta\psi_l(t)-i\bar\psi_l(t)\bar\theta +\bar\theta \theta D_l(t),
\label{expansion}
\ee
where $x_l(t), D_l(t)$ are bosonic real $d-$component fields, and $\psi_l(t), \bar\psi_l(t) -$ mutually conjugate $d-$component fermionic (anticommuting) fields. Restricting ourselves to {\it superscalar} fields,
\begin{equation}\label{scalarrr}
\varphi'_l(t'; \theta', \bar\theta')=\varphi_l(t; \theta, \bar\theta),
\end{equation}
one obtains the following rules of transformation,
\ba
& & \delta x_l(t)=i\epsilon\psi_l(t)-i\bar\psi_l(t)\bar\epsilon;\quad \delta D_l(t)=\bar\epsilon\dot{\bar\psi}_l(t)+\dot\psi_l(t)\epsilon ; \nonumber\\
& & \delta \psi_l(t)=i\bar\epsilon D_l(t)-\bar\epsilon\dot x_l(t);\quad  \delta \bar\psi_l(t)=-i\epsilon D_l(t)-\epsilon\dot x_l(t),
\label{delta}
\ea
where dot means derivative over $t.$

As usual, for construction of superinvariant action functional we need the operators of supercovariant derivatives,
\be
D=\frac{\partial}{\partial\theta}-i\bar\theta\frac{\partial}{\partial t};\quad
\bar D=\frac{\partial}{\partial\bar\theta}-i\theta\frac{\partial}{\partial t},
\label{covariant}
\ee
which allow to build the superinvariant action,
\begin{equation}\label{action}
S=\int dtL(\vec \varphi) = \int dt\int d\theta d\bar\theta\biggl(-\bar D\vec\varphi D\vec\varphi + W(\vec\varphi)\biggr)
\end{equation}
with arbitrary functional $W(\vec\varphi)$ - superpotential of the model. Expanding it around the point $x_l(t),$ integrating the action (\ref{action}) over both Grassmann variables and substituting nondynamical variables $D_l, \bar D_l$ from the corresponding equations of motion,
one obtains the action as functional of component fields and superpotential,
\begin{equation}\label{action1}\fl
S=\int dt \biggl(\dot x_l\dot x_l+i\bar\psi_l\dot{\psi}_l-i\dot{\bar\psi}_l\psi_l-(\partial_lW)(\vec x)(\partial_lW)(\vec x)-
(\partial_l\partial_m W)(\vec x)[\bar\psi_l(t), \psi_m(t)] \biggr).
\end{equation}
The corresponding {\it classical} Hamiltonian is,
\begin{equation}\label{HHH}
H_{cl}=p_lp_l+(\partial_lW)(\partial_lW)+(\partial_l\partial_mW)[\bar\psi_l, \psi_m],
\end{equation}
where $p_l$ is the momentum $p_l=\delta L / \delta \dot x_l=\dot x_l.$ According to Noether theorem (in the supersymmetric context), supercharges in the present field theory are \cite{fubini},
\begin{equation}\label{noether}
\fl Q^+_{cl}=\frac{\delta L}{\delta \dot\epsilon}=\bigl(ip_l+(\partial_lW)\bigr)\psi_l;\quad
Q^-_{cl}=\frac{\delta L}{\delta \dot{\bar\epsilon}}=\bigl(-ip_l+(\partial_lW)\bigr)\bar\psi_l=(Q^+)^{\dagger}.
\end{equation}

To present point, the model was formulated on the classical level, but now everything is prepared for its quantization (the prescriptions for canonical quantization of mixed systems both with bosonic and fermionic degrees of freedom see in \cite{1976}). After quantization, coordinates $x_l,$ momenta $p_l$ and fermionic variables $\psi_l, \bar\psi_l$ become the operators $\hat x_l, \hat p_l, b_l^+, b_l^-,$ correspondingly,
\begin{equation}\label{quantization}
[\hat p_l, \hat x_m]=-i\delta_{l m};\quad \{b_l^+, b_m^-\}=\delta_{l m};\quad \{b_l^+, b_m^+\}=\{b_l^-, b_m^-\}=0,
\end{equation}
and anticommuting operators $b_l^{\pm}$ can be interpreted as creation and annihilation operators for the system of $d$ localized {\it scalar} particles, but with {\it Fermi} statistics. The quantized super-Hamiltonian has the form,
\begin{equation}\label{quantum}
H = \hat p_l \hat p_l+ (\partial_lW)(\partial_lW)+(\partial_l\partial_mW) b_l^+b_m^-,
\end{equation}
and quantum supercharges are,
\begin{equation}\label{ssuper}
Q^{\pm}=\Bigl(\pm i\hat p_l+(\partial_l W)(\vec x)\Bigr)b_l^{\pm}.
\end{equation}
These quantum operators satisfy all relations of superalgebra (\ref{algebra}), and they provide the required generalization of one- and two-dimensional supersymmetric models of Introduction and Subsection 2.1 to the case of arbitrary dimensionality $d.$

\subsection{The structure of super-Hamiltonian in SUSY QM for arbitrary dimensionality of space}

In order to understand the structure of $d-$dimensional quantum super-Hamitlonian (\ref{quantum}) and supercharges (\ref{ssuper}), it is useful \cite{abei} to use the fermion number operator,
\begin{equation}\label{number}
N_F=b_l^+b_l^-,
\end{equation}
which commutes with $H, \, Q^{\pm}$ as follows,
\begin{equation}\label{number1}
[H,\, N_F]=0;\quad [N_F,\, Q^{\pm}]=\pm Q^{\pm}.
\end{equation}
If the basis in the Hilbert space $\cal H$ of states is chosen so that the space is a direct sum of subspaces ${\cal H}_{n}$ with fixed number $n$ of fermions $(n=0,1,2,...,d),$ the conservation law (\ref{number1}) means that super-Hamiltonian has the block-diagonal form in the fermionic Fock space,
\begin{equation}\label{structure}
H=\left(
  \begin{array}{cccc}
    h^{(0)} & 0 & 0 & 0 \\
    0 & h^{(1)} & 0 & 0 \\
    0 & 0 & ... & 0 \\
    0 & 0 & 0 & h^{(d)} \\
  \end{array}
\right)
\end{equation}
with components $h^{(n)}$ acting in the subspace ${\cal H}_{n}.$ The convenient basis in ${\cal H}_n$ is the basis of occupation numbers - $C_d^n$ vectors $|n_1, n_2,...,n_d >$ with $n_l=0,\, 1,$ and $\sum_{l=1}^{l=d} n_l=n.$ In turn, the supercharge $Q^+$ has the block structure with nonvanishing matrix $C_d^{n+1} \times C^n_d$ blocks $q^+_{n, n+1}$  below the main diagonal, and $Q^-$ - matrix $C_d^{n} \times C^{n+1}_{d}$ blocks $q^-_{n+1, n}$  above the main diagonal,
\begin{equation}\label{matrixQ}\fl
Q^+=\left(
           \begin{array}{ccccc}
             0 & 0 & 0 & 0 & 0 \\
             q^+_{0,1} & 0 & 0 & 0 & 0 \\
             0 & q^+_{1,2} & 0 & 0 & 0 \\
             0 & 0 & ... & 0 & 0 \\
             0 & 0 & 0 & q^+_{d-1,d} & 0 \\
           \end{array}
         \right);\quad
Q^-=\left(
           \begin{array}{ccccc}
             0 & q^-_{1,0} & 0 & 0 & 0 \\
             0 & 0 & q^-_{2,1} & 0 & 0 \\
             0 & 0 & 0 & ... & 0 \\
             0 & 0 & 0 & 0 & q^-_{d,d-1} \\
             0 & 0 & 0 & 0 & 0 \\
           \end{array}
         \right).
\end{equation}
The algebra (\ref{algebra}) and matrix representations (\ref{structure}), (\ref{matrixQ}) provide the following relations,
\begin{equation}\label{ffactor}
h^{(n)}=q^+_{n-1,n} q^-_{n,n-1} + q^-_{n+1,n} q^+_{n,n+1}\equiv \underline{h}^{(n)} + \underline{\underline{h}}^{(n)}
\end{equation}
(quasifactorization - compare with (\ref{h1})),
\begin{equation}\label{orthog3}
q^+_{n+1,n+2}q^+_{n,n+1}=q^-_{n-1,n-2}q^-_{n,n-1}=0;\quad \underline{h}^{(n)} \underline{\underline{h}}^{(n)}=\underline{\underline{h}}^{(n)}\underline{h}^{(n)}=0,
\end{equation}
(orthogonality - compare with the orthogonality $p_l^{\pm}q_l^{\mp}=q_l^{\pm}p_l^{\mp}=0$ in Subsection 2.1),
\ba
& &\underline{\underline{h}}^{(n+1)}q^+_{n,n+1}=q^+_{n,n+1}\underline{h}^{(n)}=
\underline{h}^{(n)}q^-_{n+1,n}=q^-_{n+1,n}\underline{\underline{h}}^{(n+1)}=0;\nonumber\\
& & \underline{h}^{(n+1)}q^+_{n,n+1}=q^+_{n,n+1}\underline{\underline{h}}^{(n)};\quad
q^-_{n+1,n}\underline{h}^{(n+1)}=\underline{\underline{h}}^{(n)}q^-_{n+1,n},
\label{intertw3}
\ea
(intertwining relations analogous to (\ref{intertw})).

Thus, each matrix component $h^{(n)},\,(n=1,...,d-1)$ of the super-Hamiltonian $H$ contains two mutually orthogonal terms
$\underline{\underline{h}}^{(n)},\, \underline{h}^{(n)}.$ These terms are intertwined with analogous terms of neighbor matrix components,
providing their spectral equivalence (again, up to zero modes of supercharges).
Due to orthogonality (\ref{orthog3}), each matrix component $h^{(n)},$ as a whole, is also intertwined with neighbor $h^{(n-1)}$ and $h^{(n+1)}.$
As for connection of the wave functions, those of $h^{(n)}$ are connected, analogously to (\ref{connection}) and (\ref{connect2}) with wave functions of neighbor Hamiltonians $h^{(n+1)},\, h^{(n-1)}.$ This scheme can be illustrated by the following diagram,
\be
[h^{(0)}] \leftrightarrows [\underline{h}^{(1)} + \underline{\underline{h}}^{(1)}] \leftrightarrows
[\underline{h}^{(2)} + \underline{\underline{h}}^{(2)}] \leftrightarrows ...  [\underline{h}^{(d-1)} + \underline{\underline{h}}^{(d-1)}]
\leftrightarrows h^{(d)}.
\label{scheme}
\ee

Thereby, the structure of elements of SUSY QM for arbitrary dimensionality of space is known. In order to obtain the explicit expressions for potentials, one has to choose any matrix representation for the fermionic creation and annihilation operators $b^{\pm}.$ In particular, choosing
\begin{equation}\label{d1}
b^+=\left(
      \begin{array}{cc}
        0 & 0 \\
        1 & 0 \\
      \end{array}
    \right);\quad
b^-=\left(
      \begin{array}{cc}
        0 & 1 \\
        0 & 0 \\
      \end{array}
    \right),
\end{equation}
the relations of the Introduction for $d=1$ are obtained. For the space dimensionality $d=2,$ the suitable representation for fermionic creation and annihilation operators $b_l^{\pm},\, l=1,2$ is $4 \times 4$ matrix,
\begin{equation}\label{d2}
b^+_1=\left(
      \begin{array}{cccc}
        0 & 0 & 0 & 0\\
        1 & 0 & 0 & 0\\
        0 & 0 & 0 & 0\\
        0 & 0 & -1 & 0\\
      \end{array}
    \right)=(b_1^-)^{\dagger};\quad
b^-_2=\left(
      \begin{array}{cccc}
        0 & 0 & 0 & 0\\
        0 & 0 & 0 & 0\\
        1 & 0 & 0 & 0\\
        0 & 1 & 0 & 0\\
      \end{array}
    \right)=(b_2^-)^{\dagger},
\end{equation}
leading \cite{abei} to the same $4 \times 4$ matrix dimensionality of the supercharges $Q^{\pm}$ and the super-Hamiltonian $H$ given by (\ref{3times3}).

\subsection{The particular case of $d=3:$ nucleon-nucleon interaction}

For physically interesting case of $d=3,$ three creation and annihilation operators $b_l^{\pm}, \, l=1,2,3,$ and the supercharges $Q^{\pm}$ are of matrix dimensionality $8 \times 8.$ The super-Hamiltonian $H$ includes \cite{abei} four components: two scalars $h^{(0)},\,h^{(3)}$ and two $3 \times 3$ matrices $h^{(1)}_{ik},\, h^{(2)}_{ik}.$ According to the general scheme of previous Subsection, there are intertwining relations in the pairs $h^{(0)} \rightleftharpoons h^{(1)}_{ik};\, h^{(1)}_{ik} \rightleftharpoons h^{(2)}_{ik};\, h^{(2)}_{ik} \rightleftharpoons h^{(3)},$ providing the relations between corresponding scalar and vector (three-component) wave functions. In this case, the same commutation and anticommutation relations of the superalgebra (\ref{algebra}) are fulfilled.

The possibility of using of this $8 \times 8$ realization of (\ref{algebra}) can be illustrated by derivation \cite{abinon1} of simple connections between different sectors of nucleon-nucleon and nucleon-antinucleon interactions in the OPEP (one-pion-exchange-potential) approximation. The general form of the Hamiltonian of interaction of two non-relativistic particles of spin $1/2$ with scalar, spin-spin and tensor forces is:
\be\fl
H(1/2\otimes 1/2)= -\frac{1}{2} \Delta^{(3)}+V_S(r)+\frac{1}{2}(1+\vec\sigma_1\vec\sigma_2)V_{\sigma}(r)+
([3(\vec\sigma_1\vec{\partial})3(\vec\sigma_2\vec{\partial})-(\vec\sigma_1\vec\sigma_2)\Delta^{(3)}]V_T)(r),
\label{onehalf}
\ee
where $\vec\sigma_1,\, \vec\sigma_2$ are Pauli matrices describing spin operators of two nucleons.
Separating (\ref{onehalf}) onto singlet spin $(S=0)$ and triplet spin $(S=1)$ parts,
\be\fl
V^{S=0}=V_S(r)-V_{\sigma}(r);\quad V_{lk}^{S=1}=(V_S+V_{\sigma}+2\Delta^{(3)}V_T)\delta_{lk}-6(\partial_l\partial_k V_T);\, l,k=1,2,3,
\label{spin}
\ee
and choosing suitably the spherically symmetric superpotential $W(r),$ one may identify the first part $V^{S=0}$ with the scalar potential of the component $h^{(0)}$ of the super-Hamiltonian:
\be
V^{S=0}=\frac{1}{2}[(\vec{\partial} W)^2(r)-(\Delta^{(3)}W)(r)]=\frac{1}{2}[(W^{\prime})^2-W^{\prime\prime}+\frac{2}{r} W^{\prime}],
\label{ident}
\ee
where prime means derivative over $r.$ Simultaneously, it is possible \cite{abinon1} to identify the triplet spin part $V^{S=1}_{lk}$ with the
$3 \times 3$ component $h^{(2)}_{lk}$ of the super-Hamiltonian, i.e. to put:
\be\fl
V_T(r)=\frac{1}{6}W(r);\quad V_{\sigma}(r)=\frac{1}{3}(\Delta^{(3)}W)(r);\quad V_S(r)=\frac{1}{2}(\vec{\partial} W)^2(r)-\frac{1}{6}(\Delta^{(3)}W)(r).
\label{tensor}
\ee

In the OPEP-approximation, the explicit form of $NN$ interaction potential is:
\be
V^{NN} = f^2(\vec\tau_1 \vec\tau_2)(\vec\sigma_1 \vec\partial)(\vec\sigma_2 \vec\partial)\frac{e^{-\mu r}}{r},
\label{NN}
\ee
where $\vec\tau_1,\, \vec\tau_2$ are isospin operators of nucleons, $f^2$ is a coupling constant, and $\mu -$ pion mass. The expression (\ref{NN}) corresponds to the specific forms of tensor, spin-spin and scalar terms:
\be
V_T^{NN}=\frac{1}{3}f^2(\vec\tau_1 \vec\tau_2)\frac{e^{\mu r}}{r};\quad V_{\sigma}^{NN}=2\Delta^{(3)}V_T^{NN};\quad V_S^{NN}=-\Delta^{(3)}V_T^{NN}.
\label{terms}
\ee
Choosing the superpotential $W$ in the super-Hamiltonian $H$ as
\be
W(r)=2(\vec\tau_1 \vec\tau_2)f^2\frac{e^{\mu r}}{r},
\label{choose}
\ee
one can identify \cite{abinon1} the Hamiltonian $H^{NN}$ with $h^{(0)}\bigoplus h^{(2)}_{lk}$ with exponential accuracy $(W^{\prime})^2/W^{\prime\prime}$ (see \cite{urrutia}).

It is important that in the OPEP-approximation, two other components $h^{(1)}_{lk}$ and $h^{(3)}$ of the super-Hamiltonian $H$ have also simple physical sense (see \cite{abinon1}): they describe the triplet spin $S=1$ and singlet spin $S=0,$ correspondingly, of nucleon-antinucleon interaction $H^{N \bar N}=h^{(1)}_{lk}\bigoplus h^{(3)}.$ Thus, just the parts of $H^{NN}$ and $H^{N\bar N}$ with definite total spin being the components of the same super-Hamiltonian satisfy the supersymmetric intertwining relations of the form (\ref{intertw3}). In case, if there were bound $NN$ states, their spectra would (almost) coincide, and we could write the relations between the wave functions $\Psi^{NN}_n$ and $\Psi^{N\bar N}_n,$ similar to (\ref{connect2}). In the present case, supersymmetrical intertwining relations allow to connect asymptotic behaviour of the wave functions of continuous spectra. The non-stationary and stationary non-relativistic scattering theory for a supersymmetrical Hamiltonian was developed in \cite{abinon1}, where the simple connections between scattering data - partial amplitudes, phase shifts and mixing parameters - of the components of super-Hamiltonian were derived.

\section{Polynomial SUSY in d=1}
\subsection{Supercharges of higher order in derivatives from ladder construction}

The supersymmetric methods of the previous section was based on the superalgebra (\ref{algebra}) which includes: supersymmetry, i.e. conservation of the supercharges $Q^{\pm},$ (quasi)factorization of the super-Hamiltonian components, and the nilpotency of supercharges which is provided by their matrix form. Just the supersymmetry (i.e. the intertwining relations) guarantees the (almost exact) isospectrality between the components of super-Hamiltonian with non-singular potentials, providing the most interesting practical results of the method. This is a reason to consider generalized SUSY with supercharges of higher order in derivatives, and therefore, without quasi-factorization of Hamiltonians in terms of components of supercharges. In the context of Darboux transformation \cite{darb}, such a generalization of the intertwining relations (\ref{intertw}) corresponds to the so called Crum transformation \cite{crum}:
\be
h^{(0)} q^+_N = q^+_N h^{(1)} , \quad q^-_N h^{(0)}= h^{(1)} q^-_N , \label{intertwN}
\ee
where  $N$-th order differential operators are:
\be
q^{\pm}_N =\sum_{k=0}^N w^{\pm}_k (x)\partial^k, \quad w^{\pm}_N
\equiv (\mp 1)^N \label{crum}
\ee
with differentiable coefficient functions. Herein let us take them real and  the pairs of
intertwining operators to be mutually Hermitian conjugated $q^{-}_N = (q^{+}_N)^\dagger$.

In the literature there are several synonyms for the higher-order
SUSY algebra: originally it was named as a
Polynomial (or Higher-derivative) one \cite{ais,acdi95},
later
a more general term of Nonlinear SUSY has
been used \cite{kliplyu00,plyush04} in  a certain relationship to a nonlinear SUSY
algebra arising in the conformal QM \cite{ivanov}. The title of $N$-fold SUSY has been also suggested in
\cite{ast00}.

One way to achieve the non-linear SUSY is in combining  of several linear ${\cal N} = 1$ SUSY QM systems in the fermion number
representation which are implemented by nilpotent supercharges of
 first order in derivatives \gl{q} built from a real super-potentials $W(x)$,
\be
\fl
q^\pm_1\equiv r^\pm = \mp\partial + \chi(x);\qquad \chi(x) \equiv (\partial W)(x)
\label{char1}
\ee
In the latter case the SUSY algebra is completed \gl{algebra} by the appropriate factorization \gl{factorization}, \gl{partner} of
the super-Hamiltonian \cite{schr,inf}.

Let us proceed by recursion:
from a simple Darboux transformation
to a ladder \cite{suku1} or a dressing chain \cite{adler} made of several simple Darboux steps. To produce the required
transformation operators the two different linear SUSY systems
may be overlapped. Namely, consider two super-Hamiltonians
\mbox{${H}_i, \ i = 1,2$}, Eq.~\gl{2times2}, respectively
two sets of supercharges $Q^\pm_i$ with coefficient functions $\chi_i = (\partial W_i)$ and
supercharge components \mbox{$r_i^\pm = \mp\partial + \chi_i $}.
Let us match two elements of super-Hamiltonians with a constant shift,
 \be
h^{(1)}_1 = h^{(0)}_2 +\lambda;\quad \chi^2_1 +\chi'_1= \chi^2_2 -\chi'_2 +\lambda .
\label{glue1}
\ee
Evidently,  the ground state energies for the above Hamiltonians  $ \bar E_1, \bar E_2$ satisfy the inequalities $ \bar E_1\geq \lambda \geq  - \bar E_2$ .
The constant shift of the super-Hamiltonian
$H_2 \rightarrow H_2 +\lambda$ does not break or change its supersymmetry but modify the superalgebra.

After matching the chain of intertwining relations \gl{intertw}
can be assembled to produce the final Hamiltonian $h^{(1)}_2$ only
from the initial one, $h^{(0)}_1$ by means of
a second-order Darboux transformation,
\be
\fl h^{(0)}_1 r^+_1 r^+_2 = r^+_1 h^{(1)}_1 r^+_2  =  r^+_1 (h^{(0)}_2 + \lambda) r^+_2 =
 r^+_1 r^+_2  (h^{(1)}_2 + \lambda) ; \quad
r^-_2 r^-_1 h^{(0)}_1  = (h^{(1)}_2 + \lambda) r^-_2 r^-_1 . \label{intertw2}
\ee
Following the spectral design  one concludes that
the two-step ladder SUSY dynamics contains the redundant information, namely,
about the intermediate Hamiltonians  $h^{(1)}_1 = h^{(0)}_2 +\lambda$. Let us remove it
 and define
the two (almost) isospectral components,
\ba
&&h^{(0)} \equiv h^{(0)}_1  + \lambda_1 =  r^+_1 r^-_1 + \lambda_1;\quad
h^{(1)} \equiv h^{(1)}_2 + \lambda_2 =  r^-_2 r^+_2 + \lambda_2;\label{def2}\\
&&r^-_1 r^+_1 + \lambda_1 =  r^+_2 r^-_2 + \lambda_2; \label{glue2}
\ea
for the super-Hamiltonian \gl{2times2} for which we have employed an
energy reference shifted by arbitrary $\lambda_{1,2}$,  $\lambda =  \lambda_2 - \lambda_1$ .
Then the intertwining relations \gl{intertwN} are identical to Eq.~\gl{intertw2}
with $q^+_2 =  r^+_1 r^+_2 = (q^-_2)^\dagger $ and the supersymmetry  $[H, Q^+_2] = [H, Q^-_2]
= 0$ is generated by the conserved nilpotent supercharges,
\be Q^+_2=\left(\begin{array}{cc}
 0 &  0\\
q^-_2 & 0
\end{array}\right)= (Q^-_2)^\dagger,\quad  (Q^+_2)^2 = (Q^-_2)^2 = 0 .
\label{char2}
\ee
Now in virtue of \gl{def2} the algebraic closure of the SUSY algebra
 is given by,
\be
 \{Q^+_2 , Q^-_2\} =   \left(\begin{array}{cc}
 r^+_1 r^+_2 r^-_2 r^-_1 & 0 \\
0 &   r^-_2 r^-_1 r^+_1 r^+_2
\end{array}\right) =   (H - \lambda_1) (H - \lambda_2), \label{polyn2}
\ee
generalizing Eq.\gl{algebra}.
Thus we have obtained the second-order Polynomial SUSY algebra
\cite{ais}-\cite{fern96} as a concise
form of isospectral deformation of a potential system
realized by a ladder
\cite{coop}--\cite{berez} or a dressing chain
\cite{adler}--\cite{shab2} of a couple of one-step
Darboux transformations
or, equivalently,
by a second-order Crum-Darboux intertwining \cite{crum} or
by a blocking of two linear SUSY with partial overlapping of
super-Hamiltonians \cite{ais} \mbox{(``weak SUSY'' \cite{smilga})},
or by a tower of para-SUSY transformations \cite{ruba}--\cite{aisv-ps} (see how to proceed in \cite{Andrianov:2004vz}) or by a reduction of spin-j parasupersymmetry \cite{9903130}.

The {\it formal}  zero-modes\footnote{ We mark them as "formal" because they are not supposed to be normalizable, just being solutions of the Schr\"odinger equation but not necessarily belonging to the set of spectrum eigenfunctions.} of intertwining operators $q^\pm_2$  form the basis of a
two-dimensional representation
of the super-Hamiltonian,
\[
h^{(0)}q^+_2 \phi^+_i (x) = 0 = q^+_2 h^{(1)}  \phi^+_i (x);\quad i = 1,2;\]
\be h^{(1)}  \phi^+_i (x) = \sum_{j=1}^2 S^-_{ij}   \phi^+_j (x);\quad
h^{(0)}  \phi^-_i (x) = \sum_{j=1}^2 S^+_{ij}   \phi^-_j (x), \label{zerom1}
\ee
due to intertwining relations \gl{intertwN}, \gl{intertw2}.
In terms of these Hamiltonian projections -- constant matrices $S^\pm$ ,
the SUSY algebra closure
takes the polynomial form \cite{ansok} (see also \cite{ast00},
\be
\left\{Q^+_2,  Q^-_2\right\} =
{\rm det} \left[E{\bf I}- S^{+}\right]_{E = H}
={\rm det} \left[E{\bf I}- S^{-}\right]_{E = H} \equiv
{\cal P}_2 (H) . \label{clos2}
\ee
Thus both matrices have the same set of eigenvalues which for the ladder
construction \gl{polyn2} consists of $\lambda_1, \lambda_2$ . As
the formal zero-mode set is not uniquely derived from \gl{zerom1} the matrices
$S^\pm$ are not necessarily diagonal. For instance, the equation
$r^+_1 r^+_2 \phi^+ (x) = 0$ has  one formal zero-mode $\phi^+_2$ obeying
$r^+_2 \phi^+_2 (x) = 0$ and another one obeying
\mbox{$r^+_1 \tilde\phi^+_1 = 0;\,
 \tilde\phi^+_1 = r^+_2 \phi^+_1 (x) \not= 0$.}
Evidently the zero-mode solution
 $\phi^+_1 (x)$ is determined up to an addition of  $\phi^+_2$. When multiplying
these linear equations by $r^-_2$ one proves with the help of
Eqs.~\gl{def2},\gl{glue2} that
\be
\fl (h^{(1)} - \lambda_2) \phi^+_2 (x) = 0;\quad (h^{(1)} - \lambda_1) \phi^+_1 (x)
= C  \phi^+_2 (x);\quad S^- = \left(\begin{array}{cc}
 \lambda_1 &  C\\
0 & \lambda_2
\end{array}\right), \label{cellgen}
\ee
where $C$ is an arbitrary real constant. If $\lambda_1 \not= \lambda_2$
then by the redefinition \mbox{$(\lambda_1 -\lambda_2) \bar\phi^+_1 \equiv
(\lambda_1 -\lambda_2) \phi^+_1 +  C  \phi^+_2 (x)$} one arrives to the diagonal
matrix  $S^-$. However in the {\sl confluent case},
$\lambda_1 = \lambda_2 \equiv \lambda,\, C\not=0$
it is impossible to diagonalize
and by  a proper normalization of the zero-mode  $\phi^+_1$
one gets the elementary Jordan cell,
\be
\fl (h^{(1)} - \lambda) \phi^+_2 (x) = 0;\quad (h^{(1)} - \lambda) \phi^+_1 (x)
=   \phi^+_2 (x);\quad S^- = \left(\begin{array}{cc}
 \lambda &  1\\
0 & \lambda
\end{array}\right). \label{cell2}
\ee
Therefore in the confluent case
the zero-mode  $\phi^+_1$ is not anymore a solution of the Schr\"odinger
equation but it is an adjoint solution  which can be derived
by differentiation of the first relation in \gl{cell2},  $\phi^+_1
=d\phi^+_2/d\lambda + c_1 \phi^+_2$ where $c_1$ is an arbitrary constant.
Still the
intermediate Hamiltonian $\tilde h =r^-_1 r^+_1 + \lambda
= r^+_2 r^-_2 + \lambda $ is well defined and therefore
the intermediate isospectral partner  $\tilde\phi^+_1 (x)$
of the zero-mode $\phi^+_1 (x)$ is a solution of Schr\"odinger equation
with the above Hamiltonian.
The analysis of the matrix  $S^+$ is similar.
Thus we convince ourselves that in general the Hamiltonian projection onto the
subspace of supercharge zero-modes is not diagonalizable but can
be always transformed into a canonical Jordan form.

To complete the description of
Polynomial SUSY algebras generated by a second-order
ladder one should take into consideration also the degenerate case
when \mbox{$\lambda_1 = \lambda_2 \equiv \lambda,\, C=0$}. For this choice
the matrix   $S^-$ is automatically diagonal and both zero modes
$\phi^+_{1,2} (x)$ are
(independent) solutions of the   Schr\"odinger
equation with  the  Hamiltonian  $h^-$. Then it can be proved \cite{ansok} that
the intertwining operator $q^+_2$ is just a linear
function of this Hamiltonian,
$q^+_2 = \lambda - h^{(1)}$. Hence the intertwining is trivial $h^{(1)} = h^{(0)}$ and
such supercharges must be ruled out due to their triviality. However for  SUSY of higher order $N \geq 3$ the
removal of such blocks from supercharges
 may lead to less-dimensional, ladder-irreducible SUSY algebras \mbox{(see Subsection 3.3).}
\subsection{Polynomial SUSY QM from a  ladder of linear SUSY systems }

Let us give the general description of the $N$-step
ladder of linear SUSY algebras which entails the Polynomial Superalgebra of $N$th-order \cite{ais}, \cite{acdi95}. We
introduce a set of first-order differential
operators for intermediate intertwinings ,
\be
r^\pm_l = \mp \partial + \chi_l(x),
\quad l = 1,...,N,
\ee
and the relevant number of
intermediate coefficient functions $\chi_l(x) = (\partial W_l)$ related to intermediate superpotentials $W_l(x)$. The set of the
initial, $h^{(0)} \equiv h_0$,
the final, $h^{(1)} \equiv h_N$ and intermediate Hamiltonians,
$ h_l = - \partial^2 + v_l (x)$ consists of Schr\"odinger operators,
nonsingular and real ones. They obey the matching
(ladder) relations,
\ba
 &&h_l \equiv r^-_l\cdot r^+_l  + \lambda_l =  r^+_{l+1}\cdot
r^-_{l+1} +
\lambda_{l+1} ,\quad l = 1,\ldots,N-1,\nonumber\\
&&
 h_0 \equiv h^{(0)}= r^+_1 \cdot r^-_1  + \lambda_1;\quad h_N \equiv h^{(1)} = r^-_{N} \cdot r^+_{N} + \lambda_{N}. \label{ladderN}
\ea
These ladder relations correspond to the (dressing) chain equations on
super-potentials, $W_l(x),\ \chi_l(x) = (\partial W_l)(x)$,
\be  v_l (x) =
(\chi_l(x))^2 +
(\chi_l(x))' + \lambda_l=
(\chi_{l+1}(x))^2 -
(\chi_{l+1}(x))' + \lambda_{l+1}
\ee
The corresponding intertwining (Darboux) transformations hold in each
adjacent pair of Hamiltonians,
\be
h_{l-1} \cdot r^+_l =  r^+_l\cdot h_{l},\quad  r^-_l\cdot h_{l-1} =
h_{l} \cdot r^-_l,
\ee
and therefore the chain of $N$ overlapping SUSY systems is properly built,
\ba
&&H_l = \left(\begin{array}{cc}
h_{l-1}& 0\\
0 & h_l
\end{array}\right),\qquad  R^-_l=\left(\begin{array}{cc}
 0 &  r^+_l\\
0 & 0
\end{array}\right)= (R^+_l)^\dagger;\nonumber\\
&&[H_l, R^+_l] = [H_l, R^-_l] = 0,\qquad
H_l - \lambda_l= \{R^+_l,R^-_l\},
\ea
Now let us omit a chain of intermediate Hamiltonians between
  $h^{(0)}$ and $h^{(1)}$ and produce the Higher-derivative $\simeq$
Polynomial
$\simeq$ Nonlinear SUSY  algebra for the super-Hamiltonian $H$ given in
Eq.\gl{2times2}. The intertwining between $h^{(0)}$ and $h^{(1)}$ is performed by
the Crum-Darboux operators,
\be
q^+_N = r^+_1 \ldots r^+_N, \quad q^-_N = r^-_N \ldots r^-_1 . \label{inop}
\ee
The SUSY symmetry
$[H, Q^+_N] = [H, Q^-_N] =0, $ is accomplished by the supercharges of the
same matrix structure \gl{char2} and the super-Hamiltonian is
represented by finite-dimensional matrices on the subspaces of
supercharge zero-modes,
\[
q^\pm_N \phi^\pm_i (x) = 0;\quad i = 1,2,\ldots,N;\]
\be h^{(0)}  \phi^-_i (x) = \sum_{j=1}^N S^+_{ij}   \phi^-_j (x);\quad
h^{(1)}  \phi^+_i (x) = \sum_{j=1}^N S^-_{ij}   \phi^+_j (x)
\label{zeromN}
\ee
due to intertwining relations \gl{intertwN}.
In terms of the constant matrices $ S^\pm$ ,
the algebraic closure is given by  a non-linear SUSY
relation
\cite{ansok, ast00},
\be
\fl \left\{Q^+_N, Q^-_N\right\} =
{\rm det} \left[E{\bf I}-{S}^{+}\right]_{E = H}
={\rm det} \left[E{\bf I}-{S}^{-}\right]_{E = H} \equiv
{\cal P}_N (H) = \prod^N_{l=1} \left(H - \lambda_l\right) . \label{polialg}
\ee
Again both matrices have the same set of eigenvalues which for the ladder
construction \gl{ladderN} consists of $\lambda_1,\ldots \lambda_N$ .
If the degenerate roots appear the normal (canonical) form of matrices
$ S^\pm$ may consist of nontrivial Jordan cells.
If all intermediate $h_l$
are hermitian, nonsingular, then $\lambda_l$ are real and each ladder
step is well defined.

However it turns out that not all higher-order intertwining operators can be built by a ladder algorithm based on linear SUSY elements with real and nonsingular intermediate Hamiltonians.

\subsection{Generation of Polynomial SUSY QM by ladder: irreducible SUSY blocks of type I, II, and III}

\hspace*{3ex} Let us elucidate the circumstances which may obstruct the SUSY ladder
decomposition of a Polynomial SUSY algebra.  The class of Polynomial
SUSY algebras can be extended admitting  complex $\lambda_l$  and singular $h_l$.
In fact, the full variety of building elements
for  non-linear SUSY can
be established within the class of
intertwining (or transformation) operators  $q^{\pm}_2$ of second-order in derivatives.
Eventually one has to find the basic set  of Crum-Darboux operators \gl{crum}
with nonsingular coefficient functions
which produce a nonsingular potential $V^{(1)}$ after intertwining  \gl{intertwN}
with the Hamiltonian possessing a smooth
initial potential $V^{(0)}$.

When taking a supercharge of second order in derivative with real
coefficient functions one always can find
 formal (not necessarily normalizable) real zero-modes of the intertwining operators $q^\pm_2$ and, further on,
 the $2\times2$  matrix representation \gl{zerom1} for the
super-Hamiltonian components $h^{(0)}, h^{(1)}$ by matrices ${S}^\pm$.
Then latter matrices
can be  obtained as real but, in general, not symmetric. Therefore the first obstruction
for the ladder decomposition may arise when the reduction to a
Jordan form does not yield real eigenvalues. For instance,
if $h^{(0)}  \phi^-_i (x) = \bar\lambda \epsilon_{ik} \phi^-_k (x)$ then
the eigenvalues of  ${S}^+ =  \bar\lambda \hat\epsilon$ are imaginary,
mutually conjugated $\pm i \bar\lambda$.  The possibility of complex pairs of
mutually conjugated roots in a Polynomial SUSY algebra
can be easily read off  from its closure \gl{clos2}
as for supercharges with real coefficients
polynomials ${\cal P}_2 (H)$ possess real coefficients.
We call this kind of {\it irreducibility} to be {\it of type I}.
Its elementary block corresponds to the polynomial
${\cal P}^{(I)}_2 (H) = (H + a)^2 + d,\quad d > 0$ and its analytical
properties have been investigated in \cite{acdi95}, \cite{Fern:2002}.
Some examples of related isospectral potentials are described
in \cite{deber2002}.

Next, one has to ensure the positivity of the SUSY algebra relation \gl{clos2}
in a particular differential realization of a
super-Hamiltonian $H$
with real non-singular potentials
and the supercharges $Q^+_N, Q^-_N$ (with $N=2$ in our case)
made of differential operators with real
coefficients.
Let the energy spectrum $E_j; j=0,1,\ldots; E_j < E_{j +1}$
of $H$ be discrete, for simplicity.
Then,
\be
 {\cal P}_N (E_j) = \langle Q_N^+\Psi_j | Q_N^+\Psi_j\rangle +
 \langle Q_N^-\Psi_j | Q_N^-\Psi_j\rangle \geq 0 , \label{ineq2}
\ee
if the action of supercharges is well defined in the Hilbert space spanned
on eigenfunctions of a super-Hamiltonian.
It can be  extended on a continuous energy spectrum
with usage of wave packets.

Thus for non-singular potentials the acceptable disposition
of polynomial roots related to zero-modes of a supercharge
ensures non-negative values of  ${\cal P} (E)$ for {\sl each} energy
  level of a Hamiltonian. Accordingly , the following positions
for polynomial real roots are acceptable(for a pair of complex,
mutually conjugated roots the positivity is obvious) .
\smallskip

\noindent
A.\quad $\lambda_{1} \leq \lambda_2 \leq E_0$ or
$\lambda_{1} = E_0;\, \lambda_2 = E_1 $.\\ The related SUSY algebra is well embedded into
 a chain or ladder realization. It is
 reducible  because one can gradually add/remove $\lambda_1$ and then $\lambda_2$
without breaking the positivity of intermediate SUSY algebra.
The coincidence of roots and energies correspond to
the isospectral transformation with deleting/inserting energy levels. For instance,
if $\lambda_{1} = E_0;\, \lambda_2 = E_1 $ then two pairs of zero-modes
of $q^\pm_2$ can be chosen as solutions of  Schr\"odinger equations
with Hamiltonians $h^{(0)}, h^{(1)}$. One can implement
\cite{suku1} the energy levels $E_0, E_1$ to appear in
any of the Hamiltonians $h^{(0)}, h^{(1)}$ but each level only once, either in $h^{(0)}$ or in $ h^{(1)}$.
\smallskip

\noindent
 B.\quad  $E_0 < \lambda_{1} < \lambda_2 \leq E_1$ or
$E_j \leq \lambda_{1} < \lambda_2 \leq E_{j+1},\, 1\leq j$.\\ A pair of real
roots is inserted between adjacent energy levels. Such an algebra
cannot be decomposed into a chain of two linear SUSY with non-singular intermediate potentials as the removal of
any of roots $\lambda_{1,2}$ immediately breaks the positivity in \gl{ineq2}.
Then the intermediate Hamiltonian
acquires inevitably a  singular potential which
usage would lead to the loss of isospectrality.
The related Darboux transformations had been known in 50ties \cite{krein}.
We call this {\it irreducibility} to be {\it of type II}.
The examples and certain theorems are given in \cite{samsirr1},
\cite{deber2002} (see also \cite{sok11} for an exhaustive analysis of this type of irreducible  Darboux transformations).
\smallskip

\noindent
 C.\quad   $E_0 < \lambda_{1} = \lambda_2 \leq E_1$ or
$E_j < \lambda_{1} = \lambda_2 \leq E_{j+1},\, 1\leq j$.\\ This is a confluent
case which seems to be obtained as a limit of the previous one. However, one-dimensional QM does not allow degenerate levels and,
besides, the matrix projection for the corresponding super-Hamiltonian
contains a non-trivial Jordan cell. Therefore we specify this case as a separate one.
This kind of {\it irreducibility} is named to be {\it of type III}. One may find
more information on the analytical properties of related potentials in
\cite{Ferna:2003}.

One can use these second-order blocks to build a $N$th-order
Polynomial SUSY system. Their general form is again given by
Eq.~\gl{polialg} if to accept the presence of complex conjugated roots
$\lambda_l$.

Meanwhile it seems that a pair of supercharge zero-modes or even
a pair of new excited energy levels of the super-Hamiltonian
can be {\it always} inserted by successive application of an appropriate ladder construction described in the previous
Section using first-order intertwining transformations between regular
Hamiltonians. But the order of
a relevant ladder of first-order transformations and
respectively of the final Polynomial SUSY will be certainly higher
than two.
We come to the problem of possible relationship between first-order reducible
and irreducible SUSY algebras having the same super-Hamiltonian.

The related question concerns the degenerate roots. These roots  are distributed between different Jordan
cells in the matrices $S^\pm$. The problem is on how many Jordan
cells
may coexist for the same eigenvalue and  what is their role in the
supercharge structure.

\subsection{Minimization of SUSY algebra for a given Hamiltonian and emergence of irreducible blocks of type II and III}

\hspace*{3ex} Let us reveal a possible redundancy in supercharges which can
be eliminated without any changes in the super-Hamiltonian.
There exists a possibility when the intertwining operators
$q^\pm_N$ and
$p^\pm_{N_1}$  for $N > N_1$ are related by a polynomial factor $F(x)$
depending on
the Hamiltonian,
\be
q^+_N =
F (h^{(0)}) p^+_{N_1}
=  p^+_{N_1}  F (h^{(1)});\quad q^-_N =
F (h^{(1)}) p^-_{N_1}
=  p^-_{N_1}  F (h^{(0)}). \la{triv}
\ee
Obviously in
this case the reduction
to the second supercharge does not result in any modifications of potentials.

Thus the problem consists in how to factorize a minimal essential part of the
supercharge and avoid numerous SUSY algebras generated by means of ``dressing''
\gl{triv}.

Let us  illustrate how it could work in the following example:\\
the matrix ${S}^-$ for the minimizable intertwining operator $k^+_3$
with Jordan cells of different size having
the same eigenvalue. It is generated by the operators,
\be
p^\pm = \mp\partial + \chi,\quad h^{(1)} =p^- p^+ + \lambda,\quad
k^+_3 = - p^+ p^- p^+ =p^+ (\lambda - h^{(1)}).
\ee
Respectively, the basis of formal zero-modes (eigen- and associated functions) for the intertwining operator $k^+_3 \phi_j = 0$ generates the non-diagonal matrix ${S}^-$,
\be
\fl \begin{array}{cccc}
\phi^+_1:\,& p^- p^+\phi^+_1 = \phi^+_2 &\longrightarrow&
h^{(1)} \phi^+_1 = \lambda \phi^+_1 + \phi^+_2;\\
\phi^+_2:\,& p^+\phi^+_2 = 0 &\longrightarrow& h^{(1)}\phi^+_2 = \lambda \phi^+_2;\\
\phi^+_3:\,& p^+\phi^+_3 \not= 0,\, p^- p^+\phi^+_3 = 0&
\longrightarrow &h^{(1)} \phi^+_3 = \lambda \phi^+_3;
\end{array}
\quad
{S}^- =\left(\begin{array}{ccc}
\lambda&1&0\\
0&\lambda&0\\
0&0&\lambda
\end{array}\right).
\ee
Thus the algebraic redundancy in the operator $k^+_3$ finds its unambiguous track in presence of two Jordan cells in the characteristic matrix ${S}^-$ with the same eigenvalues.

The supercharge components cannot be factorized
in the form \gl{triv}
if the polynomial ${\cal P}_N (x)$ in the SUSY algebra closure \gl{clos2}
does not reveal the degenerate zeros. Indeed the SUSY algebra closure
contains the square of polynomial $F(x)$, for instance,
\be
k^-_N k^+_N = F (h^{(1)}) p^-_{N_1} p^+_{N_1}  F(h^{(1)})
=  F^2 (h^{(1)}) {\cal P}_{N_1} (h^{(1)}),
\la{dzero}
\ee
where ${\cal P}_{N_1} (x)$ is a polynomial of
lower order, $N_1 \leq N-2$.
Therefore each zero of the polynomial $F (x)$ will produce a double zero
in the SUSY algebra provided by \gl{dzero}.

Thus the absence of double zeros
is sufficient to deal with the SUSY charges non-factorizable in the sense
of Eq.~\gl{triv}. However it is not necessary because the degenerate zeros
may well appear in the ladder construction giving
new pairs of isospectral potentials (see, for instance, \ci{acdi95} for the
Polynomial SUSY of second order).

Now we unravel the origin of irreducible, type-II and -III
transformations based on the strip-off factorization.
Let us consider an example of
irreducible SUSY of type II with supercharges of second order
in derivatives (see previous Subsection).  Suppose that it realizes
insertion of two new energy levels between the ground
and first excited states. Then three lowest energy
levels $E_0 < E_1 < E_2$ are of importance to study the relevant SUSY
systems: the ground state level
is degenerate between SUSY partners $h^{(0)}$ and
$h^{(1)}$, {\it i.e.} $E^{(0)}_0 = E^{(1)}_0$ whereas the two excited levels are
present only in the spectrum of $h^{(1)}$.

Certainly one can use the ladder construction \gl{ladderN}--\gl{inop}
to prepare the same
spectral pattern. For this purpose, the intertwining operators
\gl{inop}
of forth order in derivatives must be employed.
Indeed, one can prescribe
the ladder steps for $q^\pm_4$ as follows: start from a
pair of isospectral Hamiltonians with ground state energies $E_3$;
generate the level $E_0$ in the Hamiltonian $h^{(0)}$ using the
intertwining operators $r^\pm_1$, then sequentially
create in the spectrum of $h^{(1)}$ the state with energy
$E_2 < E_3$ by means of $r^\pm_2$, next the energy level $E_1 < E_2$ using
$r^\pm_3$ and finally the ground state with energy $E_0 < E_1$
exploiting $r^\pm_4$.  These elementary
steps are reflected in zero-modes of the
intertwining operators $q^+ = r^+_1 r^+_2 r^+_3 r^+_4$ and
 $q^- = r^-_4 r^-_3 r^-_2 r^-_1$. Namely the ground state of $h^{(0)}$ is
a zero-mode of $r^-_1$ ({\it i.e.} of $q^-$) whereas the states of
$h^{(1)}$ for
 $E_0, E_1, E_2$  are annihilated by the product $r^+_2 r^+_3 r^+_4$
({\it i.e.} by $q^+$) according to Eq.~\gl{zeromN}. As the ground
state energies coincide for $h^{(0)}, h^{(1)}$ the Hamiltonian
projections on the $q^\pm_4$ zero-mode space -- the matrices
${S}^\pm$ are, in general, not diagonalizable but have one rank-two
Jordan cell each. Thus, for instance,
\be
{S}^- =\left(\begin{array}{cccc}
E_0&0&0&C\\
0&E_2&0&0\\
0&0&E_1&0\\
0&0&0&E_0
\end{array}\right)\quad \Longrightarrow \quad
\tilde{S}^- =\left(\begin{array}{cccc}
E_0&C&0&0\\
0&E_0&0&0\\
0&0&E_2&0\\
0&0&0&E_1
\end{array}\right) ,  \label{cell4}
\ee
where a non-zero constant $C$ can be normalized to $C=1$.
The canonical Jordan form $\tilde {S}^-$ in \gl{cell4} is achieved by means of
re-factorization
\mbox{$q^+ = r^+_1 r^+_2 r^+_3 r^+_4 = r^+_1 \tilde r^+_2 \tilde r^+_3
 \tilde r^+_4$} so that the generation of ground state for $h^{(0)}$ is
associated now with $\tilde r^\pm_2$.
Respectively, the Polynomial
SUSY algebra shows up one degenerate root,
\be
{\cal P}_4 (H) = (H- E_0)^2(H- E_1)(H- E_2) . \label{alg4}
\ee
Evidently, this fourth-order algebra
cannot be optimized to a lower-order one because  there is no
replication of roots in different Jordan cells of matrices $\tilde {\bf
  S}^\pm$ (see the "Strip-off" theorem on minimization in the next Section).  However one may perform fine-tuning of Darboux
transformation parameters to provide the constant $C = 0$ in \gl{cell4}. This
peculiar choice provides two rank-one cells in \gl{cell4} with the same
eigenvalue $E_0$. The SUSY algebra is still given by Eq.~\gl{alg4} but
the intertwining operators reveal a redundancy,
\be
q^+_4 =  (h^{(0)} - E_0)\, q^+_2.
\ee
By construction the
left-hand side of this relation is fully factorizable in elementary
binomials $r^+_j$ with hermitian nonsingular intermediate Hamiltonians.
But in the right-hand side the operator  $q^+_2 = \tilde r^+_3
 \tilde r^+_4$ does not admit a further
factorization with a nonsingular intermediate Hamiltonian
because after removal of the redundant factor $(h^{(0)} - E_0)$ such a
factorization is forbidden by the positivity of the SUSY algebra,
Eq.~\gl{ineq2}.

One can easily extrapolate the previous argumentation onto the case of
additional
degeneracy of excited levels $E_1 = E_2$ to analyze the irreducible
SUSY of
type III. Thus
we come to the conjectures that:\\
a) the factorization \gl{inop} of intertwining operators $q^\pm_N$ is
not unique and there exist options to have more reducible ladders and less
reducible ones with a larger number of singular intermediate Hamiltonians;\\
b) irreducible algebras of type II and III can be
identified with special cases of fully reducible higher-order  algebras of ladder type
when the Hamiltonian projections ${S}^\pm$ have an appropriate
number of pairs of Jordan cells with coinciding eigenvalues.

We notice that one may save time in calculations
when instead of a more lengthy binomial ladder construction the isospectral  irreducible transformations of type II or III are exploited to embed pairs of energy levels between two excited
ones. In the next section the results of a more rigorous investigation of the relationship
between the reducible and irreducible intertwinings are presented.

\section{Rigorous  results on structure and (ir)reducibility of Polynomial SUSY}
Let us summarize the organization of Polynomial SUSY QM in a more systematic way.
In this Section the single supercharge components are analyzed.
In the general situation when the unique supercharge exists the {\it minimization} procedure is the way to factor out the redundant polynomials of Hamiltonian to leave the part essential for isospectral transformation of potentials after intertwining. In the consequent Sec. 5 devoted to hidden symmetries a more specific class of reflectionless potentials is involved and the two inequivalent supercharges are revealed. Then for lowering of their order as differential operators an additional procedure of {\it optimization} can be employed \cite{ansok} which exploits linear combinations of different intertwining operators for a given pair of Hamiltonians with coefficients polynomial in Hamiltonians. Nevertheless, eventually the two supercharges of minimal order in derivatives survive after optimization and this very procedure is substantially based on the following theorems. We notice that the two theorems of subsec.5.1 are applicable to both non-periodic and periodic potentials whereas the two theorems of Subsec.5.2 have been proven for the special class of potentials (the Sokolov class) which does not include periodic potentials.
\subsection{Basic theorems on the structure of QM with a nonlinear SUSY }

First the following theorem regulates the very structure of Polynomial SUSY in terms of the Hamiltonian projection of the kernels of intertwining operators.\\

{\bf Theorem 1 .}\\
{\it Let $\phi^\mp_n(x)$, $n=1$, \dots, $N$ be a basis in ${\rm{ker}}\,q_N^\mp$:
\be q_n^\mp\phi_n^\mp=0,\qquad q_N^-=(q_N^+)^\dagger.\ee
Then:

1) the action of the Hamiltonians  $h^{(0)}, h^{(1)}$ on the functions $\phi_n^\pm(x)$ is described by constant $N\times N$ matrices,
\be h^{(0)}\phi_n^-=\sum\limits_{m=1}^NS_{nm}^+\phi_{m}^-;\quad h^{(1)}\phi_n^+=\sum\limits_{m=1}^NS_{nm}^-\phi_{m}^+;\quad
n=1,\ldots,N;\la{hs}\ee

2) the closure of the supersymmetry algebra takes a polynomial form,
\be \{Q^+,Q^-\}=\det[E{\bf I}-{S^+}]_{E=H}=\det[E{\bf I}-{S^-}]_{E=H}
\equiv {\cal P}_N(H),\ee
where $\bf I$ is an identity matrix and $S^\pm$ is the matrix with entries
$S_{nm}^\pm$.}

{\bf Corollary  1.} The spectra of the matrices  $S^+$ and $S^-$ are equal.

Now we describe the minimization ("Strip-off") theorem.

A basis in the kernel of the intertwining operator in which the matrix
$S$ of this operator has a Jordan form  is called {\it canonical};
elements of a canonical basis are called  {\it transformation functions}.

The potentials $V^{(0)}(x)$ and $V^{(1)}(x)$ of the Hamiltonians $h^{(0)}$ and $h^{(1)}$ are interrelated by the equation \cite{matv}
\be V^{(1)}(x)=V^{(0)}(x)-2[\ln W(x)]'',\la{v2v1}\ee
where $W(x)$ is the  Wronskian of elements of an arbitrary (a canonical as well)
basis in ${\rm{ker}}\,q_N^-$.

The intertwining operators $q_N^\pm$ are called {\it  minimizable } if these operators can be presented in the form
\be q_N^+={\cal P}(h^{(0)}) p_{M}^+=p_{M}^+
{\cal P}(h^{(1)}),\quad q_N^-={\cal P}(h^{(1)}) p_{M}^-=p_{M}^-
{\cal P}(h^{(0)}),\la{minim}\ee
where $p_{M}^\pm$ are  operators of order  $M$ which intertwine the same Hamiltonians as  $q_N^\pm$  and ${\cal P}(x)$ is a polynomial of degree $(N-M)/2>0$. Otherwise the intertwining operators  $q_N^\pm$ are named as  {\it
non-minimizable}.

The following theorem contains necessary and sufficient conditions under which an intertwining operator $q_N^+ = (q_N^+)^\dagger$ is minimizable or not (a proof can be found in \cite{ansok}).

{\bf ("Strip-off") Theorem 2 on minimization  of an intertwining operator $q_N^+$ (and equivalently $q_2^-$ ) } \\
{\it An intertwining operator $q_N^+$ can be presented in the form
\be q_N^+= p_{M}^+\prod\limits_{l=1}^m(\lambda_l-h^{(1)})^{\delta k_l}, \ee
where $p_{M}^+$ is a non-minimizable operator intertwining the same Hamiltonians as   $q_N^+$ (so that $p_{M}^+ h^{(1)}=h^{(0)}
p_{M}^+$),
 if and only if  a Jordan form of the matrix  $S^+$ of the operator
$q_N^+$ has $m$ pairs (and no more) of Jordan cells with equal eigenvalues $\lambda_l$ such that, for the $l$-th pair,  $\delta k_l$ is  an order of the smallest cell and  $k_l+\delta k_l$ is an order  of the largest cell . In this case,
$M=N-2\sum_{l=1}^m \delta k_l = \sum_{l=1}^n k_l$, where the  $k_l$, $m+1\leqslant
l\leqslant n$ are orders of the remaining {\it unpaired} Jordan cells.}\\

{\bf Remark 1.} A Jordan form of the  matrix  $S^+$ of the intertwining operator
$q_N^+$ cannot have more than two cells with the same eigenvalue
 $\lambda$; otherwise  ${\rm{ker}}(\lambda
-h^{(1)})$ includes more than two linearly independent elements.

{\bf Corollary 2.} Jordan forms of the matrices  $S$ of the operators $q_N^+$ and $q_N^-$ coincide up to permutation of Jordan cells.

{\bf Remark 2.} The Theorems 1 and 2 are valid for a broad class of potentials including the periodic ones.

If a Jordan form of the matrix
$S$ of an intertwining operator  has  cells of order higher than one, then
the corresponding canonical bases contains not only formal solutions of the Schr\"odinger equation but also formal associated functions, which are defined as follows \cite{naim}.

A function $\psi_{n,i}(x)$ is called a {\it formal associated function of $i$-th order} of the Hamiltonian  $h$ for a spectral value
$\lambda_n$ if
\be
(h-\lambda_n)^{i+1}\psi_{n,i}\equiv0,\quad \mbox{and}\quad
(h-\lambda_n)^{i}\psi_{n,i}\not\equiv 0 . \label{canbas1}\ee
The term "formal" emphasizes that this function is not necessarily normalizable (not necessarily belongs to $L_2(\mathbb R)$) and therefore it does not belong to the energy spectrum and, accordingly, is not included into the spectral resolution of identity. In particular,
an associated function  $\psi_{n,0}$ of zero order is a formal eigenfunction of
 $h$ (not necessarily a normalizable solution of the homogeneous Schr\"odinger equation). They appear in the subspace of zero-mode solutions for irreducible intertwining operators and play role in building S-matrix of the Hamiltonian projection.

\subsection{Classification of really (ir)reducible SUSY trans\-for\-ma\-tions}
Assume that the intertwining operators
 $q_N^\pm$ are represented as products of the intertwining operators  $k_{N-M}^\pm$ and $ p_M^\pm$, $0<M<N$ so that
\ba &&\fl q_N^+=p_M^+k_{N-M}^+,\quad q_N^-=k_{N-M}^- p_M^-;\qquad
p_M^+h_M=h^{(0)} p_M^+,\quad p_M^- h^{(0)}=h_M p_M^-;\nonumber\\
 &&\fl k_{N-M}^+h^{(1)}=h_M k_{N-M}^+,\quad k_{N-M}^- h_M=h^{(1)}k_{N-M}^-,\quad h_M=-\partial^2+v_M(x), \la{fakt2}\ea
where the coefficients
 $k_{N-M}^\pm$ and $p_M^\pm$
as well as the potential  $v_M(x)$ may be complex and/or singular.  The Hamiltonian $h_M$ is called  {\it intermediate } with respect to  $h^{(0)}$ and  $h^{(1)}$. In this case, by  Theorem 1, the spectrum of the matrix  $S$ of the operator
$q_N^\pm$ is a union of the spectra of the matrices  $S$ for the operators
$k_{N-M}^\pm$ and $p_M^\pm$.

The intertwining operators
$q_N^\pm$ are called {\it reducible} if these operators can be presented as  products of two nonsingular intertwining operators (with real coefficients)
$k_{N-M}^\pm$ and $p_M^\pm,\ 0 < M < N$ so that Eqs.
\gl{fakt2} are valid and the  intermediate Hamiltonian $h_M$ has a real nonsingular potential.
Otherwise $q_N^\pm$ are called {\it irreducible}.

Irreducible, non-minimizable, intertwining operators of second order
with real coefficients are divided into  three types:
 \cite{Andrianov:2004vz}.

{\it An irreducible intertwining operator of I type } is a differential intertwining operator with real coefficients for which eigenvalues of the matrix
 $S$ have nontrivial imaginary parts and are mutually complex conjugate.

{\it An irreducible intertwining operator of the II type } is a differential intertwining operator $q_2^+$ (or equivalently $q_2^-$ )  of second order with real coefficients  such that:

(1) eigenvalues  of the matrix $S^+$ of the  operator $q_2^+$ are real and different;

(2) both elements  $\varphi^+_1(x)$ and $\varphi^+_2(x)$ of a canonical basis of ${\rm{ker}}\,q_2^+$ have zeros.

{\it An irreducible intertwining operator of the III type } is a differential intertwining operator $q_2^+$ (or equivalently $q_2^-$ )  of second order with real coefficients   such that:

(1)  the eigenvalues  $\lambda_{1,2}$ of the matrix $S^+$ of  the operator $q_2^+$ are equal, $\lambda_1=\lambda_2$;

(2) a canonical basis in ${\rm{ker}}\,q_2^+$ consists of  formal eigenfunctions,
$\varphi^+_{10}(x)$, and associated  functions, $\varphi^+_{11}(x)$, of the Hamiltonian $h^{(1)}$ which assemble into a Jordan cell,
$$h^{(1)}\varphi^+_{10}=\lambda_1\varphi^+_{10},
\qquad(h^{(1)}-\lambda_1)\varphi^+_{11}=\varphi^+_{10};$$

(3) $\varphi^+_{10}(x)$ has at least one root.

Other types of irreducible non-minimizable intertwining operators of second order do not exist.

Now let us state (in a concise form) two Theorems which characterize reducibility of intertwining operators of any order (proven in \cite{anso1,sok1,sok11}):\\

{\bf Theorem 3. (on reducibility of "dressed" non-minimizable intertwining operators)}\\
It asserts that for any non-minimizable intertwining operator with {\it real spectrum of the matrix}
 $S$, one can "dress" it, i.e. find and multiply by an  appropriate polynomial of the Hamiltonian, so that the resulting (certainly  minimizable) intertwining operator can be factorized into a product of intertwining operators solely of first order with intermediate real non-singular Hamiltonians. Thus any Polynomial SUSY algebra with real
 roots can be embedded into a higher order algebra corresponding to a ladder of linear SUSY.

{\bf Theorem 4. (on complete reducibility of non-minimizable intertwining operators)} \\
It asserts that any non-minimizable intertwining operator with {\it arbitrary spectrum of the matrix}
 $S$ can be factorized into a product of intertwining operators of first order and {\it irreducible second-order intertwining operators} of the
I, II and III type with intermediate real non-singular Hamiltonians. Still this factorization is not necessarily unique and, in particular, it may contain more or less blocks of first order.

These theorems have been formulated at full length and rigorously proven in \cite{anso1,sok1,sok11} for the class $K$ (Sokolov class) of potentials $V(x)$ such that:

1) $V(x)$ is a real-valued function from  $C_{\mathbb R}^\infty$;

2)  there exist numbers $R_0>0$, $C > 0$   (they
depend on $V(x)$) such that the inequality $\inf_{|x|\geqslant R_0}V(x)+C>0$ takes place for any $|x|\geqslant R_0$;

3) the functions
\be
\bigg(\int\limits_{\pm R_0}^x\sqrt{|V(x_1) + C|}dx_1\bigg)^2
\bigg({{|V'(x)|^2}\over{|V(x)+C|^3}}+{{|V''(x)|}\over{|V(x)+C|^2}}\bigg)
\ee
are bounded for  $x\geqslant R_0$ and $x\leqslant -R_0$, respectively.
It can be proven \cite{anso1} that the set $K$ is closed under intertwining of Hamiltonians.

The last condition  is not very
restrictive: for
example, it is fulfilled (for $x\ge R_0$),  for potentials: \hfill \hfill \linebreak
1)\qquad $V(x)=ax^\gamma[1+o(1)], \qquad
a>0,\quad\gamma>0;$\\
2)\qquad $V(x)=V_0+ax^{-\gamma}[1+o(1)], \qquad
V_0>0,\quad a\in\Bbb{R},\quad\gamma>0;$\\
3)\qquad $V(x)=ax^{\alpha}e^{bx^\beta}[1+o(1)], \qquad
a>0,\quad b>0,\quad\alpha\in\Bbb R,\quad\beta>0;$\\
4)\qquad $V(x)=V_0+ax^{\alpha}e^{-bx^\beta}[1+o(1)],\qquad
V_0>0,\,a\in\Bbb{R},\,b>0,\,\alpha\in\Bbb
R,\,\beta>0.$\\
A similar statement holds also for $x\le-R_0$. \\
However periodic potentials do not belong to this class $K$.


\section{ Hidden symmetry in One-dimensional SUSY QM}
 \subsection{Simplest SUSY with Hidden symmetry}
\hspace*{3ex}
The emergence of hidden symmetries happens to be an intrinsic feature of  SUSY algebra being related to so called central charges. Specifically in SUSY QM there are classes of potentials in super-Hamiltonians which structure is dictated by a hidden symmetry. In two (and more) dimensions the variety of such isospectral QM systems is rather typical and their spectra and eigenfunctions reveal partial or complete solvability accounted for by SUSY separation of variables (see Sections 8 and 9).

However hidden symmetries exist also in one-dimensional SUSY QM \cite{acin2000} if the same super-Hamiltonian conserves several non-trivial (non-minimizable and optimized\cite{ansok}) supercharges. Following the algorithm of minimization on the variety of supercharges one can apply the optimization process\cite{ansok} and prove that the maximal number of independent and non-minimizable (in the terminology of the previous Section) supercharges is two. Thus
the two pairs of conserved
supercharges $K^+, K^- $  and $P^+, P^-$  may form
two SUSY algebras for a hermitian Super-Hamiltonian $H$.

Let us examine the algebraic structure of the simplest non-linear SUSY
with two non-minimizable supercharges,
\ba
k^\pm &\equiv & \partial^2 \mp 2f(x)\partial + \tilde b(x) \mp f'(x) ; \no
p^\pm &\equiv & \mp\partial + \chi(x), \la{exgen}
\ea
induced by a complex supercharge
of second order in derivatives (see the preprint version of \cite{acin2000} and \cite{ansok}).
The supersymmetries \gl{extsusy}
generated by $K^+, K^- $  and $P^+, P^-$ prescribe that
\ba
V^{(0),(1)} &=&
\mp 2f' + f^2 + \frac{f''}{2f} - \left(\frac{f'}{2f}\right)^2 -
\frac{d}{4f^2} -a,\no
\tilde b &=& f^2 - \frac{f''}{2f} +\left(\frac{f'}{2f}\right)^2 +
\frac{d}{4f^2}, \label{Kint}\\
V^{(0),(1)} &=& \chi^2 \mp \chi',\label{Pint}
\ea
where $\chi, f$ are real functions and $a, d$ are real constants.
The related superalgebra closure for $K^+, K^- $  and $P^+, P^-$
takes the form,
\be
\{K^+, K^- \} = (H + a)^2 + d,\quad  \{P^+, P^-\} = H.
\la{secor}
\ee
The compatibility of two supersymmetries is achieved by solutions
of the following equations
\be
\chi = 2 f,\quad f^2 + \frac{f''}{2f} -
\left(\frac{f'}{2f}\right)^2 - \frac{d}{4f^2} -a = \chi^2 =
 4 f^2. \la{2equ}
\ee
Eq.~\gl{2equ} represents a nonlinear second-order differential equation
which solutions are parameterized by two integration constants. Therefore
the existence of two SUSY reduces substantially
the class of potentials for which they may appear.
Evidently Eq.~\gl{2equ} can be integrated into the first-order one,
\be
(f')^2 = 4 f^4 + 4 a f^2 + 4 G_0 f - d \equiv \Phi_4(f),
\la{firstor}
\ee
where $G_0$ is a real constant.

The solutions of this equation are elliptic functions which
 can be found in the implicit form,
\be
\int^{f(x)}_{f_0} \frac{df}{\sqrt{\Phi_4(f)}} = \pm (x - x_0),
\ee
where $f_0$ and $x_0$ are real constants.

They are nonsingular if:\\
a)  $\Phi_4(f)$ has three different real roots and the double root
$\beta/2$
is either the maximal one or a minimal one,
\be
 \Phi_4(f) = 4 (f - \frac{\beta}{2})^2
\left((f + \frac{\beta}{2})^2 - (\beta^2 -
\epsilon)\right),\quad
 0 < \epsilon < \beta^2. \la{polyn}
\ee
Then there exists a relation
between constants $a, d, G_0$ in terms of
coefficients $\beta,\epsilon$,
\be
a = \epsilon - {3 \beta^2\over 2} < 0,\quad G_0 = \beta
(\beta^2 - \epsilon),
\quad d = \beta^2  \left({3 \beta^2\over 4} - \epsilon \right) .
\la{defa}
\ee
The constant $f_0$ is
taken between the double root and a nearest simple root.\\
b)  $\Phi_4(f)$ has two different real double roots which corresponds
in \gl{polyn}, \gl{defa} to
$G_0 = 0,\quad \beta^2 = \epsilon > 0, \quad a = -\epsilon/2,\quad
d = -\epsilon^2/4$. The constant $f_0$ is
taken between the roots.

The corresponding potentials $V^{(0)},V^{(1)}$ are well known \ci{refl,refl1} to be
reflectionless, with one bound state at the energy $ (\beta^2 -
\epsilon)$ and
with the continuum spectrum starting from $ \beta^2$.
Respectively the scattering wave function is proportional to $\exp(ikx)$
with $k = \sqrt{E - \beta^2 }$.

In the case a) the potentials coincide in their form
and differ only  by shift in the coordinate (``Darboux Displacement''
\cite{Fernandez:2002wh}),
\be
V^{(0),(1)} =  \beta^2 -
\frac{2\epsilon }{\mbox{\rm ch}^2 \left(\sqrt{\epsilon}(x -
x^{(0),(1)}_0)\right)},\quad x^{(0),(1)} =x_0 \pm \frac{1}{4\sqrt{\epsilon}}
\ln\frac{\beta - \sqrt{\epsilon}}{\beta + \sqrt{\epsilon}},
\la{caseb}
\ee
and  in the case b) one of the potentials can be taken constant,
\be
V^{(0)} = \beta^2,\quad V^{(1)} =  \beta^2 \left(1 -
\frac{2}{\mbox{\rm ch}^2 \left(\beta (x - x_0)\right)}\right),
\la{casec}
\ee
For these potentials one can elaborate
the relations of extended SUSY algebra.

The initial algebra is given by the
relations \gl{secor}. It must be completed by the mixed anti-commutators containing symmetry
operators,
\be
\{ K^+, P^-\} = \{K^-, P^+\}^\dagger =
{\cal B} (H) + {\cal E} (H), \la{mixed}
\ee
where the first term is Hermitian and in general (see the next Subsection) a polynomial of the
super-Hamiltonian and
the second one is anti-Hermitian and not a polynomial of $H$. In the example under consideration the first
symmetry operator turns out to be constant, $ {\cal B} (H) = G_0$
after taking into account \gl{exgen} and \gl{firstor}.
Meanwhile the second
operator reads,
\be
{\cal E} (H) = \left[{\bf I}\,\partial^3 - \Bigl(a {\bf I} +
\frac32 {\bf V}(x)\Bigr) \partial
- \frac34 {\bf V}' (x)\right],
\la{oddrep}
\ee
in the notations $
H \equiv -\partial^2 {\bf I} + {\bf V}(x)$.
By construction the operator ${\cal E} (H)$ realizes a new symmetry
for the super-Hamiltonian.
Directly from Eq.~\gl{oddrep}
one derives  that,
\be
-{\cal E}^2 (H) = H \left[ (H + a)^2 + d\right]
 -  G_0^2  =  (H - E_b)^2 (H -  \beta^2),\la{quadr}
\ee
where $E_b = \beta^2 - \epsilon $ is the energy of a bound state.
Thus (some of) the zero modes of ${\cal E} (H)$ characterize either bound states or
zero-energy states in the continuum.
We remark that in the case b)
only the Hamiltonian $h^{(1)}$ has a bound state. Accordingly the physical
zero modes of
 ${\cal E} (H)$ may be either degenerate [case a), broken SUSY] or
 non-degenerate [case b), unbroken SUSY].

The square root in \gl{quadr} can be established  from the
analysis of scattering (transmission) coefficients,
\be
{\cal E} (H) = i (E_b - H) \sqrt{H -  \beta^2}.\la{root}
\ee
An unambiguous determination of this square root needs to specify the space of asymptotic "incoming" states on which this operator acts, i.e. the scattering condition which selects out the direction of scattering either from the left to the right or in the opposite direction. These two spaces do not overlap. In \gl{root} the first type of scattering condition has been selected: from the left to the right with the asymptotic limit of an incoming state, $x\to -\infty, \Psi(k,x) \to \exp(ikx), k>0$. For the alternative selection one should change the sign of square root.
We notice that
the symmetry operator \gl{oddrep}, \gl{root} is irreducible, {\it i.e.}
 the binomial $(E_b - H)$ cannot be removed. Indeed the
elimination of this binomial would lead to an essentially nonlocal
operator.

When taking Eq.~\gl{root} into account one finds (on the same space of scattering states) the mixed anti-commutators
of the extended  SUSY algebra \gl{mixed} in a non-polynomial form,
\be
\{ K^+, P^-\} = \{K^-, P^+\}^\dagger =
G_0 - i (H - E_b) \sqrt{H -  \beta^2}. \la{nonpol}
\ee
Thus the ``central charge'' of the extended SUSY is built of the elements
\gl{secor} and \gl{nonpol} and  cannot be diagonalized by a unitary
rotation with elements polynomial in $H$. Therefore the SUSY algebra is extended in the class of differential operators of finite order. The existence of polynomial relations between commuting operators can be in fact  established from the Burchnall-Chaundy theorem \cite{burchaun} as it was noticed recently in \cite{period-6}.
 \subsection{General case: one-dim SUSY algebra with
 Hidden symmetry}
Let us examine the very possibility to have
several supercharges for the same super-Hamiltonian. We remind
that a number of supercharges can be produced trivially with the help of
multiplication
on a polynomial of the Hamiltonian (see Section 4). Certainly such
supersymmetries are absolutely equivalent for the purposes of spectral
design and one must get rid of them. It was proved in \cite{ansok} that
the infinite set of possible supercharges for a given super-Hamiltonian can be always optimized so
that no more than two nontrivial supercharges remain and that an
optimal set of two real supercharges $K^\pm$ and $P^\pm$ with components of minimal order in derivatives contains
one operator of odd-order in derivatives and another one of even-order. They can be used to
generate all possible conserved supercharges $Q_i$ by ``dressing'' their components with polynomials of the Hamiltonians,
\be
q^\pm_i = \alpha_i^\pm k^\pm + \beta_i^\pm p^\pm .
\ee
Thus in one-dimensional (scalar) QM one may have the ${\cal N} = 1,2$ SUSY only.

Correspondingly suppose that the super-Hamiltonian $H$ admits two
supersymmetries with supercharges $K^\pm$ and $P^\pm$ made of differential
intertwining operators of order $N$ and $N_1$ respectively,
\be
[H, K^\pm] = [H, P^\pm]
= 0 . \la{extsusy}
\ee
The second pair of supercharges $P^\pm$
is assumed to be made of
differential operators of lower order $N_1 < N$.

To close the algebra one has to include
all anti-commutators between supercharges, {\it i.e.}
the full algebra based on two pairs of supercharges $K^\pm$ and $P^\pm$
with real intertwining operators. Two supercharges generate
two Polynomial SUSY,
\be
\left\{K^+, K^-\right\} = \tilde{\cal P}_N (H),\quad
\left\{P^+, P^- \right\} = \tilde{\cal P}_{N_1} (H). \la{2alg}
\ee

The closure of the extended, ${\cal N} =2$ SUSY algebra is given by
\ba
\left\{P^-, K^+ \right\} & \equiv & {\cal R}
= \left(\begin{array}{cc}
 p^+_{N_1} k^-_{N} &  0\\
0 & k^-_{N} p^+_{N_1}
\end{array}\right),\no
\left\{K^-, P^+\right\} &\equiv& {\cal R}^\dagger
= \left(\begin{array}{cc}
 k^+_{N} p^-_{N_1} &  0\\
0 & p^-_{N_1} k^+_{N}
\end{array}\right). \la{roper}
\ea
Evidently the
components of operators ${\cal R},\, {\cal R}^\dagger$
are differential
operators of $N + N_1$ order commuting with the Hamiltonians $h^{(0)}, h^{(1)}$,
hence they form symmetry operators ${\cal R},\, {\cal R}^\dagger$ for the super-Hamiltonian.
 However, in general, they are not
polynomials of the Hamiltonians $h^\pm$ and these symmetries impose certain
constraints on potentials.

All four operators
$\tilde{\cal P}_N (H),\,  \tilde{\cal P}_{N_1} (H),\, {\cal R},\, {\cal R}^\dagger$
commute each to other. The Hermitian matrix characterizing  this
 ${\cal N}=2$ SUSY,
\ba
{\cal Z} (H) = \left(\begin{array}{cc}
 \tilde{\cal P}_N (H) & {\cal R}  \\
{\cal R}^\dagger & \tilde{\cal P}_{N_1} (H)
\end{array}\right), \quad \mbox{\rm det} \left[{\cal Z} (H)\right] =
 \tilde{\cal P}_N \tilde{\cal P}_{N_1} -  {\cal R} {\cal R}^\dagger = 0, \la{centr}
\ea
is degenerate. Therefore it seems that the two supercharges are not
independent and by their redefinition
(unitary rotation) one might reduce the extended SUSY to an ordinary
${\cal N}=1$ one. However such rotations cannot be global (constant) and must
use non-polynomial, pseudo-differential operators as ``parameters''.
Indeed, the operator components of the ``central charge'' matrix ${\cal Z} (H)$
have different order in derivatives. Thus, globally the extended nonlinear
SUSY
deals with two sets of supercharges but when they act on
a given eigenfunction of the
super-Hamiltonian $H$ one could, in principle, perform the energy-dependent
rotation and
eliminate a pair of supercharges. Nevertheless this reduction can be
possible only after the constraints on potentials have been resolved.

Let us find the formal relation between
the symmetry operators ${\cal R}, \,{\cal R}^\dagger$ and the super-Hamiltonian.
These operators can be decomposed into a hermitian and an anti-hermitian
parts,
\be
{\cal B}\equiv \frac12({\cal R} +{\cal R}^\dagger) \equiv \left(\begin{array}{cc}
b^+ & 0\\
0 & b^-
\end{array} \right),\quad
{\cal E} \equiv \frac12 ({\cal R}^\dagger - {\cal R})\equiv  \left(\begin{array}{cc}
e^+ & 0\\
0 & e^-
\end{array} \right). \la{herm}
\ee
The
operator ${\cal B}$ is a differential operator of even order and therefore
it may be a polynomial of the super-Hamiltonian $H$. But if the
operator ${\cal E}$ does not vanish identically
it is a differential operator of {\it odd} order and
 cannot be realized by a polynomial of $H$.

It can be proven \cite{ansok} that the hermitian operator ${\cal B}$ is indeed a
polynomial of the super-Hamiltonian of the order $N_b \leq N -1$.
Let us use it to unravel the super-Hamiltonian content of the operator ${\cal E}$,
\be
{\cal E}^2 (H) = \tilde{\cal P}_N (H)
\tilde{\cal P}_{N_1} (H) - {\cal B}^2 (H), \la{secsym}
\ee
which follows directly from \gl{centr} and \gl{herm}. As
the (nontrivial) operator ${\cal E} (H)$
is a differential operator of odd order $N_e$ it may have only a
realization non-polynomial in $H$ being a square root of \gl{secsym} in
an operator sense. An unambiguous determination of this square root needs to specify the space of asymptotic "incoming" states on which this operator acts, i.e. the scattering condition (see the previous Subsection). The symmetry operator is certainly non-trivial if the sum of orders
$N + N_1$ of the operators $k^\pm_N$ and $p^\pm_{N_1}$ is odd and therefore
$N_e = N + N_1$.

The existence of a
nontrivial symmetry operator ${\cal E}$ commuting with the
super-Hamiltonian results in common eigenstates which however are not necessarily
physical wave functions. In general they may be combinations of two solutions
of the Shr\"odinger equation with a given energy, the physical and
unphysical ones. But if  the
symmetry operator ${\cal E}$  is
anti-Hermitian in the
Hilbert space spanned on the eigenfunctions of the super-Hamiltonian $H$ then
both operators have a common set of physical wave functions. This fact imposes quite rigid
conditions on  partner potentials $V^{(0),(1)}$. It can be proven \cite{ansok} that they
 inevitably belong to the
class of transparent or reflectionless ones \ci{refl,refl1}.
Such a
  symmetry has relations to the Lax method in the soliton theory
\cite{matv,leble}.

As the
symmetry operator ${\cal E}$  is
anti-Hermitian its eigenvalues are imaginary but, by construction,
its coefficients are real. Since the
wave functions of bound
states of the system $H$ can always be chosen to be real we conclude that they
must be zero-modes of the operator ${\cal E} (H)$,
\be
{\cal E} (H) \psi_i = {\cal E} (E_i) \psi_i = 0,\quad
  \tilde{\cal P}_N (E_i)\tilde{\cal P}_{N_1} (E_i) - {\cal B}^2 (E_i) = 0,
\la{zeroeq}
\ee
which represents the algebraic equation on bound state
energies of a system possessing two supersymmetries. Among solutions
of \gl{zeroeq} one reveals
also a zero-energy state at the bottom of
continuum spectrum.
\section{Shape invariance for SUSY related potentials}
\subsection{Shape invariance in linear SUSY }
 One of the goal of SUSY QM design consists in search for  exactly solvable models of Quantum Mechanics. For one-dimensional Schr\"odinger equation, a list of exactly solvable is well known: harmonic oscillator, Coulomb, Morse, P\"oschl-Teller, Scarf, Eckart potentials etc \cite{dabrowska}. Each of such potentials has its own history of solution, but all of them were reproduced in the framework of a single algebraic procedure of Factorization Method \cite{inf} in the middle of last century. The related method was formulated in the context of modern Supersymmetric Quantum Mechanics with the help of new notion - shape invariance - introduced by L.E.Gendenshtein in \cite{gend}.

The basic steps of standard shape invariance in {\it absence of spontaneous SUSY breaking} \cite{gend}, \cite{mallow} -- \cite{mallow-2}, \cite{cooper95} are the following. Let us consider a parametric family of one-dimensional superpartners $h^{(0)}(a), \, h^{(1)}(a)$ and first order supercharges $q^{\pm}(a)$ depending on a parameter $a.$  We say that the Hamiltonians are shape invariant when in addition to supersymmetrical intertwining relations (\ref{intertw}) they have the property (analogous to "cyclic" in \cite{shab1,shab2}),
\be
h^{(1)}(a)=h^{(0)}(\bar a) + {\cal R}(a),
\label{shape1}
\ee
where $\bar a=\bar a(a)$ is a modified value of parameter depending on $a$ and $ {\cal R}(a)$ is a ($c$-number) function of $a$ taken positive for definiteness.
Then the absence of spontaneous breaking of supersymmetry for all values of $a$ implies that just the lowest eigenvalue $E_0(a)$ of $h^{(0)}(a)$ vanishes, and the corresponding eigenfunctions $\Psi^{(0)}_0(a)$ are normalizable zero modes of $q^-(a).$  This property allows to solve the models (i.e. to find the spectrum and all bound state wave functions) algebraically.

To perform it we start from
\be
h^{(0)}(\bar a)\Psi^{(0)}_0(\bar a)= E_0(\bar a)\Psi^{(0)}_0(\bar a)=0.
\label{sch}
\ee
Consider the relation (\ref{shape1}) to obtain,
\be
h^{(1)}(a)\Psi^{(0)}_0(\bar a)={\cal R}(a)\Psi^{(0)}_0(\bar a).
\label{tildesch}
\ee
We notice that $\Psi^{(0)}_0(\bar a)\equiv \Psi^{(1)}_0(a)$
has no nodes and therefore is the ground state wave function of $h^{(1)}(a).$
The combination of (\ref{intertw}) and (\ref{tildesch}) yields,
\be
h^{(0)}(a) q^+(a)\Psi^{(0)}_0(\bar a)=
{\cal R}(a) q^+(a)\Psi^{(0)}_0(\bar a).
\label{RR}
\ee
Provided $q^+(a)\Psi^{(0)}_0(\bar a)$ is normalizable, we have generated an excited state of $h^{(0)}(a)$. These steps can be repeated up to the last step, where the resulting wave function $\Psi$ will not be normalizable anymore. There are notorious cases (oscillator-like potentials) where the spectrum is not bounded from above. The simplest and famous case is the harmonic oscillator with $q^\pm = \mp \partial + \lambda x$  and ${\cal R}(a) = 2\lambda$. By the cyclic algorithm one easily reconstructs the energy spectrum $E^{(0)}_n = 2n\lambda .$

It is clear that the isospectrality of $h^{(0)}(a)$ and $h^{(1)}(a)$ (up to the only zero mode $\Psi^{(0)}_0(a) )$ implies that there is no eigenvalue of $h^{(0)}(a)$ between zero and the ground state energy $E^{(1)}_0(a)$ of $h^{(1)}$. This observation can be used  to  prove that after a number of iterations one gets the entire spectrum of $h^{(0)}(a).$ Thus the shape invariance or cyclic method leads to spectrum generating algebras which allow to find algebraically all energy eigenvalues and eigenfunctions of shape invariant Hamiltonians in one-dimensional SUSY QM - complete solvability.

We notice that the shape invariance relation (\ref{shape1}) combined with SUSY factorization of Hamiltonians (\ref{factorization}), (\ref{partner}) entails the conventional Heisenberg algebra,
\be
[q^-, q^+] =  {\cal R}(a) ,
\label{heis0}
\ee
to supply $q^\pm$ with the meaning of creation and annihilation operators.
\subsection{Intertwining with shift and higher order
shape-invariance}
\hspace*{3ex}
The particular construction of higher order
shape-invariance can be realized \cite{acin2000} by breaking of hidden symmetry (\ref{extsusy}) for two Polynomial SUSY algebras (\ref{2alg})
,
\ba
\fl h^{(0)} k_N^+ &=& k_N^+ h^{(1)} ;\quad  k_N^+ k_N^- = \tilde{\cal P}_N (h^{(0)});\quad  k_N^- k_N^+ = \tilde{\cal P}_N (h^{(1)}); \label{inttM} \\
\fl h^{(0)} p_{N_1}^+ &=& p_{N_1}^+ (h^{(1)} + 2\lambda);\quad  p_{N_1}^+ p_{N_1}^- = \tilde{\cal P}_{N_1} (h^{(0)});\quad  p_{N_1}^- p_{N_1}^+ = \tilde{\cal P}_{N_1} (h^{(1)} + 2\lambda);  \label{inttq}
\ea
with the help of shift on a positive constant ${\cal R}(a) \equiv 2\lambda$.
In the sector of one of partner Hamiltonians (for example, $h^{(0)}$)  we obtain,
\be
h^{(0)} {\bf a}^+ = {\bf a}^+ (h^{(0)} + 2\lambda) ,\quad  {\bf a}^-  h^{(0)}=  (h^{(0)} + 2\lambda) {\bf a}^- \label{shape}
\ee
if introducing the product operators ${\bf a}^+ \equiv p_{N_1}^+ k_N^- =  ({\bf a}^-)^\dagger$ .
We will call such a Hamiltonian as "higher order
shape-invariant".
One can also work the other way around and start from (\ref{shape})
with a differential operator ${\bf a}^+$ of $N_1 + N$ order in derivatives, factorize the latter as a product
$p_{N_1}^+ k_N^-$ with nonsingular operator multipliers and find an auxiliary super-partner Hamiltonian $h^{(1)}$ to obtain (\ref{inttM}).

Eq.(\ref{shape}) is a spectrum generating, ladder ("dressing chain" \cite{shab1,shab2}) equation where ${\bf a}^+$ plays the role of
generalized creation operator which provides an excitation energy
of $2\lambda .$ In order to study the spectrum it is essential to determine
normalizable zero modes of ${\bf a}^-$ and ${\bf a}^+.$ The former ones describe the lowest lying
levels of the system for positive $2\lambda$ and one has to apply recursively
the operator ${\bf a}^+$ to them in order to generate the excitation spectrum.
The energies of the zero modes can be obtained from vanishing the average of the operator product
${\bf a}^+{\bf a}^- .$ In Nonlinear (Polynomial) SUSY this product can be  evaluated algebraically because
\cite{acdi95}
\be
{\bf a}^+{\bf a}^- = p_{N_1}^+ k_N^-k_N^+p_{N_1}^- = p_{N_1}^+ \tilde{\cal P}_N (h^{(1)}) p_{N_1}^-=\tilde{\cal P}_N (h^{(0)} - 2\lambda)\tilde{\cal P}_{N_1} (h^{(0)})
, \label{prod1}
\ee
see (\ref{2alg}) .
In contrast to the simple harmonic oscillator, one has also the possibility
of  zero modes of the operator ${\bf a}^+$. Accordingly in this case
the relevant operator product reads
\be
{\bf a}^-{\bf a}^+ = \tilde{\cal P}_N (h^{(0)})\tilde{\cal P}_{N_1} (h^{(0)} +2\lambda) ,\label{prod2}
\ee
where the polynomials contain zeros some of which correspond to zero modes of the operator ${\bf a}^+$.
Thus (\ref{shape}) provides a
connection between different levels of $h^{(0)}$ by a given shift, which is the simplest
realization of the notion of shape-invariance \cite{gend}.
Though some properties of the spectrum, like normalizable zero modes of ${\bf a}^{\pm}$,
will depend on the explicit product structure of ${\bf a}^{\pm},$
the equidistant excitation spectrum  and corresponding wave functions can be
mainly
obtained algebraically .

The generalized (or deformed) Heisenberg algebra is finally built by the following closure,
\be\fl
[{\bf a}^+,{\bf a}^-] = \tilde{\cal P}_N (h^{(0)} - 2\lambda)\tilde{\cal P}_{N_1} (h^{(0)}) - \tilde{\cal P}_N (h^{(0)})\tilde{\cal P}_{N_1} (h^{(0)} +2\lambda) \sim 2\lambda {\cal P}_{N+N_1-1} (h^{(0)}).\label{defheis}\ee
It is possible to study the certain consequences of (\ref{shape}) algebraically, using the deformed Heisenberg algebra  (\ref{defheis})
without taking into account the specific definition of the operators
${\bf a}^{\pm}$ although the physical spectrum is essentially based on normalizable zero modes of ${\bf a}^\pm$.

The important comment is in due. Taking into account the minimization recipe from Subsection 3.4 and the Theorem 2 from Subsection 4.1 one can find in certain cases that the order of operators $p_{N_1}^\pm$ and $k_N^\pm$ can be lowered without changing the potential in the Hamiltonian $h^{(0)}$.  However the appropriate factorization of redundant polynomials of the Hamiltonian can be typically realized only at the level of individual SUSY algebras and, in general, cannot be extended to the deformed Heisenberg algebra  (\ref{defheis}). Thus at the formal level there may be several algebras (\ref{defheis}) with the same spectrum pattern.

\subsection{Intertwining with shift and second order
shape-invariance}
\hspace*{3ex}
Let's elucidate the particular case when  ${\bf a}^{\pm}$
is of second order, in which
we can obtain a generalized singular harmonic oscillator which is also
shape-invariant.
Correspondingly we consider in
(\ref{inttM}), (\ref{inttq}) $k^\pm, p^\pm$ of first order.
We can thereby
explore that they  can
lead to nontrivial consequences.  The superpotentials which solve (\ref{inttM}),
(\ref{inttq}) are  growing linearly for large $x$ and behave
like $1/x$ at the origin just leading to two singular potentials \cite{gango}
$V^{(0)}(x)$ and $V^{(1)}(x)$ as follows,
\be\fl
V^{(0)}(x) = \frac{\rho(\rho - 1)}{x^2} + \frac{\lambda^2x^2}{4} +\lambda (\rho+\frac12);\qquad
V^{(1)}(x) = \frac{\rho(\rho + 1)}{x^2} + \lambda^2x^2 +\lambda (\rho-\frac12).
\label{sing}
\ee
These potentials are shape-invariant in the standard sense \cite{gend}
and belong to the class of algebraically solvable models,
because $V^{(1)}(x;\rho , \lambda) = V^{(0)}(x;\rho + 1 , \lambda) - 2\lambda .$
On the half-line with suitable boundary conditions for
wave functions \cite{acdi95} at the origin the above potentials can be interpreted
as  radial harmonic oscillators for integer values of $\rho$. However this system makes sense also for arbitrary real $\rho$.

The algebraic properties of these
systems are based on the following factorization,
\be\fl
h^{(0)} = k^+ k^- =p^+ p^- + \lambda (2\rho + 1);\quad h^{(1)}=  k^- k^+ = p^- p^+ + \lambda (2\rho - 1),\label{brokenhidd}
\ee
satisfying the relations (\ref{inttM}),
(\ref{inttq}) . The intertwining operators are given by,
\be
k^+ = (k^-)^\dagger = - \partial - \frac{\rho}{x} - \frac{\lambda x}{2};\quad p^+ = - \partial - \frac{\rho}{x} + \frac{\lambda x}{2} .
\ee
In terms of the product operators ${\bf a}^+ \equiv p^+k^-$ and
${\bf a}^- \equiv k^+p^-$  one obtains an algebra suggestive of a generalization
of the standard harmonic oscillator algebra,
\be
[ h^{(0)}, {\bf a}^\pm ] =\pm 2\lambda {\bf a}^\pm ,
\label{alg}
\ee
where the raising/lowering ("creation/annihilation") operators ${\bf a}^\pm$  can be presented in the form convenient for further evaluations,
\be
{\bf a}^+ = ({\bf a}^-)^\dagger = \exp\left\{\frac{\lambda x^2}{4}\right\}\left(-\partial^2 +\frac{\rho(\rho-1)}{x^2}\right) \exp\left\{- \frac{\lambda x^2}{4}\right\} .
\ee
Similar results hold for $h^{(1)}.$

Meanwhile the crucial polynomial algebra describes the individual products,
\be\fl
{\bf a}^+{\bf a}^- =  (h^{(0)} - 2\lambda) \big(h^{(0)} - \lambda (2\rho + 1)\big);\quad  {\bf a}^- {\bf a}^+ =
h^{(0)}\big(h^{(0)} - \lambda (2\rho - 1)\big) . \label{2polin}
\ee
The zeros of polynomials in \gl{2polin} may indicate the zero modes ("vacuum states") of the annihilation, ${\bf a}^-$ or creation, ${\bf a}^+$ operators. In the case under discussion it happens that the appropriate normalizable zero modes appear for the operators $k^+$ and $p^-$,
\be\fl
k^+ \psi_{0,1} = 0,\quad \psi_{0,1} = x^{-\rho}\exp\left\{ -\frac{\lambda x^2}{4}\right\};\qquad
p^- \tilde\psi_{0,2} = 0,\quad \tilde\psi_{0,2} = x^{\rho}\exp\left\{ -\frac{\lambda x^2}{4}\right\},
\ee
which entails the two zero modes for the annihilation operator,
\be\fl
{\bf a}^- \psi_{0,i} = 0 ;\quad \psi_{0,1} = x^{-\rho}\exp\left\{ -\frac{\lambda x^2}{4}\right\};\quad \psi_{0,2} = x^{1-\rho}\exp\left\{ -\frac{\lambda x^2}{4}\right\} .
\ee
Accordingly they are associated to two zeros of the first polynomial in \gl{2polin},
\be
 (h^{(0)} - 2\lambda) \psi_{0,1} = 0 = \big(h^{(0)} - \lambda (2\rho + 1)\big)\psi_{0,2} .
\ee
Another couple of zeros for the second polynomial in \gl{2polin} are not related to any normalizable solution.

 Eventually the oscillator algebra \gl{alg} generates one ladder of equidistant levels for $\rho \geq     \frac32$ or, equivalently for $\rho \leq - \frac12$  and two independent ladders of equidistant levels for $ - \frac12 < \rho < \frac12$ or, equivalently, $ \frac12 < \rho < \frac32$ . For $\rho = \frac12$ these two ladders coincide. The corresponding spectra of $h^{(0)}$ are: $E_{n,1} = 2\lambda (n+1)$ or $E_{n,2} = \lambda (2\rho +1 + 2n)$ for $n = 0,1,\ldots$.

The generalized Heisenberg algebra for this sort of oscillator reads,
\be
[ {\bf a}^+, {\bf a}^- ] = -4\lambda h^{(0)} + 2\lambda^2 (2\rho +1). \label{heis2}
\ee

There might be an (superficial) impression of the algebra \gl{alg} as to be a broken hidden symmetry  of the ${\cal N} = 2$ SUSY \gl{2alg}, \gl{roper}. However the contraction $\lambda \rightarrow 0$ results in the system with ${\bf a}^\pm \sim h^{(0)}$ and continuum spectrum
and therefore  no any nontrivial hidden symmetry emerges. One can find the deformed Heisenberg algebra \gl{heis2} arising in a more complicated setting of superconformal mechanics \cite{corrolmo}.

A {\it nontrivial}, shape-invariant, breaking of the ${\cal N} = 2$ SUSY arises for the third-order spectrum-generating algebra.

\subsection{Intertwining with shift and third order shape-invariance}
\hspace*{3ex} Let us now consider the intertwining operator $k^{\pm}$ to be
a reducible or irreducible operator of
second order but $p^{\pm}$ still of first order like in \gl{exgen} \cite{acin2000}, \cite{iof2004}.
Accordingly they are parameterized by\ba
k^\pm &\equiv & \partial^2 \mp 2f(x)\partial + \tilde b(x) \mp f'(x) ; \label{M^+}\\
p^\pm &=& \mp \partial + \chi(x) .
\label{q^+}
\ea
The solutions of (\ref{inttM}) are unchanged in respect to Eqs.(\ref{Kint}) of Subsection 5.1 but Eq.(\ref{inttq}) implies a shift for the
potential $V^{(1)} (x)$ in (\ref{Pint}). Thus
the consistency equations for $\chi(x)$ and $f(x)$ are modified.
The first Eq.(\ref{2equ}) becomes:
\be
\chi(x) = 2f(x) + \lambda x , \label{W}
\ee
where a  possible integration constant can be ignored because of
a shift of $x,$ which fixes the origin of coordinate.
Then the potential $V^{(0)}(x)$ can be written from (\ref{Kint}) as:
\be
V^{(0)}(x) = -2f^{\prime}(x) + 4f^2(x) + 4\lambda x f(x) + \lambda^2x^2 -
\lambda . \label{VV}
\ee
The coefficient in the intertwining operator $k^\pm$ reads,
\be
\tilde b = - \left( 2f^2 + 4\lambda x f + 4 \lambda^2 x^2 +a\right) .
\ee
In the second Eq. \gl{2equ} one has to replace $\chi^2$ in accordance with \gl{W}, i.e. $f(x)$ satisfies the following equation,
\be
f^{\prime\prime} = \frac{f^{\prime 2}(x)}{2f(x)} + 6f^3(x) +
8\lambda xf^2(x) + 2(\lambda^2 x^2 - \lambda +a) f(x) +
\frac{d}{2f(x)}.    \label{painleve}
\ee
The equation (\ref{painleve}) can be transformed by the substitution
$f(x)\equiv 1/2 \sqrt{\lambda} g(y); \,\, y\equiv \sqrt{\lambda}x$ to
the Painleve-IV equation \cite{painl1},
\be
g^{\prime\prime} =
\frac{g^{\prime 2}(y)}{2g(y)} + \frac{3}{2} g^3(y) +
4 yg^2(y) + 2(y^2 - \alpha)g(y) - \frac{b}{2g(y)}.
           \label{Painleve+}
\ee
where
\be
\alpha \equiv 1 - \frac{a}{\lambda};\quad b\equiv -\frac{4d}{\lambda^2} .
\label{ab}
\ee
This equation has been studied intensively in the
last years \cite{ferpain},\cite{marquette1}.In what follows  we will focus mainly
on asymptotic properties of its solutions which will determine the
asymptotics of potentials (\ref{VV}) and the normalizability of eigenfunctions.

The spectrum generating operators ${\bf a}^+ = p^+ k^- = ({\bf a}^-)^\dagger$ form the polynomial algebra,
\be
{\bf a}^+ {\bf a}^- = h^{(0)}\left[(h^{(0)} + a - 2\lambda)^2 +d\right];\quad {\bf a}^- {\bf a}^+ = [h^{(0)} +2\lambda]\left[(h^{(0)} + a)^2 +d\right] .\label{nonl3}
\ee
We show how one can derive the spectrum
from (\ref{shape}) and (\ref{alg2}) if normalizable zero modes
of the annihilation operator ${\bf a}^-$ exist. We stress that this algebraic
method is very powerful now since the explicit form of the potential
is known only in terms of Painleve transcendents.

The equation for zero modes of ${\bf a}^-$ reads:
\be
{\bf a}^- \Psi^{(0)}_k = k^+p^-\Psi^{(0)}_k = 0  \label{zero}
\ee
where $k$ labels the normalizable solutions.

In accordance to the non-linear SUSY algebra (\ref{nonl3}),
the equation for eigenvalues $E^{(0)}_k $ is,
\be
E^{(0)}\cdot \bigl[(E^{(0)} + a - 2\lambda)^2 + d\bigr]=0 ,
\label{energy}
\ee
and has at most three real solutions.

Concerning the generalized Heisenberg algebra \cite{ferhus2}, the modification in respect to (\ref{alg}) in the
previous Subsection is given by,
\be
[ {\bf a}^+, {\bf a}^-] = -2\lambda \bigl( 3(h^{(0)})^2 +(4 a - 2\lambda)h^{(0)} + a^2 + d
\bigr)
\label{alg2}
\ee
and similarly for $h^{(1)}.$

\subsection{Spectrum patterns for third order
shape-invariance}
We give now a short description of typical spectrum patterns (the details can be found in \cite{acin2000}), \cite{iof2004}. Below on, the case of equal asymptotics of f(x) at $\pm\infty$ is discussed.
\begin{itemize}
\item Three normalizable zero modes of $ {\bf a}^{-} $ and,
respectively, three equidistant sequences of levels may arise only for
$ \lambda > \sqrt{-d} $\, if\, $-\sqrt{-d}>a >\sqrt{-d}-2\lambda $ \,
or \,
$2\lambda -\sqrt{-d} \geq a > \sqrt{-d}.$ The corresponding solutions
of Painleve-IV equation must have the leading asymptotics
$ f(x)\sim -\lambda x/3 $ with subleading oscillations
$ cos \bigl(\lambda x^{2}/\sqrt{3} + o(x) \bigr). $

\item Two normalizable zero modes of $ {\bf a}^{-} $ (and two sequences of
levels) may exist for different solutions of Painleve-IV equation.
\begin{itemize}
\item Namely, for $ \lambda >\sqrt{-d} $ the solution with asymptotics
$ f(x)\sim -\lambda x/3 $ provides the fall off of
$ \Psi^{(0)}_{-}(x), \, \Psi^{(0)}_{+}(x) $ if
$ a < \sqrt{-d}-2\lambda  $ and of
$ \Psi^{(0)}_{0}(x), \, \Psi^{(0)}_{+}(x) $ if
$ 2\lambda +\sqrt{-d}> a >2\lambda -\sqrt{-d}. $
\item For $ \lambda >\sqrt{-d} $ the solution with asymptotics
$ f(x)\sim -\sqrt{-d}/2\lambda x $ generates two sequences of levels
starting from $\Psi^{(0)}_{0}(x), \, \Psi^{(0)}_{-}(x) $ if
$ 2\lambda n +\sqrt{-d}< - a <2\lambda (n+1)-\sqrt{-d}; \, n=0,1,2... $
\item For $ \lambda < \sqrt{-d} $ two sequences (from $\Psi^{(0)}_{0}(x)$
and $\Psi^{(0)}_{+}(x)$) are generated for two possible asymptotics
$ f(x)\sim\sqrt{-d}/2\lambda x $ and $ f(x)\sim -\lambda x/3 $ if
$ 2\lambda +\sqrt{-d}> a >\sqrt{-d} . $
\end{itemize}
\item For $ a \geq 2\lambda +\sqrt{-d} $ and arbitrary positive
$ \lambda $
one sequence of equidistant levels will be realized
with ground state $ \Psi^{(0)}_{0}(x) $ and one of three asymptotics:
$ f(x)\sim -\lambda x/3, \,\, \sim\pm\sqrt{-d}/2\lambda x .$
\begin{itemize}\item For
$ f(x)\sim -\lambda x/3 $ only one sequence of levels
will be realized also for $ \lambda=\sqrt{-d}, \,\, a =\pm\sqrt{-d} $
(starting from $ \Psi^{(0)}_{0}(x) $), for $ \lambda =\sqrt{-d}, \,\,
a < \sqrt{-d} $ (starting from $ \Psi^{(0)}_{-}(x) $) and for
$ \lambda <\sqrt{-d},\,\, a \leq 2\lambda +\sqrt{-d} $
(starting from $ \Psi^{(0)}_{+}(x) $).
\item For asymptotic behavior
$ f(x)\sim -\sqrt{-d}/2\lambda x $ and $ \lambda >\sqrt{-d} $ the
spectrum consists of one sequence of levels if
$ |a +2\lambda n|<\sqrt{-d},\,\,n=0,1,2,... $ (starts from
$ \Psi^{(0)}_{-}(x) $) or if $ a = -\sqrt{-d} $ (starts from
$ \Psi^{(0)}_{0} $).
\item Last, for $ f(x)\sim \sqrt{-d}/2\lambda x $
one obtains one sequence with ground state $ \Psi^{(0)}_{+}$ if
$ \lambda\leq\sqrt{-d} $ and $ a <\sqrt{-d} $ or if
$ \lambda >\sqrt{-d} $ and $ |a +2\lambda n| < \sqrt{-d},\,\,
n=0,1,2,... $
\end{itemize}
For specific values of parameter $ a$ the
normalizable zero mode of $ {\bf a}^{+} $ can cause  truncation of one of two
sequences ( with ground state $ \Psi^{(0)}_{0}(x) $ ).

\item Let us also
remark that particular value $ \alpha = -2\lambda -\sqrt{-d},$ {\it i.e.}
$ E^{(0)}_{+}=0, E^{(0)}_{-}<0, $
just corresponds to the case when Painleve-IV equation (\ref{Painleve+})
has a class of particular solutions which coincide with solutions
of the Riccati equation $ g'(y)=g^{2}(y)+2yg(y)+\sqrt{b}.$ Substituting
this Riccati equation into  Eq.(\ref{VV}) one finds that potential
is the pure harmonic oscillator: $ V_{1}(x)=\lambda^{2}x^{2}-\lambda .$
\item One additional singlet state  satisfies the equation
$$ {\bf a}^{+}\Psi^{(0)}_{0}(x) = {\bf a}^{-}\Psi^{(0)}_{0}(x) = 0 .$$
For $ \lambda > \sqrt{-d}\quad ( \lambda < \sqrt{-d} ) $
it occurs when $ a = \pm\sqrt{-d} $ and
$ \Psi^{(0)}_{0}(x) = \Psi_{2, 1}(x) $ which entails the equation:
\be
f'(x) = -2f^{2}(x) - 2\lambda xf(x) \mp \sqrt{-d} \label{equ}
\ee
with asymptotic behavior $ f(x) \sim \mp \sqrt{-d}/2\lambda x . $
The spacing between the two ground states
$ \Psi^{(0)}_{0}(x) $ and $ \Psi^{(0)}_{-}(x) $ is:
$ \Delta E = 2\lambda \mp 2\sqrt{-d} .$

\item The doublet representation
$ \bigl( \Psi^{(0)}_{0}(x), \,\, {\bf a}^{+} \Psi^{(0)}_{0}(x) \bigr) $
of the spectrum generating algebra (\ref{shape}), (\ref{alg2}) is
built on solutions of the equation:
$$
({\bf a}^{+})^{2}\Psi^{(0)}_{0}(x) = {\bf a}^{-} \Psi^{(0)}_{0}(x) = 0 .
$$
It may hold when $ a = -2\lambda - \sqrt{-d}$ for arbitrary
positive value of $ \lambda $ and when $ a = -2\lambda + \sqrt{-d}$
for $ \lambda > \sqrt{-d}$. It is equivalent to
$ {\bf a}^{+}\Psi^{(0)}_{0}(x) = \Psi_{1, 2}(x) ,$ which is satisfied
when $ f(x) $ obeys the following equation:
\ba
& &8\lambda f^{2}(x)\bigl( f'(x) + 2f^{2}(x) + 2\lambda x f(x) -
2\lambda \mp \sqrt{-d} \bigr) =
\bigl( f'(x) + 2f^{2}(x) +  \nonumber\\
& &2\lambda x f(x) \mp \sqrt{-d} \bigr)\cdot
\biggl[\, \bigl( f'(x) + 2f^{2}(x) + 2\lambda x f(x) -2\lambda
\mp \sqrt{-d} \bigr) \nonumber\\
& &\bigl( -f'(x) + 2f^{2}(x) + 2\lambda x f(x) -2\lambda
\mp \sqrt{-d} \bigr) - 4\lambda (\lambda \pm \sqrt{-d}) \biggr] .
\label{*}
\ea
One can show that all solutions of this equation fulfill the Painleve-IV
equation (\ref{painleve}). These solutions have the asymptotics
$ f(x) \sim \pm \sqrt{-d}/2\lambda x $ and cannot have any (pole)
singularity. The spectrum consists of a doublet $ (0,\,\,2\lambda) $
and infinite sequence
$ E_{n}=\pm2\sqrt{-d} + 2(n+2)\lambda , \,\, n=0,1,2... $
\end{itemize}
Recently  some higher-order generalizations of shape-invariance have been elaborated leading to the Painleve-V etc. equations\cite{painhigh},\cite{carbal}. As well new solutions of Painleve-IV with complex parameters based on the third-order shape invariance have been found \cite{ferncomp}.
\section{Nonlinear SUSY for non-stationary Schr\"odinger equations}
\subsection{Linear SUSY  and
hidden symmetry}
\hspace*{3ex}  In this Section our aim is to elucidate that many of the nonlinear SUSY
constructions illustrated before can be implemented also in the Schr\"odinger
time dependent framework \cite{bagsa96}, \cite{tdi}, \cite{bagsam1}, \cite{Finkel98}.

The first- and higher-order
intertwining operators and the corresponding relations between non-stationary
one-dimensional Schr\"odinger operators can be introduced straightforwardly.
But,  as compared to the stationary case, already the first-order
intertwining relations  imply some hidden
symmetry \footnote{Symmetry properties of the time dependent Schr\"odinger equation
were studied in general in \cite{boyer}, \cite{nikitin} (and references therein).} which leads to a specific quantum dynamics when the
evolution is described by quantum orbits and results
in the $R$-separation of variables\cite{tdi}.
In turn second-order intertwining operators \cite{tdi},\cite{Beckers:1998np}
and the corresponding non-linear SUSY  give rise to the quantum motion
governed by the spectrum generating algebras.
We start with the {\em
  non-stationary Schr\"odinger operator}
\begin{equation}
 {\cal S}[V]=i\partial_t+\partial^2_x - V(x,t)\;.
  \label{S}
\end{equation}
Here $\partial_t=\partial/\partial t$ and
$\partial_x=\partial/\partial x$ denote the partial derivatives with respect
to time and position.  When they are applied to some function $f$, the following notations for these derivatives are used: $\dot{f}(x,t)=(\partial_t f)(x,t)$ and $f'(x,t)=(\partial_x
f)(x,t)$.

The intertwining operator of first order \cite{tdi} is given by
\begin{equation}
  \label{q+}
  q_t^+ =\xi_0(x,t)\partial_t + \xi_1(x,t)\partial_x + \xi_2(x,t)
\end{equation}
with, in general, complex-valued functions $\xi_0,\xi_1$ and $\xi_2$.The
possibility of a
complexification of the intertwining (Darboux) (first and also higher order)
operator
was emphasized by \cite{Beckers:1998np}.

For the  Schr\"odinger operator (\ref{S})
the intertwining relation reads
\begin{equation}
  \label{inS}
  {\cal S}[V^{(0)}]q_t^+ = q_t^+ {\cal S}[V^{(1)}]\;,
\end{equation}
where the functions $\xi_i$ ($i=0,1,2$) and  $V^{(0),(1)}$
are  not independent of each other. It can be also represented in the
SUSY form Eqs.~\gl{2times2},\gl{algebra} when the stationary Hamiltonians $h^{(0),(1)}$
are extended to the Schr\"odinger operators ${\cal S}[V^{(0),(1)}]$, then
\be
[\hat{\cal S}_t, Q^-_t ] = 0, \quad  \hat{\cal S}_t = \left(\begin{array}{cc}
{\cal S}[V^{(0)}]& 0\\
0 & {\cal S}[V^{(1)}]
\end{array}\right),\quad  Q^-_t = \left(\begin{array}{cc}
 0& q_t^+\\
0 & 0
\end{array}\right). \label{susytime}
\ee
After inserting  (\ref{S}) and the
intertwining operator (\ref{q+}) into relation (\ref{inS}) it can be
derived that
$\xi_0$ and $\xi_1$  may depend only on time, i.e.\ $\xi_0'=0=\xi_1'$. Then the
assumption
that $\xi_0$ does not vanish identically would entail
that the
potential difference $V^{(0)}- V^{(1)}$ does depend on time only. This is a rather
trivial case
and therefore let us take $\xi_0\equiv 0$ .
Making the appropriate choice of variables
$\xi_1(t)=e^{i\beta(t)}\rho(t)$ and
$\xi_2(x,t)=e^{i\beta(t)}\rho(t)\omega'(x,t)$ with real $\beta$,
positive $\rho$ and  complex $\omega$ functions one finds
\begin{equation}
  \label{V1andV2:1}
  \begin{array}{l}
   V^{(0)}(x,t)=\omega'\,^2(x,t)+\omega''(x,t) - i \dot{\omega}(x,t) +
   \alpha(t)-\dot{\beta}(t)+i\dot{\rho}(t)/\rho(t)\;,\\[2mm]
   V^{(1)}(x,t)=\omega'\,^2(x,t)-\omega''(x,t) - i \dot{\omega}(x,t) + \alpha(t)\;,
  \end{array}
\end{equation}
where $\alpha$ is a time-dependent complex-valued integration constant.
One may set \cite{tdi}
$\beta\equiv 0$ without loss of generality. Furthermore, one may also remove
$\alpha\rightarrow 0$ as being absorbed in $\omega$ by the
shift
$\omega\to \omega -i\int dt\,\alpha$. Finally, we are left with
\begin{equation}
  \label{V1andV2:2}
  \begin{array}{l}
   V^{(0)}(x,t)=\omega'\,^2(x,t)+\omega''(x,t) - i \dot{\omega}(x,t) +
   i\dot{\rho}(t)/\rho(t)\;,\\[2mm]
   V^{(1)}(x,t)=\omega'\,^2(x,t)-\omega''(x,t) - i \dot{\omega}(x,t)\;.
  \end{array}
\end{equation}
Here the super-potential $\omega$ is not arbitrary as the
potentials are assumed to be real. It could  be achieved
by taking a stationary real super-potential. However this choice leads to
the stationary SUSY QM discussed previously . Therefore, we
consider a complex super-potential
\begin{equation}
  \omega(x,t)= \omega_R(x,t) + i \omega_I(x,t)
\label{chi}
\end{equation}
with real functions $\omega_R$ and $\omega_I$. The reality condition
  $\mbox{Im } V^{(0)}=\mbox{Im }V^{(1)}=0$ is fulfilled if
\begin{equation}
  \label{h_and_g}
  2(\omega_R)''+\dot{\rho}/\rho=0\;,\qquad 2(\omega_R)'(\omega_I)'-
(\omega_I)''-\dot{\omega_R}=0\;,
\end{equation}
which can be integrated to
\begin{equation}
  \begin{array}{rcl}
\omega_I(x,t)&=&\displaystyle
      -\frac{1}{4}\,\frac{\dot{\rho}(t)}{\rho(t)}\,x^2+
            \frac{1}{2}\,\rho(t)\dot{\mu}(t)x +\gamma(t)\;, \\
  \omega_R(x,t)&=&\displaystyle
      \frac{1}{2}\,\ln\rho(t)+K\Bigl(x/\rho(t)+\mu(t)\Bigr)\;,
  \end{array}\label{g_and_h}
\end{equation}
where $\mu$ and $\gamma$ are arbitrary real functions of time and $K$ is
an arbitrary real function of the variable $y=x/\rho +\mu$. In terms of
these functions the final form of the two partner potentials is
\begin{eqnarray}
\fl V^{(0),(1)} (x,t)=\frac{1}{\rho^{2}(t)}\left[K'^2(y)\pm K''(y)\right]
\nonumber\\-
  \frac{\ddot{\rho}(t)}{4\rho (t)}\,x^{2}+\left(\dot{\rho}(t)\dot{\mu}(t)+
  \frac{\rho(t)\ddot{\mu}(t)}{2}\right)x-\frac{\rho^2(t)\dot{\mu}^2(t)}{4}
  +\dot{\gamma}(t)
\label{V_1/2}
\end{eqnarray}
and the intertwining operator reads
\begin{equation}
  q_t^ +(x,t)=\rho(t)\partial_x+K'\Bigl(x/\rho(t)+\mu(t)\Bigr)-
  \frac{i}{2}\left(\dot{\rho}(t)x-\rho^2(t)\dot{\mu}(t)\right).
\label{q+_2}
\end{equation}
Let us demonstrate that the non-stationary Schr\"odinger
equation ${\cal S}[V^{(0),(1)}]\psi^{(0),(1)} =0$
with potentials given in Eq.~(\ref{V_1/2}) (which is equivalent to the
intertwining (\ref{inS})) admits a separation of variables. Indeed, after the transformation
\begin{equation}
\label{trafo}
\fl  y=x/\rho(t) + \mu(t),\quad
  \psi^{(0),(1)}(x,t)=\frac{1}{\sqrt{\rho(t)}}\,e^{-i\omega_I(x,t)}\phi^{(0),(1)}(y,t)
\equiv \Omega (x,t) \phi^{(0),(1)}(y,t)
\end{equation}
this Schr\"odinger equation becomes quasi-stationary \cite{Bluman96}
\begin{equation}
  i\rho^{2}(t)\partial_{t} \phi^{(0),(1)}(y,t)=
  \left[-\partial_{y}^{2} + K'^{2}(y) \pm K''(y)\right]\phi^{(0),(1)}(y,t)\;,
\label{SEiny}
\end{equation}
which is obviously separable in $y$ and $t$. Hence, the solutions of the
original Schr\"odinger equations have the general form
$\psi(x,t)=\Omega(y,t)Y(y)T(t)$ which
is known as the $R$-separation of variables \cite{miller}. In other words,
for any pair of
Schr\"odinger operators ${\cal S}[V^{(0),(1)}]$, which admits a first-order intertwining
relation (\ref{inS}) there exists a transformation (\ref{trafo}) to some new
coordinate in which the potentials become stationary (see also \cite{Finkel98}).

 This $R$-separation of variables is certainly
related to the existence of a symmetry operator for the
super-Hamiltonian.
First,
one can directly verify the adjoint intertwining relation for  real potentials,
\begin{equation}
  q_t^ -{\cal S}[V^{(0)}]={\cal S}[V^{(1)}]q_t^ -
\label{inSadjoint}
\end{equation}
where
\begin{equation}\fl
q_t^ -\equiv(q_t^ +)^\dagger=
  -\rho(t)\partial_x+K'\Bigl(x/\rho(t)+\mu(t)\Bigr)+
  \frac{i}{2}\left(\dot{\rho}(t)x-\rho^2(t)\dot{\mu}(t)\right).
\end{equation}

Then from (\ref{inS}), \gl{susytime} and (\ref{inSadjoint}) we obtain the closure of
the SUSY algebra,
\begin{equation}
\left\{Q_t, \bar Q_t\right\} = {\cal R}_t,\quad
  \Bigl[\hat{\cal S}_t, {\cal R}_t\Bigr]=0\;,\quad \bar Q_t =
  (Q_t)^\dagger\, ,
\end{equation}
where the symmetry operator ${\cal R}_t$ has the following components
\ba
\fl R_t^\pm = q_t^\pm q_t^\mp =  -\rho^2(t)\partial^2_x +
\frac{i}{2}\left(\rho \dot{\rho}(t)\{x,\partial_x\}-
2\rho^3(t)\dot{\mu}(t)\partial_x\right)\nonumber\\ \fl +
\left[
  K'\Bigl(x/\rho(t)+\mu(t)\Bigr)\right]^2
 \pm K''\Bigl(x/\rho(t)+\mu(t)\Bigr) + \frac{1}{4}
\left(\dot{\rho}(t)x-\rho^2(t)\dot{\mu}(t)\right)^2\nonumber\\\fl
\lo= \exp\{-i\omega_I (x,t)\} \left[-\partial_{y}^{2} +
  K'^{2}(y) \pm K''(y)\right] \exp\{i\omega_I (x,t)\}.
\ea
Thus the quasi-stationary Hamiltonians in Eq.~\gl{SEiny} are just
unitary
equivalent to the symmetry operators $R^\pm_t$. It means that the
supersymmetry entails the separation of variables because it provides a
new symmetry for the super-Hamiltonian. As consequence the quantum
dynamics splits in orbits with a given eigenvalue of the symmetry operator.


\subsection{Second-order intertwining for stationary potentials:
spectrum generating algebra}

\hspace*{3ex} Let us relate a
pair of Schr\"odinger operators ${\cal S}[V^{(0)}]$ and ${\cal S}[V^{(1)}]$ by second-order
(intertwining) operators of the form,
\begin{equation}
  q_t^{+}(x,t) = G(x,t)\partial_{x}^{2} - 2F(x,t)\partial_{x} + B(x,t)\;.
\label{old18}
\end{equation}
We will explore the existence of the time-dependent SUSY charges with
appearance of the spectrum generating (oscillator like) algebras for
corresponding Hamiltonians.

As in the first-order case it can be shown \cite{tdi}
that the inclusion of an additional
term of first order in $ \partial_{t} $ leads to the trivial solutions
when the difference $V^{(0)}- V^{(1)}$ depends on the time $t$ only. Furthermore, from the intertwining relation
(\ref{inS}) with the above $q_t^ +$ one can conclude that the function $G$ cannot
depend on $x$ and it is even
possible to exclude a phase. In other words, without loss of generality
 $G(x,t)\equiv g(t)$ and accordingly, from now on, an intertwining operator can be reduced to,
\begin{equation}
  q_t^{+}(x,t) = g(t)\partial_{x}^{2} - 2F(x,t)\partial_{x} + B(x,t)\;.
\label{18+}
\end{equation}
In \cite{tdi} particular solutions of the intertwining
relation (\ref{inS})  were constructed with $q_t^ {+}$ as given above .
In this section we shall analyze
 the solutions
of the intertwining relation (\ref{inS}) for the case where both potentials
$ V^{(0)}$ and $V^{(1)}$ are stationary, {\it i.e.} do not depend on $t$.

One class of such solutions is known from
\cite{acdi95}. Assuming a supercharge $ q_t^{+} $
with real coefficient functions independent on $t$, one finds that
the corresponding solutions of (\ref{inS})  coincide with those of
the stationary intertwining relations $(-\partial^2_x + V^{(0)}(x))q^{+}(x) =
q^{+}(x)(-\partial^2_x + V^{(1)}(x))$.

Here we are interested in more general solutions of (\ref{inS}) when
operators $ q_t^{+}$ depend  on $t$,
\begin{equation}
  (i\partial_t -h^{(0)})q_t^+(x,t)=q_t^+(x,t) (i\partial_t -h^{(1)})\;,
\label{inH}
\end{equation}
with stationary Hamiltonians $h^{(0),(1)}=-\partial_x^2+V^{(0),(1)}(x)$ but
time-dependent intertwining operators.

Let us employ the suitable ansatz with simple $ t$-dependence in (\ref{18+}),
\begin{equation}
  q_t^ {+}(x,t) = k^{+}(x) + A(t) p^{+}(x)\;,
  \label{20}
\end{equation}
where,
\begin{equation}
  k^{+}(x)\equiv \partial_{x}^{2} - 2f(x)\partial_{x} + \tilde b(x)- f' (x) ,\qquad
  p^{+}(x)\equiv - \partial_{x} + \chi(x) . \label{20a}
\end{equation}
Here all functions besides $A$ are considered to be real.  We
  also assume $A\not\equiv 0$.
With this ansatz the intertwining relation (\ref{inH}) can be shown \cite{tdi}
to yield,
\begin{equation}
  \begin{array}{l}
i\dot A=2\,\tilde m + 2\,m\, A\;,\\[2mm]
k^{+}h^{(0)} - h^{(1)}k^{+} -  = 2\,\tilde m\, p^{+}\;,\\[2mm]
h^{(0)}p^{+} - p^{+}h^{(1)} = 2\, m\, p^{+}\;,
  \end{array}\label{25}
\end{equation}
with real constants $ \tilde m$ and $m$.

We find it interesting to focus on the case $m\neq 0$ to explore
certain spectrum generating algebras.
 The first equation in (\ref{25}) immediately leads to
\be
A(t) = m_0e^{-2imt} - \tilde m/m \label{time}
\ee
with a real $m_0$ and,
\begin{equation}\fl
  q_t^ {+}(x,t) = \partial_{x}^{2} - \biggl(2f(x) + \frac{\tilde m}{m}
\biggr)\partial_{x} + \tilde b(x) -  f'(x) + \frac{\tilde m}{m}\,\chi(x) - m_0e^{-2imt} p^{+}(x)\;.
\end{equation}
Without loss of generality we may set $\tilde m = 0$
because a non-vanishing $\tilde{m}$ may always be absorbed via a proper
redefinition of $f$ and $\tilde b$, {\it i.e.} of the operator $k^+$.

As a consequence, the second relation in
(\ref{25}) leads to a second-order intertwining between $h^{(0)}$ and $h^{(1)}$.
This has already been considered in \cite{acdi95} and it was found that
the potentials $V^{(0)}$, $V^{(1)}$ and the function $\tilde b$ can be expressed in terms of
$f$ and two arbitrary real constants $a$ and $d$ as in Subsect.5.1, Eq.\gl{Kint},
\begin{equation}
  \begin{array}{l}
\displaystyle
V^{(0),(1)}(x) = \mp 2 f'(x) + f^{2}(x) + \frac{f''(x)}{2f(x)}
         - \frac{f'^{\, 2}(x)}{4f^{2}(x)}- \frac{d}{4f^{2}(x)} - a\;,\\[2mm]
\displaystyle
\tilde b(x) = f^{2}(x) - \frac{f''(x)}{2f(x)}
       + \frac{f'^{\, 2}(x)}{4f^{2}(x)} + \frac{d}{4f^{2}(x)}\; .
  \end{array}\label{27}
\end{equation}
The corresponding second-order SUSY algebra generated by the
supercharge $K$ is similar to \gl{secor}.

One may find some similarities between the present intertwining
algebra \gl{25} and the extended SUSY relations discussed in Section
5. But we emphasize that now for $m\not= 0$ the last relation in
\gl{25} does not generate a second SUSY. Rather it creates the
equivalence of relatively
shifted spectra of two Hamiltonians $h^{(0)}$ and $h^{(1)}$ which is typical for
spectrum generating algebras emerging for the shape-invariant Hamiltonians (Sect.6). Specifically,
\be
p^+ p^- = h^{(0)} - m + c;\quad p^- p^+ = h^{(1)} + m +c, \label{shifted}
\ee
where $c$ is a real constant.
Therefore the
reflectionless potentials found in Section 5 are produced only in the limit
of $m = 0$.

The genuine spectrum generating algebra for stationary Hamiltonians
$h^{(0),(1)}$ can be derived from Eq.~\gl{25}
\begin{equation}
  \begin{array}{l}
[h^{(0)}, {\bf a}_{(0)}^-] = - 2m {\bf a}_{(0)}^- \;,\quad {\bf a}_{(0)}^-  \equiv
k^{+} p^{-}\, ,\\[2mm]
[h^{(0)}, {\bf a}_{(0)}^+] =  2m {\bf a}_{(0)}^+\;,\quad {\bf a}_{(0)}^+ \equiv p^{+}
k^- \, , \\[2mm]
[h^{(1)},  {\bf a}_{(1)}^- ] = 2m  {\bf a}_{(1)}^-  \;,\quad {\bf a}_{(1)}^- \equiv
k^{-}p^{+} \, , \\[2mm]
[h^{(1)},{\bf a}_{(1)}^+] = - 2m {\bf a}_{(1)}^+ \;, \quad {\bf a}_{(1)}^+ \equiv p^{-}
k^+ \, ,
  \end{array}\label{sga}
\end{equation}
where $p^-=(p^+)^\dagger$, $k^-=(k^+)^\dagger$ and ${\bf a}_{(0,1)}^- = ( {\bf a}_{(0,1)}^+)^\dagger $.

The closure of this spectrum generating algebra  is a polynomial
deformation of Heisenberg algebra \cite{ferhus2} (see Subsect. 6.4),
\be
[{\bf a}_{(0,1)}^+ , {\bf a}_{(0,1)}^- ] = P^{(0,1)} (h^{(0,1)}).
\ee
The explicit form of the polynomials $P^{(0,1}$ can be obtained with
the help of Eqs.~\gl{25} and \gl{shifted} for $\tilde m = 0$. For
instance,
\ba
{\bf a}_{(0)}^+  {\bf a}_{(0)}^- =  (h^{(0)} -m +c)\, \Big[(h^{(0)} -2m +a)^2 +d\Big];\nonumber\\
{\bf a}_{(0)}^-  {\bf a}_{(0)}^+=  (h^{(0)} +m +c)\, \Big[(h^{(0)} +a)^2 +d\Big] , \label{gener1}
\ea
where the notations from Eqs.~\gl{secor} and \gl{shifted} are
employed.
The polynomials $P^{(0,1)}$
turn out to be different for the isospectral partners
$h^\pm$,
\ba
\fl P^{(0)} (h^{(0)}) = - 6m (h^{(0)})^2 + 4m(2m - 2a -c)h^{(0)} -2m[a^2 + d +
2(a-m)(c-m)];\nonumber\\
\fl P^{(1)} (h^{(1)}) = - 6m (h^{(1)})^2 - 4m(2m + 2a +c) h^{(1)} -2m[a^2 + d +
2(a+m)(c+m)] . \label{gener2}
\ea
Hence the two spectrum generating algebras are, in general, different
that is essentially due to the shift in intertwining relations
\gl{shifted}. When comparing to the equations in Subsection 6.4 we find that the generating algebras  \gl{gener1}, \gl{gener2} are in fact originating from the third-order shape invariance.

There is a formal discrete symmetry between their constants and
Hamiltonians $h^{(0)}, a, c \Longrightarrow - h^{(1)}, - a, -c$.

 The intertwining
relation (\ref{inH}) and its adjoint give rise to the symmetry operators
$q_t^ {+}q_t^ {-}$ and $q_t^ {-} q_t^ {+}$ for $(i\partial_{t} - h^{(0)})$ and
$(i\partial_{t} - h^{(1)})$, respectively.
Using Eqs.~\gl{secor},\gl{time},\gl{shifted} and after elimination of
polynomials
of the Hamiltonians $h^{(0,1)}$
these symmetry
operators may be reduced to,
\begin{equation}
  \begin{array}{l}
R^{(0)}(x,t) =m_0\left[ e^{2imt} {\bf a}_{(0)}^- + e^{-2imt} {\bf a}_{(0)}^+ \right]\;,\\[2mm]
R^{(1)}(x,t) = m_0 \left[e^{2imt} {\bf a}_{(1)}^-  + e^{-2imt} {\bf a}_{(1)}^+ \right]\;.
  \end{array}\label{30}
\end{equation}
 As our potentials
do not depend on time the time derivatives $\dot R^{(0,1)}(x,t)$
of Hermitian symmetry operators $R^{(0,1)}(x,t)$ form the independent set of
hermitian symmetry operators which do not commute between themselves.
Similar
results have also been obtained in \cite{nikitin}
using a different approach. We see that the non-stationary SUSY
delivers the  time-dependent hidden symmetry operators \cite{Andrianov:2004vz},
\be
e^{2imt}{\bf a}_{(0,1)}^- = \frac{1}{2m_0} R^{(0,1)}(x,t) -\frac{i}{4m m_0} \dot R^{(0,1)}(x,t),
\ee
which give the entire set of
spectrum generating algebras previously found for the third-order shape invariance.  Thus hidden symmetries for nonstationary Schr\"odinger super-Hamiltonians give the universal framework for both stationary hidden symmetries and for spectrum generating algebras of shape invariant SUSY systems.
\section{Polynomial SUSY QM in $d=2$}

\subsection{Supercharges of second order in derivatives: generalities}

It was shown in Section 2, that multidimensional generalization of SUSY QM includes both scalar and matrix Hamiltonians, and some physical examples of matrix problems, incorporated into SUSY QM, were given in Subsections 2.2 and 2.5. In general, the construction does not provide any intertwining and therefore any direct relations between spectra and wave functions of two scalar components of super-Hamiltonian, if we are not interested in models with separation of variables. But such opportunity can be opened by applying and development of the ideas of polynomial SUSY of Section 3 for higher dimensionality of space. In particular, this approach turned out \cite{d5} -- \cite{d3}, \cite{IV-1}, \cite{d7}, \cite{IV-2} -- \cite{d10}, \cite{d12} -- \cite{d15} to be very fruitful for study of nontrivial spectral problems in the case of $d=2$ (see also \cite{ioffe1}, \cite{ioffe2}).

Thus, in order to get rid of matrix component of two-dimensional super-Hamiltonian, we shall explore
the supercharges of second order in partial derivatives.
The simplest variant of second order supercharges - of reducible form $q^{\pm}=(q_i^{\pm})(\tilde q_i^{\pm})$ -
is not very promising: the intertwined partner Hamiltonians differ by a trivial constant only, and both
of them admit the separation of variables (see details in \cite{d5}, \cite{d6}).
By this reason, here we are interested in general irreducible second order components of supercharges,
\ba
q^+ = g_{ik}{(\vec x)}\partial_i \partial_k + C_i(\vec x )\partial_i +
B(\vec x);\quad q^-\equiv (q^+)^{\dagger},
\label{gench}
\ea
with real "metrics" $g_{ik}(\vec x)$ and coefficient functions $C_i(\vec x), B(\vec x).$
The familiar intertwining relations
\be
h^{(0)}q^+=q^+h^{(1)};\quad q^-h^{(0)}=h^{(1)}q^-. \label{intertww}
\ee
for two scalar two-dimensional Hamiltonians of Schr\"odinger form,
\be
h^{(0), (1)}=-\Delta^{(2)}+V^{(0), (1)}(\vec x) \label{HHHH}
\ee
can be rewritten as a system of nonlinear partial differential equations for $g_{ik}(\vec x),$ the coefficient functions $C_i(\vec x),\, B(\vec x)$
and potentials $V^{(0)(1)}(\vec x).$ The general solution for the "metrics" $g_{ik}(\vec x)$ is \cite{d5}, \cite{d6},
\be\fl
g_{11} =a x_2^2 + a_1 x_2 +  b_1;
\,\,g_{22} = a x_1^2 + a_2 x_1 + b_2; \,\,g_{12}
=-\frac{1}{2}(2a x_1 x_2 + a_1 x_1 + a_2 x_2) + b_3,  \label{metrics}
\ee
where $(a, a_i, b_i = Const).$
Thus, the senior in derivative part of supercharges
belongs to the $E(2)$ - universal enveloping algebra \cite{miller}.
We distinguish four different (inequivalent) classes in second derivatives \cite{d5}, \cite{d6},
\ba
&&q^{(1)+} = \gamma (P_1^2+P_2^2) + C_i \partial_i + B; \label{4.6}\\
&&q^{(2)+} = \alpha P_1^2 + (\alpha + \gamma ) P_2^2 + C_i \partial_i + B; \label{4.7}\\
&&q^{(3)+} = \gamma \{J_3,P_1\} + \alpha (P_1^2+P_2^2) + C_i \partial_i + B; \label{4.8}\\
&&q^{(4)+} = \gamma J_3^2 + \beta P_1^2 + \alpha P_2^2 + C_i\partial_i + B, \label{4.9}
\ea
where $J_3$ and $P_1,\, P_2$ are generators of rotations and translations, correspondingly,
and $\gamma \not = 0$.

The solutions (\ref{metrics}) must be inserted into six other equations which followed from the intertwining relations (\ref{intertww}),
\ba
&&\fl\partial_iC_k(\vec x ) +
\partial_kC_i(\vec x) + \Delta^{(2)} g_{ik}(\vec x) - (V^{(0)}(\vec x) -
V^{(1)}(\vec x))g_{ik}(\vec x) = 0; \label{sys1}\\
&&\fl\Delta^{(2)}C_i(\vec x) + 2\partial_iB(\vec x) + 2 g_{ik}(\vec x
)\partial_k V^{(1)}(\vec x) - (V^{(0)}(\vec x) - V^{(1)}(\vec x
))C_i(\vec x)=0;\label{sys2}\\ &&\fl\Delta^{(2)} B(\vec x) +
g_{ik}(\vec x)\partial_k\partial_i V^{(1)}(\vec x) + C_i(\vec x
)\partial_i V^{(1)}(\vec x) - (V^{(0)}(\vec x) - V^{(1)}(\vec x
))B(\vec x) = 0.\label{sys3}
\ea
It is clear that we have no chances to solve these equations with general form (\ref{metrics}) of $g_{ik}(\vec x).$
Therefore, one is forced to consider the simplest cases of constant metrics $g_{ik}$
looking for particular solutions for functions $C_i(\vec x),\, B(\vec x),\, V^{(0)}(\vec x),\, V^{(1)}(\vec x).$

The first sample of constant metrics is the elliptic (Laplace) one corresponding to (\ref{4.6}), $g_{ik}=\delta_{ik}.$
In this case, the system (\ref{sys1}) - (\ref{sys3}) is
essentially simplified, so that all coefficient functions can be found analytically \cite{d5}, \cite{d6}.
In particular, the combination $C\equiv C_1+iC_2$ depends only on $z=x_1+ix_2,$ and it has the specific form,
\begin{equation}\label{C}
C^2(z)=az^2+bz+c,
\end{equation}
where $a$ is real and $b,\, c$ complex constants. All other functions can be also found explicitly, but depending on values of constants in (\ref{C}), both potentials allow $R-$separation \cite{miller} of variables - in elliptic (for $b=0,\, a\neq 0$), parabolic (for $a=0,\, b\neq 0$) or polar ($b=c=0,\, a\neq 0$) coordinates. The situation with both $a\neq 0,\, b\neq0$ is reduced to the first variant by a suitable constant shift of coordinates. The case of polar coordinates just corresponds to reducible supercharges $q_i^{\pm}\tilde q_i^{\pm}.$ Thus, for all possible values of coefficients the class of two-dimensional problems with elliptic metrics of supercharges can be reduced to two one-dimensional models, and it will not be considered below.

\subsection{Hyperbolic (Lorentz) metrics}

\subsubsection{Supercharges with twisted reducibility.}

The constant metrics of hyperbolic (Lorentz) type, $g_{ik}=diag(+1, -1)$ gives much more interesting models \cite{d6}, \cite{d2}, \cite{d1}. One specific class of such solutions of intertwining relations (\ref{intertww}) with the so called supercharges with twisted reducibility.
\begin{equation}\label{twist}
q^-=(q^+)^{\dagger}=q_l^+(\sigma_3)_{lk}\tilde q_k^-=(-\partial_l+\partial_l W(\vec x))(\sigma_3)_{lk}(+\partial_k+
  \partial_k\widetilde W(\vec x)),
\end{equation}
was studied in papers \cite{IV-1}, \cite{IV-2}, \cite{IV-3}. In (\ref{twist}), $W (\vec x), \widetilde W(\vec x)$ are two different functions (superpotentials), $\sigma_3$ is the Pauli matrix, and summation over $k,l=1,2$ is implied.
This form of supercharges can be considered as a generalization of the simplest reducibility for elliptic metrics, mentioned in the beginning of
previous Subsection. In contrast to that case, the form (\ref{twist}) leads to very nontrivial models, which do not allow the standard separation of variables. This construction of supercharges can be interpreted as obtained after gluing of two systems of two-dimensional SUSY QM of Subsection 2.1 with first order supercharges with different superpotentials $W,\, \widetilde W,$ correspondingly. These systems are glued by their matrix components of super-Hamiltonian: these components are taken coinciding but up to an unitary constant matrix rotation by $\sigma_3.$ In such a case, both Hamiltonians $h^{(0)}=\tilde q_l^+\tilde q^-_l$ and $h^{(1)}=q^+_lq_l^-$ are quasifactorized in terms of $\tilde q_i^{\pm}$ and $q_i^{\pm},$ correspondingly.

It was shown in \cite{IV-1}, \cite{IV-2} in a general form that
Eqs.(\ref{intertww}) and (\ref{twist}) lead to the following representation
for superpotentials $W, \tilde W$ in terms of four functions
$\mu_{1,2,\pm},$
$$ \begin{array}{ll}
\chi & =  \mu_1(x_1) + \mu_2(x_2)+ \mu_+(x_+) + \mu_-(x_-), \nonumber\\
\widetilde{\chi} & =  \mu_1(x_1) + \mu_2(x_2)- \mu_+(x_+) -
\mu_-(x_-),\nonumber
\end{array} $$
where the light cone variables $x_\pm = x_1 \pm x_2$ were introduced. These functions $\mu_{1,2,\pm}$ have to satisfy the following equation,
\be
\mu'_1(x_1)\left[ \mu'_+(x_+) + \mu'_-(x_-) \right] + \mu'_2(x_2) \left[
\mu'_+(x_+) - \mu'_-(x_-) \right] = 0, \label{mumu}
\ee
where $\mu'(x)$ means derivative over the argument.
By substitutions $\phi \equiv \mu',$ it is evident that we deal with a purely functional equation with no derivatives,
\begin{equation}
\phi_1(x_1)\left[ \phi_+(x_+) + \phi_-(x_-) \right] = - \phi_2(x_2) \left[
\phi_+(x_+) - \phi_-(x_-) \right]. \label{phimain}
\end{equation}

Each solution of Eq.(\ref{phimain}) gives the corresponding solution of the intertwining relations (\ref{intertww}) for the potentials $V$,
$\widetilde{V}$ and for the supercharges $q^{\pm},$
\begin{eqnarray}
 &&\fl V^{(0)}(\vec x)=\Bigl(\phi_1^2(x_1) - \phi'_1(x_1)\Bigr) + \Bigl(\phi_2^2(x_2)
- \phi'_2(x_2)\Bigr) + \Bigl(\phi_+^2(x_+) + \sqrt{2}\phi'_+(x_+)\Bigr) +\nonumber\\
 &&\fl\phantom{ V^{(0)}(\vec x)=}+\Bigl(\phi_-^2(x_-) + \sqrt{2}\phi'_-(x_-)\Bigr),\label{potphiold1}\\
&&\fl V^{(1)}(\vec x)= \Bigl(\phi_1^2(x_1) - \phi'_1(x_1)\Bigr) + \Bigl(\phi_2^2(x_2)
- \phi'_2(x_2)\Bigr) + \Bigl(\phi_+^2(x_+) - \sqrt{2}\phi'_+(x_+)\Bigr) +\nonumber\\
&&\fl\phantom{ V^{(0)}(\vec x)=}+\Bigl(\phi_-^2(x_-) - \sqrt{2}\phi'_-(x_-)\Bigr),\label{potphiold2}\\
&&\fl Q^{\pm} = \partial_1^2 - \partial_2^2 \pm \sqrt{2}\Bigl(\phi_+(x_+) +
\phi_-(x_-)\Bigr)\partial_1 \mp\sqrt{2} \Bigl(\phi_+(x_+) -
\phi_-(x_-)\Bigr)\partial_2 - \nonumber\\
&&\fl\phantom{Q^{\pm} =} -\Bigl(\phi_1^2(x_1) - \phi'_1(x_1)\Bigr) + \Bigl(\phi_2^2(x_2) -
\phi'_2(x_2)\Bigr) +2\phi_+(x_+)\phi_-(x_-).\nonumber
\end{eqnarray}
Thus, in order to find the systems with intertwining (\ref{intertww}) by
supercharges of the form (\ref{twist}), it is necessary to solve
(\ref{phimain}). This equation seems to be rather complicated, but it
appeared to be solvable in a general form (see details in \cite{IV-3}).
In particular, it was shown that $\phi_{1,2}(x)$ are defined from solutions of the first order nonlinear differential equation,
\begin{equation}
({\phi}')^2_{1,2}=a\phi_{1,2}^4 + b\phi_{1,2}^2 + c, \label{ell}
\end{equation}
where $a,b,c$ are arbitrary real constants. All solutions of this equation
can be expressed in terms of elliptic functions, and they are described
for different ranges of parameters, for example, in Appendix B of \cite{perelomov}. In turn,
the functions $\phi_{\pm}$ are defined uniquely by solution of (\ref{ell}): they also satisfy
equation of the form (\ref{ell}), but with different constant coefficients. Depending on the values of $a, b, c,$
solutions $\phi$ can be functions of their argument with finite or infinite period. Corresponding models were studied in detail in \cite{IV-3} (periodic potentials) and in \cite{IV-1}, \cite{IV-2} (potentials with infinite period).

\subsubsection{Irreducible supercharges.}

Let us consider now the most general supercharges with hyperbolic (Lorentz) metrics without any kind of reducibility \cite{d5}, \cite{d6}, \cite{d2}, \cite{ioffe1}. The light cone variables $x_{\pm}\equiv x_1\pm x_2$ are again the most convenient to look for the solution of intertwining relations. In particular, it follows from the system (\ref{sys1}) - (\ref{sys3}) that the combinations $C_{\pm}\equiv C_1 \mp C_2$ depend only on one argument:
\be
C_+=C_+(x_+),\quad C_-=C_-(x_-). \label{Cx}
\ee
Then, the whole system (\ref{sys1}) - (\ref{sys3}) can be rewritten in a very compact form of two equations for functions $C_{\pm}(x_{\pm})$ and an auxiliary function $F(\vec x)$,
\ba
&&\partial_-(C_- F) = -\partial_+(C_+ F);\label{first}\\
&&\partial_+^2 F =
\partial_-^2 F.\label{second}
\ea
It is easy to write the general solution $F(\vec x)$ of Eq.(\ref{second}): it is expressed in terms of two arbitrary functions,
\be
 F=F_{1}(x_{+}+x_{-}) + F_{2}(x_{+}-x_{-}),\label{F}
 \ee
both are defined up to an arbitrary real constant, $F_1\rightarrow F_1+\gamma ,\, F_2\rightarrow F_2-\gamma .$
The potentials $ V^{(1),(2)}(\vec x)$ and the function $ B(\vec x) $ are expressed in terms of four functions $F_1(2x_1),\,F_2(2x_2)$ and $C_{\pm}(x_{\pm}),$ which must satisfy the only equation (\ref{first}),
\ba
&&\fl V^{(0),(1)}=\pm\frac{1}{2}(C_+' + C_-') + \frac{1}{8}(C_+^2 + C_-^2) + \frac{1}{4}\biggl( F_2(x_+-x_-) - F_1(x_+ + x_-)\biggr);\label{potential}\\
&&\fl B=\frac{1}{4}\biggl( C_+ C_- + F_1(x_+ + x_-) + F_2(x_+ - x_-)\biggr), \nonumber
\ea
where $C'$ means derivative in its argument. Although Eq.(\ref{first}) has a compact form, it is very nontrivial being the functional-differential equation: the functions each depend on specific argument. As a rule, equations of this kind have no regular recipe for solution, and this is the case for Eq.(\ref{first}). Usually, one may try to solve such equation starting from different suitable anzatses for functions $C_{\pm}(x_{\pm}), F(\vec x)$.

1) Let us choose $C_- = 0,$ then from (\ref{first}) one obtains
$F=\phi (x_-) / C_+(x_+).$ After inserting into Eq.(\ref{second})
the separation of variables is possible, and particular solution
reads,
\ba
C_+(x_+)& =& \frac{1}{\delta_1 \exp(\sqrt{\lambda}
\cdot x_+) + \delta_2 \exp(-\sqrt{\lambda}\cdot
x_+)};\nonumber\\
F_{1,2}(2x)& =& \delta_1\sigma_{1,2}
\exp(2\sqrt{\lambda} x) + \delta_2\sigma_{2,1}
\exp(-2\sqrt{\lambda} x),\nonumber
\ea
where the Greek letters -- arbitrary constants: depending on sign of $\lambda$ they may be real/complex.

2) Let us consider the anzats with factorized function $F(\vec x),$
$F = F_+(x_+)\cdot F_-(x_-)$. Then from Eq.(\ref{first}),
\be
C_{\pm} = \frac{\nu_{\pm}}{F_{\pm}} \pm \frac{\gamma}{F_{\pm}}\int
\limits^{x_{\pm}}F_{\pm}dx'_{\pm},\nonumber
\ee
and there are two options to fulfill the condition (\ref{second}), i.e. $F(\vec x) = F_1(2x_1)+F_2(2x_2)$,
\be
\fl a)\quad F_{\pm}(x_{\pm}) =
\epsilon_{\pm}x_{\pm},\qquad b)\quad F_{\pm} = \sigma_{\pm}
\exp(\sqrt{\lambda}\cdot x_{\pm}) + \delta_{\pm}
\exp(-\sqrt{\lambda}\cdot x_{\pm}).\nonumber
\ee
Corresponding potentials can be found according to
Eq.(\ref{potential}), being similar to ones obtained in
\cite{hietarinta} in quite different approach.

3) Let us start now from the general solution of (\ref{first}),
\be
F = L\biggl(\int\frac{dx_+}{C_+} - \int\frac{dx_-}{C_-}\biggr)/
(C_+C_-).\label{10}
\ee
Then Eq.(\ref{second}) gives the
functional-differential equation for the functional $ L(A_+-A_-)$
with $A^{\prime}_{\pm}\equiv 1/C_{\pm}(x_{\pm}),$
\be
\fl \biggl(\frac{A_+'''}{A_+'} - \frac{A_-'''}{A_-'}\biggr)L(A_+ -
A_-) + 3 (A_+'' + A_-'')L'(A_+ - A_-) + (A_+'^2 - A_-'^2)L''(A_+ -
A_-) = 0, \label{11}
\ee
where $ L^{\prime} $ denotes the
derivative of $ L $ with respect to its argument.
If we take functions $ A_{\pm} $ such that
$A_{{\pm}}'' = \lambda^2 A_{\pm},\,\lambda = \mbox{const},$
Eq.(\ref{11}) will become ordinary differential
equation for $L$ with independent variable $(A_+-A_-).$
It can be easily solved,
\ba
L(A_+ - A_-) = \alpha
(A_+ - A_-)^{-2} + \beta,\nonumber
\ea
where $A_{\pm}=\sigma_{\pm}exp(\lambda x_{\pm}) +
\delta_{\pm}exp(-\lambda x_{\pm})$ with $
\sigma_{+}\cdot\delta_{+} = \sigma_{-}\cdot\delta_{-}$ and $
\alpha , \beta $ - real constants. For $ \lambda^{2}>0, $
choosing $\sigma_{\pm}=-\delta_{\pm}=k/2$ or
$\sigma_{\pm}=+\delta_{\pm}=k/2,$
we obtain (up to an arbitrary shift in $ x_{\pm} $) two
particular solutions,
\ba
3a)\quad A_{\pm} = k \sinh (\lambda x_{\pm}),\quad 3b)\quad
A_{\pm} = k \cosh (\lambda x_{\pm}).\nonumber
\ea
Then (\ref{10}) leads to,
\ba
&&\fl 3a)\quad F_1(2x) = \frac{k_1}{\cosh^2(\lambda x)} + k_2\cosh(2\lambda x);\label{12}\\
&&\fl\phantom{ 3a)\quad } F_2(2x) = \frac{k_1}{\sinh^2(\lambda x)} + k_2\cosh(2\lambda x);\quad
C_{\pm} = \frac{k}{\cosh(\lambda x_{\pm})},\quad
k\not=0,\nonumber\\
&&\fl 3b)\quad F_1(2x) = - F_2(2x) =
\frac{k_1}{\sinh^2(\lambda x)} + k_2\sinh^2(\lambda x),\quad
C_{\pm} = \frac{k}{\sinh(\lambda x_{\pm})},\quad
k\not=0.\label{13}
\ea
For $ \lambda^{2}<0 $ hyperbolic functions
must be substituted by trigonometric ones.

We have to remark that the case $\lambda^2=0,$ i.e. $A_{\pm}^{\prime\prime}=0,$
is not of interest, leading to trivial superpartners. However, choosing
in (\ref{13}) $\lambda\to 0, k,\, k_1,\, k_2^{-1}\,\to 0 $ simultaneously,
so that $\lambda^2\sim k_1\sim k_2^{-1}\sim k^2,$ we obtain the solution,
\be
F_1(2x) = - F_2(2x) = \tilde k_1 x^{-2} + \tilde k_2 x^2, \quad
C_{\pm} = \frac{\tilde k}{x_{\pm}}. \label{14}
\ee
One can check that (\ref{first}) is also satisfied by
\be
F_1(2x) = - F_2(2x) = k_1 x^2 + k_2 x^4, \quad C_{\pm} =
\pm\frac{k}{x_{\pm}}. \label{15}
\ee

4) Starting again from (\ref{10}), it is convenient to pass on to
new variable functions $ C_{\pm}\equiv\pm f_{\pm}/f_{\pm}^{\prime}.$
Then $ F $ in (\ref{10}) is represented
in the form $ F=U(f_{+}f_{-})f_{+}^{\prime}f_{-}^{\prime}$ with an
arbitrary\footnote{Due to Eq.(\ref{second}), the function $F$ should
be additionally representable in the form $F = F_1(2x_1)+F_2(2x_2)$.}
function $ U.$ After substitution in (\ref{second}) one obtains
the functional-differential equation,
\ba\fl
(f_+'^2 f_-^2 - f_+^2 f_-'^2)U''(f) + 3 f
\biggl(\frac{f_+''}{f_+} - \frac{f_-''}{f_-}\biggr)U'(f) +
\biggl(\frac{f_+'''}{f_+'} - \frac{f_-'''}{f_-'}\biggr)U(f) = 0,
\quad f\equiv f_+f_-.\nonumber
\ea
For particular form of functions $f_{\pm}=\alpha_{\pm}exp(\lambda x_{\pm}) +
\beta_{\pm}exp(-\lambda x_{\pm}),$ this equation becomes
an ordinary differential equation for $U$ with independent
variable $f.$ Its solution is $ U=a+4bf_{+}f_{-} $
$( a,b -$real constants). Then functions
\ba
&& \fl F_1(x) =
k_1(\alpha_+\alpha_-\exp(\lambda x) + \beta_+\beta_-\exp(-\lambda
x))\nonumber\\
&&\fl \phantom{F_1(x) =}+ k_2(\alpha_+^2\alpha_-^2\exp(2\lambda x) +
\beta_+^2\beta_-^2\exp(-2\lambda x)),\nonumber\\
&&\fl -F_2(x) =
k_1(\alpha_+\beta_-\exp(\lambda x) + \beta_+\alpha_-\exp(-\lambda
x)) \nonumber\\
&&\fl \phantom{F_1(x) =} +
 k_2(\alpha_+^2\beta_-^2\exp(2\lambda x) +
\beta_+^2\alpha_-^2\exp(-2\lambda x)),\nonumber\\
&&\fl C_{\pm} = \pm
\frac{\alpha_{\pm}\exp(\lambda x_{\pm}) + \beta_{\pm}\exp(-\lambda
x_{\pm})} {\lambda(\alpha_{\pm}\exp(\lambda x_{\pm}) -
\beta_{\pm}\exp(-\lambda x_{\pm}))} \label{16}
\ea
(with $k_1\equiv a\lambda^2, \, k_2\equiv 4b\lambda^2$)
are real solutions of (\ref{first}), (\ref{second}), if $
\alpha_{\pm}, \beta_{\pm} $ are real for $ \lambda^{2}>0,$ and $
\alpha_{\pm}=\beta_{\pm}^{*} $ for $ \lambda^{2}<0.$

5) To find a next class of solutions it is useful to rewrite (\ref{first}) in terms of $ x_{1,2},$
\ba\fl
(F_1(2x_1) + F_2(2x_2))\partial_1(C_+ +
C_-) + F_1'(2x_1)(C_+ + C_-) + F_2'(2x_2)(C_+ - C_-) = 0.\nonumber
\ea
Among known particular solutions the most compact one is,
\be
C_+(x) = C_-(x) = a x^2 + c,\quad F_1(2x_1) = 0,\quad F_2(2x_2) =
\frac{b^2}{x_2^2}.\label{18}
\ee

After inserting these solutions (\ref{12}) - (\ref{18}) into the general formulas (\ref{potential}), one obtains
the analytical expressions for potentials. Their explicit form can be found in \cite{d2}.

Two additional classes of particular solutions of the system (\ref{sys1}) - (\ref{sys3}) were obtained analogously by means of suitable ansatzes for the case of degenerate metrics $g_{ik}=\mbox{\rm diag}(1,0)$ ( see \cite{d2}) and the case of deformed hyperbolic metrics $g_{ik}=\mbox{\rm diag}(1, -a^2)$ with $a \neq 0$ (see \cite{d9}).

\subsection{Integrability}

Similarly to the case of one-dimensional Polynomial SUSY Quantum Mechanics (Section 3), partner two-dimensional Hamiltonians $h^{(0)}$ and $h^{(1)}$
can be considered as components of the diagonal super-Hamiltonian $H,$ which are intertwined by components $q^{\pm}$ of the off-diagonal  supercharges $Q^{\pm}.$ Thus, the operators $H,\, Q^{\pm}$ constitute the deformed SUSY algebra: both supercharges are nilpotent operators, and they commute with the super-Hamiltonian. The difference is in the third relation of SUSY algebra: while in one-dimensional case anticommutator of $Q^{\pm}$ produced the second order polynomial of $H,$ in two-dimensional situation it gives some fourth order diagonal operator \cite{d5}, \cite{d6}, \cite{d1},
\be
\{Q^+, \, Q^-\}= \hat R;\quad \hat R =\left(
                                                  \begin{array}{cc}
                                                    R^{(0)} & 0 \\
                                                    0 & R^{(1)} \\
                                                  \end{array}
                                                \right)=\left(
                                                          \begin{array}{cc}
                                                            q^+q^- & 0 \\
                                                            0 & q^-q^+ \\
                                                          \end{array}
                                                        \right).
                                                \label{R}
\ee
In general, $R^{(0)},\, R^{(1)}$ are not expressed in terms of $h^{(0)},\, h^{(1)}$ in this case, but nevertheless they are related. Indeed,
as it follows directly from the intertwining relations (\ref{intertww}), they are the fourth order symmetry operators of corresponding systems,
\begin{equation}\label{symm}
    [h^{(0)},\, R^{(0)}] = [h^{(1)},\, R^{(1)}] = 0.
\end{equation}
In general, these symmetry operators might be expressed as functions of $h^{(i)},\, i=0,1,$ similarly to the one-dimensional case. But investigation \cite{d6}, \cite{d2}, \cite{d1} shows that this is not the case: two opportunities can be realized depending on the metrics $g_{ik}$ in the supercharges.

1) For the case of elliptic (Laplacian) metrics $g_{ik}=\delta_{ik}$ (and only for this one), the fourth order operators $R^{(i)},\, i=0,1$ are expressed in terms of Hamiltonians $h^{(i)}$ and the second order operators $\widetilde R^{(i)},$ which also commute with $h^{(i)},$
\be
R^{(i)}=(h^{(i)})^2+\eta h^{(i)} + \widetilde R^{(i)},\quad [h^{(i)},\, \widetilde R^{(i)}] = 0
\label{HR}
\ee
with $\eta = Const .$ The explicit forms of differential operators $R^{(i)}$ and of corresponding Hamiltonians $h^{(i)}$ depend on the values of constants in (\ref{C}), and they are given in \cite{d6}, \cite{d1}. It was shown that these Hamiltonians allow the standard procedure of $R-$separation of variables \cite{miller} in polar, parabolic or elliptic coordinates.

2) For all other allowed metrics (\ref{metrics}), the fourth order operators $R^{(i)},\, i=0,1$ can not be reduced to any symmetry operators
of lower order. This was shown in the general form in \cite{d6}, and it follows also from the results of L.P.Eisenhart \cite{eisenhart}, where the exhaustive list of systems which are amenable to separation of variables was given (just in Cartesian, polar, elliptic and parabolic coordinates).

Thus, for all possible metrics (\ref{metrics}) each member of the variety of two-dimensional Hamiltonians $h^{(i)},$ which satisfy the intertwining relations (\ref{intertww}), has the symmetry operator $R^{(i)},$ and therefore is completely integrable. In general, this fact does not provide the solvability of the models, but some of them are partially or completely solvable (see the next Section).

In papers \cite{d1}, \cite{d2}, the classical limit of Polynomial two-dimensional SUSY QM was considered. In particular, the prescription how to look for the integrals of motion (classical analogues of symmetry operators) of fourth order in momenta for given classical Hamiltonian was formulated. Namely, it is necessary to find such the complex classical functions $q^+_{cl}(\vec x, \vec p) = (q^-_{cl}(\vec x, \vec p))^{\dagger}$ in the phase space, that they satisfy the following relations:
\be
\{q^+_{cl}, h_{cl}\}=if(\vec x)q^+_{cl};\quad \{q^-_{cl}, h_{cl}\}=-if(\vec x)q^-_{cl},
\label{class}
\ee
where $\{\cdot ,\cdot \}$ are Poisson brackets, and $f(\vec x) -$ an arbitrary real function of $\vec x.$ If such real function exists, the classical Hamiltonian $h_{cl}$ obeys the factorizable integral of motion $I_{cl}=q^+_{cl}q^-_{cl}$ of fourth order in momenta,
\be
\{h_{cl}, I_{cl}\}=0.
\label{clin}
\ee
We see that the identification of classical integral of motion from the quantum SUSY algebra is unambiguous. However the opposite is not always true: a quantum anomaly may arise \cite{kliplyu00}.
\section{SUSY separation of variables for $d=2$}

The idea of shape invariance was very productive in one-dimensional SUSY Quantum Mechanics (see Section 6 and references therein). In the present Section, the two-dimensional analogue of that shape invariance will be exploited effectively \cite{d3}, \cite{IV-1}, \cite{ioffe1}, \cite{d7}, \cite{d10}, \cite{ioffe2}, \cite{d12}, \cite{d13}, \cite{d14}, \cite{d15} for two-dimensional Hamiltonians which satisfy the supersymmetrical intertwining relations (\ref{intertww}). In contrast to one-dimensional case, there are essential differences in $d=2$: in general, the ground state of $h^{(1)}(a)$ does not coincide with zero modes of second order supercharge $q^+(a),$ the value $E_0(a)\neq 0$, and also many zero modes of $q^+$ exist. Up to these crucial differences, all other actions can be repeated. It was shown \cite{d3}, \cite{IV-1}, \cite{ioffe1}, \cite{d15} that such direct generalization of the procedure above may provide only partial (quasi-exact) solvability of the model (see the Subsection 5.2). But for some models \cite{d10}, \cite{d12}, \cite{ioffe2}, \cite{d14}, \cite{kliplyu00}, \cite{0105135} it will help to solve the model completely (see Subsection 5.3).

The idea to consider the quasi-exactly-solvable models - the intermediate class between exactly solvable and unsolvable analytically models - was introduced on 80-ties in \cite{razavy} -- \cite{kamran}. In particular, the series of papers by A.Turbiner, A.Ushveridze and M.Shifman was devoted to the elegant algebraic method of construction of one-dimensional quasi-exactly-solvable (and sometimes, of exactly solvable) quantum models. In general, this method works beyond the supersymmetry, but both approaches can be combined in one-dimensional case as in \cite{shifman3}. This approach is applicable to two-dimensional problems as well, but only in curved spaces with nontrivial metrics \cite{shifman1}.

The realization of such opportunities to solve non-trivial two-dimensional models partially, or even completely, would be very important since only one regular method to solve analytically the Schr\"odinger equation for two-dimensional models is known: reduction to a pair of one-dimensional problems by means of the procedure of separation of variables \cite{miller}. This method can be used for very restrictive class of models, and full classification of models which allowed separation of variables was given by L.P.Eisenhart \cite{eisenhart}: four possibilities exist - Cartesian, polar, elliptic and parabolic coordinates. The general form of potentials amenable to separation of variables is also known explicitly up to arbitrary functions of one variable. And analytical solution is possible only if these functions belong to the list of exactly solvable one-dimensional potentials. All these Hamiltonians $H$ are integrable: the symmetry operator $R$ of second order in derivatives (in momenta) exists, $[H, R]=0.$
Besides models amenable to separation of variables, the class of so called Calogero-like models \cite{calogero}, \cite{ghosh} is known as well. They describe the specific forms of pairwise interaction of $N$ particles on a line, and they are solvable by means of special transformation of variables which leads to a separation of variables. Recently, new classes of solvable two-dimensional models were built in \cite{ttw}, \cite{quesne2}, \cite{marquette}, but all these models are superintegrable and amenable to separation of variables.

In two following Subsections, we shall present two special procedures of SUSY-separation of variables in two-dimensional models, which do not allow to use the conventional separation of variables. Subsection 9.2 presents the first procedure of SUSY separation of variables where variables are separated in the supercharge. It leads to QES models, and the specific model of two-dimensional Morse potential illustrates this method. In Subsection 9.3 the second procedure of SUSY separation of variables is given where variables are separated in one of partner Hamiltonians. In the case of the same Morse model, but with particular values of parameter, it allows to solve the model completely, i.e. to find analytically the whole spectrum and all wave functions.

\subsection{SUSY-separation of variables I: QES models}

The first variant of SUSY-separation of variables is realized when the Hamiltonians $h^{(0)},\, h^{(1)}$ do not allow standard separation of variables, but the supercharge $q^+$ does allow \cite{d3}, \cite{ioffe1}, \cite{IV-1}, \cite{IV-2}, \cite{IV-3}, \cite{ioffe2}, \cite{d15}.
The general scheme is the following. Let's suppose that we know zero modes of $q^+,$
$$
q^+\Omega_n (\vec x)=0;\quad n=0,1,...,N;\qquad
q^+ \vec\Omega (\vec x)=0.
$$
Due to intertwining relations (\ref{intertww}), the Hamiltonian $h^{(1)}$ obey the important property: the space of zero modes $\Omega_n$ is closed under the action of $h^{(1)},$
$$
h^{(1)}\vec\Omega (\vec x) = \hat C \vec\Omega (\vec x)
\nonumber
$$
with constant matrix $\hat C.$ If this matrix is known, and if it can be diagonalized,
$$
 \hat B \hat C = \hat\Lambda \hat B;\quad \hat\Lambda =
 diag(\lambda_0,\lambda_1,...,\lambda_N),
$$
the eigenvalues of $h^{(1)}$ can be found algebraically,
$$
h^{(1)} (\hat B\vec\Omega (\vec x)) = \hat\Lambda(\hat B\vec\Omega (\vec x)).
$$
Thus, to realize this scheme of construction some number of energy values and corresponding wave functions, we need:

- to find zero modes $\Omega_n(\vec x)$;

- to find constant matrix $B,$ such that $\hat B \hat C = \hat\Lambda \hat B.$

As for the first step, the zero modes can be obtained by using the special similarity transformation (not unitary!), which
removes the terms linear in derivatives from $q^+$,
$$
\tilde q^+ = e^{-\chi (\vec x)} q^+ e^{+\chi (\vec x)} =
 \partial_1^2 -
\partial_2^2 + \frac{1}{4}(F_1(2x_1) + F_2(2x_2));
$$
$$
\chi (\vec x) = -\frac{1}{4}\bigl( \int
C_+(x_+)dx_+ + \int C_-(x_-)dx_- \biggr). \nonumber
$$
Now, $\tilde q^+$ allows separation of variables for arbitrary solution of intertwining relations, and we obtain the first variant of new procedure - SUSY-separation of variables. Similarly to the conventional separation of variables, separation of variables in the operator $\tilde q^+$ itself
does not guarantee solvability of the problem.

The next task is to solve two one-dimensional problems,
\ba
(-\partial_1^2
-\frac{1}{4}F_1(2x_1))\eta_n(x_1)&=&\epsilon_n\eta_n(x_1); \nonumber\\
(-\partial_2^2
+\frac{1}{4}F_2(2x_2))\rho_n(x_2)&=&\epsilon_n\rho_n(x_2). \nonumber
\ea

Three remarks are appropriate now.

Remark 1: the same similarity transformation of $h^{(1)}$ does not lead to operator amenable to separation of variables.

Remark 2: the normalizability of $\Omega_n$ has to be studied attentively due to non-unitarity of the similarity transformation.

Remark 3: we have no reasons to expect exact solvability of the model, but quasi-exact-solvability can be predicted.

As for the matrix $\hat B,$ it must be found by some specific procedure. Such procedure was used in example which will be presented below.

In principle, the first scheme of $SUSY-$ separation of variables can be used for arbitrary models satisfying intertwining relations by supercharges with Lorentz metrics. The list of solutions of intertwining relations is already rather long (see Subsection 8.2.2), and it may increase in future.
The main obstacle is analytical solvability of one-dimensional equations, obtained after separation of variables in the operator $\tilde q^+.$

Below we describe briefly such a model which can be considered as the generalized two-dimensional Morse potential,
\ba
&&\fl C_+=4a\alpha;\quad C_-=4a\alpha\cdot\coth \frac{\alpha x_-}{2};\nonumber\\
&&\fl f_i(x_i) \equiv   \frac{1}{4}
F_i(2x_i)=-A\biggl(e^{-2\alpha x_i} - 2 e^{-\alpha
x_i}\biggr);\quad i=1,2
\nonumber\\
&&\fl V^{(0),(1)}= \alpha^2a(2a \mp 1)\sinh^{-2}\biggl(\frac{\alpha x_-}{2} \biggr) +
4a^2\alpha^2\nonumber\\
&&\fl\phantom{V^{(0),(1)}=}+ A \biggl[e^{-2\alpha
x_1}-2 e^{-\alpha x_1} + e^{-2\alpha x_2}-2 e^{-\alpha
x_2}\biggr];
\ea
where $A>0, \alpha >0, a-$real.

To explain the name, we present the potentials in the form,
$$V(\vec x)= V_{Morse}(x_1)+V_{Morse}(x_2)+ v(x_1,x_2),$$
where first two terms are just one-dimensional Morse potentials, and the last term mixes variables $x_1, x_2.$

The solutions of one-dimensional Schr\"odinger equations with Morse potentials $V_{Morse}(x)$ are well known \cite{landau}, and the zero modes can be written \cite{d3}, \cite{ioffe1} as,
\be\fl
\Omega_n(\vec x) = \biggl(\frac{\alpha}{\sqrt{A}}\cdot\frac{\xi_1\xi_2}{|\xi_2 -\xi_1|}\biggr)
^{2a}\exp(-\frac{\xi_1+\xi_2}{2}) (\xi_1\xi_2)^{s_n}
\cdot F(-n, 2s_n +1; \xi_1) F(-n, 2s_n +1; \xi_2);\nonumber
\ee
\be\fl
\xi_i\equiv  \frac{2\sqrt{A}}{\alpha}\exp(-\alpha x_i); \quad
s_n=\frac{\sqrt{A}}{\alpha}-n-\frac{1}{2} > 0.
\nonumber
\ee
The conditions of normalizability and of absence of the "fall to the center" are,
$$
a \in
(-\infty,\, -\frac{1}{4}-\frac{1}{4\sqrt{2}}); \quad s_n =
\frac{\sqrt{A}}{\alpha}-n-\frac{1}{2} > -2a >0
$$

To obtain the matrix $\hat C$ explicitly, one must act by $h^{(1)}$ on $\Omega_n.$ The matrix turns out to be triangular, and therefore, the energy eigenvalues coincide with its diagonal elements,
$$
E_k=c_{kk}=-2(2a\alpha^2s_k-\epsilon_k).
$$

To find a variety of wave functions is a more difficult task. For that it is necessary to
find all elements of $\hat C$ and all elements of matrix $\hat B.$
The recurrent procedure for the case of two-dimensional Morse potential was given in \cite{d3}, \cite{ioffe1}.
This variety can be enlarged by means of shape invariance property (\ref{shape1}) of the model, which can be easily checked.
Similarly to one-dimensional shape invariance, each wave function constructed by SUSY-separation of variables
leads to a set of additional wave functions, which can be written as,
\ba
&&h^{(0)}(a)\biggl[ q^-(a)q^-(a-\frac{1}{2})...q^-(a-\frac{M-1}{2})\Psi(a-\frac{M}{2}) \biggr] =
\nonumber\\
&&\biggl(E_0(a-\frac{M}{2})
+ {\cal R}(a-\frac{M-1}{2})+ ... + {\cal R}(a)\biggr)\nonumber\\
&&\times\biggl[ q^-(a)q^-(a-\frac{1}{2})...q^-(a-\frac{M-1}{2})\Psi(a-\frac{M}{2}) \biggr] \nonumber
\ea
with integer $M.$
Analogous approach works for the two-dimensional generalization of P\"oschl-Teller model \cite{IV-1}, for some two-dimensional periodic
potentials \cite{IV-3} and for the two-dimensional generalization of Scarf II potential \cite{d15}.

\subsection{SUSY-separation of variables II: Exact solvability.}

Among all known solutions of two-dimensional intertwining relations with second order supercharges
a subclass exists \cite{d10}, \cite{d12}, \cite{d8}, \cite{ioffe2}, \cite{d14}, where one of intertwined Hamiltonians $h^{(1)}$ is amenable to
standard separation of variables due to specific choice of parameters of the model. Its superpartner $h^{(0)}$ still does not allow conventional separation of variables.

The scheme will be described below for the same specific model - two-dimensional generalization of Morse potential,
\ba\fl
V^{(0),(1)}=
\alpha^2a(2a \mp 1)\sinh^{-2}\biggl(\frac{\alpha x_-}{2} \biggr) +
4a^2\alpha^2 + A \biggl[e^{-2\alpha
x_1}-2 e^{-\alpha x_1} + e^{-2\alpha x_2}-2 e^{-\alpha
x_2}\biggr]\nonumber
\ea
Let's choose $a_0=-1/2$ in order to vanish the mixed term in $V^{(1)},$ and therefore, the Hamiltonian $h^{(1)}$ allows the conventional separation of variables. Moreover, after separation of variables each of obtained one-dimensional problems is again exactly solvable: they coincide with one-dimensional problems of the previous Subsection, where they occured in a different context (separation of variables in $\tilde q$).

The discrete spectrum of this one-dimensional model is,
$$
\epsilon_n=-\alpha^2s_n^2;\quad s_n\equiv\frac{\sqrt{A}}{\alpha}-n-\frac{1}{2} >0;\quad n=0,1,2,\ldots ,
$$
Wave functions are expressed in terms of confluent hypergeometric functions,
$$
\eta_n(x_i) = \exp(-\frac{\xi_i}{2}) (\xi_i)^{s_n}
F(-n, 2s_n +1; \xi_i);\quad\,\,\xi_i\equiv \frac{2\sqrt{A}}{\alpha}\exp(-\alpha x_i).
$$

Due to separation of variables, the two-dimensional problem with $h^{(1)}(\vec x)$
is exactly solvable. Its energy eigenvalues are,
$$
E_{n,m}=E_{m,n}=\epsilon_n+\epsilon_m,
$$
being two-fold degenerate for $n\neq m.$
The corresponding eigenfunctions can be chosen as
symmetric or (for $n\neq m$) antisymmetric combinations,
$$
\Psi^{(1)\, S,A}_{E_{n,m}}(\vec x) = \eta_n(x_1)\eta_m(x_2)\pm\eta_m(x_1)\eta_n(x_2).
$$

Our initial aim here is to solve completely the problem for $h^{(0)}(\vec x)$ with $a_0=-1/2.$
The main tool is again the SUSY intertwining relations, i.e. isospectrality of
$h^{(0)}$ and $h^{(1)}$ but up to zero modes and singular properties of $q^{\pm}.$
In general, we may expect three kinds of levels of $h^{(0)}(\vec x),$

(i). The levels, which coincide with $E_{nm}.$ Their wave functions can be
obtained from $\Psi^{(1)}$ by means of $q^+.$

(ii). The levels, which were absent in the spectrum of $h^{(1)}(\vec x)$, if some wave functions
of $h^{(0)}(\vec x)$ are simultaneously the zero modes of $q^-.$ Then the second intertwining relation
would not give any partner state among bound states of $h^{(1)}(\vec x).$

(iii). The levels, which were also absent in the spectrum of $h^{(1)}(\vec x)$,
if some wave functions of $h^{(0)}(\vec x)$ become nonnormalizable after action of operator $q^-.$

We have to analyze these three classes of possible bound states of $h^{(0)}$ one after another.

{\bf (i).} The first SUSY intertwining relation gives the two-fold degenerate wave functions of $h^{(0)}$
with energies $E_{nm}, \,\,\Psi^{(0)}_{E_{nm}}=q^+\Psi^{(1)}_{E_{nm}}.$
But $q^+$ includes singularity on the line $x_1=x_2,$ therefore the
normalizability of $\Psi^{(0)}_{E_{n,m}}$ depends crucially on the behavior of
$\Psi^{(1)}_{E_{n,m}}$ on the line $\xi_1=\xi_2.$ One can check that only antisymmetric
functions $\Psi^{(1)}$ survive, i.e. only symmetric $\Psi^{(0)}$ survive.
This fact can be demonstrated \cite{d10} both by direct calculation and by indirect method - by means of
symmetry operator $R^{(0)}.$

The indirect method explores that the symmetry operator $R^{(0)}=q^-q^+$ for $a_0=-1/2$
can be written in terms of one-dimensional Morse parts $h_1(x_1),\,h_2(x_2)$ of the Hamiltonian $h^{(1)}=h_1(x_1)+h_2(x_2) ,$
$$
R^{(0)}=\biggl(h_1(x_1)-h_2(x_2)\biggr)^2 + 2\alpha^2\biggl(h_1(x_1)+h_2(x_2)\biggr)+\alpha^4.
$$
Therefore,
$$
R^{(0)}\Psi^{(0) A}_{E_{n,m}}(\vec x)=r_{n,m}\Psi^{A}_{E_{n,m}}(\vec x);\,
r_{n,m}=\alpha^4[(n-m)^2-1][(s_n+s_m)^2 - 1],
$$
and
$$
\|\Psi^{(1) S}_{E_{n,m}}\|^2=
\langle\Psi^{(0) A}_{E_{n,m}}\mid q^-q^+\mid \Psi^{(0) A}_{E_{n,m}}\rangle=
r_{n,m}\|\Psi^{(0) A}_{E_{n,m}}\|^2.
$$

For $n=m,$ wave functions $\Psi^{(0) S}_{E_{n,n}}$ vanish identically by trivial reasons.
It is clear now that wave functions $ \Psi^{(0) S}_{E_{n,n\pm 1}}$ also vanish.
For all other $n,m,$ functions $\Psi^{(0) S}_{E_{n,m}}$ have positive and finite norm, and there is no degeneracy of these levels.

{\bf (ii).} These possible bound states of $h^{(0)}$ are the
normalizable zero modes of $q^-.$ The variety of such zero modes is known from \cite{d3}: they exist only for positive values of $a$
$$
a \in (\frac{1}{4}+\frac{1}{4\sqrt{2}}\, ,\, +\infty ),
$$
which does not contain the value $a_0=-1/2.$ Thus, no normalizable bound states of this class exist for $h^{(0)}.$

{\bf (iii).} We have to study an opportunity that $q^-$ destroys normalizability of some
eigenfunctions of $h^{(0)}.$ It could occur due to singular character
of $q^-$ at $x_1 = x_2.$
The analysis was performed  \cite{d10} in suitable coordinates. It shows that
$q^-$ is not able to transform normalizable wave function
to nonnormalizable. Therefore, the third class of possible wave functions
$h^{(0)}$ does not exist too.

Summing up, the spectrum of $h^{(0)}$ with $a_0=-1/2$ consists only of the bound states with energies $E_{nm}$ for $|n-m|>1.$
This spectrum is bounded from above by the condition of positivity of $s_n, s_m:\,\, n,m < \sqrt{A}/\alpha - 1/2.$
The corresponding wave functions are obtained analytically \cite{d10}.

The results above can be expanded to the whole hierarchy of Morse potentials with $a_k=-(k+1)/2$ with $k=0,1,...$ by means of shape invariance property. Let's denote elements of the hierarchy as $h^{(0)}(\vec x; a_k),\, h^{(1)}(\vec x; a_k).$
All these Hamiltonians are also exactly solvable due to shape invariance of the model,
$$
h^{(0)}(\vec x; a_{k-1})=h^{(1)}(\vec x; a_{k});\,\, k=1,2,\ldots .
$$
This means that the following chain (hierarchy) of Hamiltonians can be built,
\be\fl
h^{(1)}(\vec x; a_0)\div h^{(0)}(\vec x; a_0)=h^{(1)}(\vec x; a_1)\div h^{(0)}(\vec x; a_1)=
\ldots \div h^{(1)}(\vec x; a_{k-1})=h^{(0)}(\vec x; a_{k})\div h^{(0)}(\vec x; a_{k}),\nonumber
\ee
where the sign $\div $ denotes intertwining by $q^{\pm}(a_i).$

In the general case, the functions
\be\fl
\Psi^{(0)}_{E_{n,m}}(\vec x; a_{k})=q^+(a_{k})\Psi^{(1)}_{E_{n,m}}(\vec x; a_{k})
=q^+(a_{k})q^+(a_{k-1})\ldots q^+(a_0)\Psi^{(1) A}_{E_{n,m}}(\vec x; a_0)\nonumber
\ee
(if normalizable) are the wave functions of $ h^{(0)}(\vec x; a_k)$ with energies $E_{n,m}=-\alpha^2(s_n^2+s_m^2).$
The symmetries of wave functions alternate and depend on the length of chain. This is true but up to zero modes of
operators $q^+.$

It is necessary to keep under the control normalizability of $\Psi$ and zero modes of $q^+.$
This control is performed algebraically by means of identity, which must be fulfilled up to a function of $H,$
$$R^{(1)}(a_{k})=R^{(0)}(a_{k-1}).$$
Actually, the following equation holds,
\be\fl
q^-(a_{k})q^+(a_{k})=q^+(a_{k-1})q^-(a_{k-1})+\alpha^2(2k+1)
\biggl[2h^{(0)}(\vec x; a_{k-1})+\alpha^2(2k^2+2k+1)\biggr].\nonumber
\ee
These relations allowed to evaluate the norms of wave functions.
The result is the following. The spectra of Hamiltonians
$h^{(0)}(\vec x; a_k)$ are not degenerate.
They consist of the bound states with energies $E_{n,m},$
with indices $|n-m|>k+2,$
and their wave functions $\Psi^{(0)}_{E_{n,m}}(\vec x; a_k)$ were given
analytically above.

The procedure of this Subsection was reproduced fully for two other two-dimensional models which are not amenable to standard separation: generalized P\"oschl-Teller \cite{d12} and Scarf II \cite{d14} potentials. The full energy spectra and corresponding wave functions were also built analytically by the second variant of SUSY-separation of variables.

\section{Perspectives, some applications and missing points}

\hspace*{3ex} The purpose of this review has been
to elucidate the recent progress in Nonlinear SUSY realization by intertwining Darboux
transformations for a community working on methods of isospectral design and their applications. We wanted also to
draw an attention to
new (quasi-)exactly solvable potential systems in two dimensions non-separable by a conventional choice of coordinates.
Summarizing our experience we describe
the general SUSY QM
as governed by the extended nonlinear SUSY algebra
with ${\cal N}$ pairs of nilpotent supercharges $Q_j,\,\, \bar Q_j = Q_j^t
= Q_j^\dagger $ and a number of hermitian hidden-symmetry differential
operators
$R_\alpha = R_\alpha^\dagger,\quad [R_\alpha, R_\beta] = 0;\quad 0\leq
\alpha,\beta \leq M$.

Not all the research areas that are linked to the subject of the present review have been covered in the main text.
A number of extensions and applications of Nonlinear SUSY QM have been found and still
there remains a plenty of open questions and challenges to be solved. Below we list some of these directions and mention few missing topics.
\begin{itemize}
\item
The method of multidimensional supersymmetric intertwining relations and shape invariance turned out to be useful for study of
different variants of Calogero-like models of $N$ particles on a line, including the models with internal degrees of freedom and
with pairwise interactions based on any root systems \cite{freedman}, \cite{minahan} -- \cite{neelov1} (see the recent review \cite{ghosh}).
\item
One-dimensional supersymmetric models with matrix superpotentials and supercharges \cite{matrix2} -- \cite{tanaka-matrix} have been used either for description of motion of spin particles
in external fields  \cite{pauli-eq} -- \cite{pauli-eq-2}, \cite{crom},  \cite{IKNN},  \cite{Cooper:wr} -- \cite{jakubsky} or to study scattering of particles with strong coupling of channels \cite{Cooper:wr},\cite{Cooper:wr-2}, \cite{matrix1} -- \cite{samsonov9} with practical application of the results to specific processes or to investigate spectral problem for the Bogoliubov-de Gennes system \cite{dunneplyu}.
Such systems are not fully investigated in searching of extended SUSY systems with hidden
symmetries. It is clear that  a comprehensive understanding of irreducible
building blocks for spectral design analogously to the scalar case is not yet achieved.
Also, matrix models with higher dimensionality of space have to be studied. To start with, two-dimensional models with matrix superpotential were studied in \cite{d-matrix}.
\item  A natural question appears in respect to the classical formulation of nonlinear SUSY whether a canonical quantization scheme similar to that of Subsect.2.3 exists in this case. A corresponding pseudoclassical Lagrangian for the particular one-dimensional model was observed in \cite{9903130} and investigated in a more general setting in \cite{kliplyu00},\cite{jakubs}.
It was found that quantization of one-dimensional nonlinear supersymmetry faces a problem of the quantum anomaly. Thus the problem arises to identify quantum anomaly in the two- and higher dimensional case. On the other hand the anomaly mentioned above is originated from noncommutativity of momenta and coordinates (ordering problem) which is unambiguously resolved in our recipe of supercharge identification and therefore does not affect any of the results presented in the review.

\item
Coherent states in the framework of SUSY QM models were built in \cite{samsonov2} -- \cite{hussin1}, but only for one-dimensional systems \cite{hussin2}.
\item
Formulation of multidimensional SUSY QM in arbitrary curvilinear coordinates was developed in \cite{aiz} (see also \cite{mateos1} -- \cite{iran}). This formalism could be useful for investigation of SUSY design of different cosmological and brain world models.
\item Generalization of Polynomial SUSY QM for higher (first of all, $d=3$) space dimensions seems to be very interesting (see \cite{d11}).
\item Effective mass SUSY QM is a rather popular generalization of conventional SUSY QM for the Schr\"odinger equation with
coordinate-dependent mass. This sort of models may have various applications in physics of solid state and nano devices as well as in cosmology. The well developed SUSY approach to effective mass QM is rather promising for solution of different problems in physics. Some of suitable references are \cite{quesne1} -- \cite{Midya:2012ti}.
\item The periodic potentials were studied actively in the framework of SUSY QM as well, see
\cite{braden} -- \cite{plyus13}, \cite{ferneg}  on one-dimensional models and \cite{IV-3} for  two-dimensional quasi-exactly-solvable models with periodic potentials. Certainly, spectral design of periodic potential systems is promising, especially, in two-dimensional QM related to new nanoscale materials like graphene.
\item The embedding of shape invariant potentials and their eigenfunctions into so called Discrete Quantum Mechanics  has been analyzed recently \cite{spirid-discr}, \cite{sasaki-discr}.
\item The generalizations of conventional SUSY QM to the models with complex potentials \cite{d4}, \cite{d8}, \cite{complex1} -- \cite{pseudo} provide an effective tool in study of non-Hermitian \cite{berry} -- \cite{Olkhovsky:2010zz} and, especially, PT symmetric  \cite{bender-complex} -- \cite{bender-complex-5} Quantum Mechanics with real energy spectrum (see reviews \cite{bender-rev}, \cite{mostafa-rev}). Spectral design of non-Hermitian systems with non-diagonalizable Hamiltonians was investigated both in one-dimensional \cite{AAA-non-diag,sigm1,sok-nond} and two-dimensional \cite{osc} cases. PT-symmetric periodic potentials with SUSY have been investigated in \cite{plyushPT}.
\item The polynomial SUSY in two dimensions has already brought a number of examples of new type of irreducible SUSY with hidden symmetries of higher-order in derivatives \cite{d5}, \cite{d6}. An interesting sample of nonlinear supersymmetry with holomorphic supercharges for a two-dimensional fermion in external magnetic field was studied in\cite{0105135}. One may expect a variety of new types of irreducible SUSY for third-order (and higher-order) supercharges as well as new discoveries in two and three dimensions.
\item The perspective direction for investigations is the interrelations between the inverse scattering problem and methods of SUSY QM \cite{mnieto} -- \cite{spabay}.
\item Hidden and exotic nonlinear supersymmetries have been found in some models as, for instance, in pure parabosonic systems \cite{9903130} and kink-antikink cristal \cite{mirror}. There are also some connections of nonlinear SUSY with Conformal SUSY mechanics \cite{leiva}.
\item The SUSY QM approach happens to be applicable and useful for building of different brane-world  models with extra space dimension \cite{AAA-soldati} -- \cite{Nagasawa:2011mu}.
\end{itemize}

\ack
Many results presented in this review were obtained together with our colleagues Nikolay V. Borisov, Francesco Cannata, Jean-Pierre Dedonder, 
Michael I. Eides, Sengul Kuru, Juan Mateos Guilarte, Georg Junker, Javier Negro, Luismi Nieto, David N. Nishnianidze, Vyacheslav P. Spiridonov. We are grateful to all of them for very fruitful collaboration. We thank also our former and present students Andrey V. Sokolov, Alexey I. Neelov\footnote{Deceased.}, Pavel A. Valinevich, Ekaterina V. Krupitskaya. A.A. acknowledges the partial financial support from projects FPA2010-20807, 2009SGR502, CPAN (Consolider CSD2007-00042). M.V.I. is grateful to Dr. V.V.Grigorian, whose skills have enabled him to complete successfully  this work.
\bigskip


\end{document}